\documentclass[fleqn,usenatbib]{mnras}

\usepackage[T1]{fontenc}

\DeclareRobustCommand{\VAN}[3]{#2}
\let\VANthebibliography\thebibliography
\def\thebibliography{\DeclareRobustCommand{\VAN}[3]{##3}\VANthebibliography}

\usepackage{graphicx}	% Including figure files
\usepackage{amsmath}	% Advanced maths commands
\usepackage{amssymb}	% Extra maths symbols

\usepackage{newtxtext,newtxmath}
\usepackage{multirow}
\usepackage{geometry}
\usepackage{caption}
\usepackage{rotating}
\usepackage{subcaption}
\usepackage[table]{xcolor}
\usepackage{setspace}

\newcommand{\lya}{Ly$\alpha$ }

\title[eBOSS DR16 prefers early reionization]{Tantalizing Evidence of Reionization Relics in the eBOSS DR16 Ly$\boldsymbol{\alpha}$ Forest Correlations: a Preference for Early Reionization}

\author[Zheng et al. ]{
Yifan Zheng,$^{1}$ Paulo Montero-Camacho,$^{2}$\thanks{E-mail: pmontero@pcl.ac.cn (PMC)} Zheng Cai$^{1,2}$ and Yi Mao$^{1}$\thanks{E-mail: ymao@tsinghua.edu.cn (YM)} 
\\
$^{1}$Department of Astronomy, Tsinghua University, Beijing 100084, China\\
$^{2}$Department of Mathematics and Theory, Peng Cheng Laboratory, Shenzhen, Guangdong 518066, China\\
}

\date{Accepted XXX. Received YYY; in original form ZZZ}

\pubyear{2025}

\begin{document}
\label{firstpage}
\pagerange{\pageref{firstpage}--\pageref{lastpage}}
\maketitle

\begin{abstract}
Cosmic reionization of \ion{H}{I} leaves enduring relics in the post-reionization intergalactic medium, potentially influencing the Lyman-$\alpha$ (Ly$\alpha$) forest down to redshifts as low as $z \approx 2$, which is the so-called ``memory of reionization'' effect. Here, we re-analyze the baryonic acoustic oscillation (BAO) measurements from \lya absorption and quasar correlations using data from the extended Baryonic Oscillation Spectroscopic Survey (eBOSS) Data Release 16 (DR16), incorporating for the first time the memory of reionization in the \lya forest. Three distinct scenarios of reionization timeline are considered in our analyses. We find that the recovered BAO parameters ($\alpha_\parallel$, $\alpha_\perp$) remain consistent with the original eBOSS DR16 analysis. However, models incorporating reionization relics provide a better fit to the data, with a tantalizing preference for early reionization, consistent with recent findings from the James Webb Space Telescope. Furthermore, the inclusion of reionization relics significantly impacts the non-BAO parameters. For instance, we report deviations of up to $3\sigma$ in the \lya redshift-space distortion parameter and $\sim7\sigma$ in the linear \lya bias for the late reionization scenario. Our findings suggest that the eBOSS \lya data is more accurately described by models that incorporate a broadband enhancement to the \lya forest power spectrum, highlighting the importance of accounting for reionization relics in cosmological analyses.  
\end{abstract}


%%%%%%%%%%%%%%%%%%%%%%%%%%%%%%%%%%%%%%%%%%%%%%%%%%

%%%%%%%%%%%%%%%%% BODY OF PAPER %%%%%%%%%%%%%%%%%%

\section{Introduction}
Cosmology has advanced at a rapid pace in recent decades, yet some major uncertainties remain. Among the most critical within the $\Lambda$CDM model are the nature of dark matter and dark energy, which together comprise approximately 95 percent of the Universe \citep{2020A&A...641A...6P}. Over the past two decades, the Lyman-$\alpha$ forest (Ly$\alpha$) has solidified itself as the current primary cosmological probe of the Universe in the $2 \lessapprox z \lessapprox 4$ range \citep{2000ApJ...543....1M,2002ApJ...581...20C}. It has contributed to advances in our understanding of dark matter \citep{2005PhRvD..71f3534V,2017JCAP...06..047Y,2020JCAP...04..038P} and dark energy \citep{2021PhRvD.103h3533A,2024arXiv240403002D}, further cementing its role in modern cosmological studies.

The \lya forest is composed of a series of absorption troughs occurring blueward of the \lya emission peak in the emitted radiation of high-redshift background quasars \citep[for a review, see][]{1998ARA&A..36..267R}. These absorption features arise from neutral hydrogen clouds in the intergalactic medium (IGM), which absorb \lya radiation in their rest frames. Since these clouds trace the underlying dark matter distribution, the \lya forest is then a biased tracer of the total matter distribution in the post-reionization era \citep{2004MNRAS.354..684V,2019JCAP...07..017C}. However, the forest's sensitivity to the thermal state of the IGM -- primarily through the abundance of neutral hydrogen -- makes it susceptible to astrophysical systematics that can modify the thermal evolution of the IGM at $z > 2$. 

Typically, the thermal evolution of the IGM is modeled by a tight temperature-density relation, expressed as a power law $T = T_0 \Delta^{\gamma - 1}$, where $T_0$ and $\gamma$ are free parameters \citep{1997MNRAS.292...27H}. This relation arises as a consequence of the scaling of the recombination rate with temperature, and it is facilitated by both adiabatic expansion and Compton cooling \citep{2016MNRAS.456...47M}\footnote{Also, see \cite{2024MNRAS.528.5845W} for the predominant role of the neutral hydrogen fraction on the temperature-density relation.}. For cosmological parameter inference, these thermal parameters must be marginalized over to account for uncertainties in the IGM's thermal state \cite[e.g.,][]{2019JCAP...07..017C}.  

The Dark Energy Spectroscopic Instrument \citep[DESI;][]{2022AJ....164..207A} is already transforming the landscape of observational cosmology \citep{2023arXiv230606308D,2023arXiv230606307D}, providing the astrophysics community with valuable \lya spectra for Baryon Acoustic Oscillations (BAO) studies and highly-anticipated broadband analyses \citep{2022arXiv220307491V}. Conservatively, the \lya forest can be used to measure the scale of the BAO signal \citep[e.g.,][]{2020ApJ...901..153D,2024arXiv240403001D}, usually by separating the analysis into a peak and a smooth component for the correlation function \citep[e.g.,][]{2014JCAP...05..027F}. Optimistically, the forest offers a wealth of additional cosmological information than just the BAO standard ruler \citep{2021MNRAS.506.5439C,2021MNRAS.508.1262M,2024arXiv240910617S}\footnote{Note that the dark matter constraints of \cite{2024arXiv240910617S} do not account for the evolution of \ion{He}{III} in the intergalactic medium; see the revised constraints in Figure 4 of \cite{2025arXiv250315595K}.} but so far limited access to shape information beyond the one-dimensional \lya power spectrum has been achieved\footnote{See \cite{2024JCAP...05..088A,2024MNRAS.533.3756D} for recent efforts aimed at unlocking the three-dimensional \lya forest power spectrum.}.  

Regardless of the desired observational program, to accurately model the \lya forest, several astrophysical and instrumental systematics need to be taken care of. For example, spectra with broad absorption lines \citep{2019ApJ...879...72G,2022MNRAS.511.3514E, 2023arXiv230405855A,2023arXiv230903434F}, continuum fitting \citep{2012AJ....143...51L,2023ApJS..269....4S}, spectra with damped Ly$\alpha$ systems \citep{2018MNRAS.476.1151P,2021MNRAS.507..704H,2022ApJS..259...28W}, relativistic effects \citep{2016JCAP...02..051I,2020JCAP...04..006L}, relics from \ion{H}{I} reionization \citep{2019MNRAS.487.1047M,2019MNRAS.490.3177W,2022MNRAS.509.6119M} and \ion{He}{II} reionization \citep{2013MNRAS.435.3169C,2017ApJ...841...87L,2020MNRAS.496.4372U,2024MNRAS.535.1035M}, self-shielding of minihalos \citep{2023arXiv230904129P}, and UV clustering \citep{2014PhRvD..89h3010P,2014MNRAS.442..187G,2019MNRAS.487.5346T,2023MNRAS.520..948L}. Moreover, instrumental systematics such as fiber collisions and sky residuals \citep{2024arXiv240403003G} also need to be corrected. 

Recent efforts have focused on using the \lya forest as much more than just a BAO probe, recovering some of the 3D shape information in the correlation function \citep[e.g.,][]{2021MNRAS.506.5439C,2022arXiv220912931C}, and highlighted the potential gains by combining the Alcock-Paczy\'nski effect \citep{1979Natur.281..358A} and BAO for cosmological analysis of the \lya correlation function \citep{2022arXiv220913942C,2023MNRAS.518.2567G}. However, access to the information stored in the broadband shape of the correlation function does not come for free. The main advantage of BAO analysis is that contaminating the location of the peak is difficult since the contaminant must rely on physical effects that have a preferred scale that aligns with the BAO \citep[e.g.,][]{2018MNRAS.474.2173H,2020PhRvD.102b3515G}. However, broadband cosmological information does not have the same protection. In particular, the Alcock-Paczy\'nski effect, for instance, can cause an increase or decrease in the amplitude of the monopole of the correlation function, which could be confused with the effect of \ion{H}{I} reionization relics  \citep[see Figure 4 of][]{2023MNRAS.520.4853M}. 

In this work, we explore an effect that can introduce broadband contamination to the \lya forest correlation function, namely the imprints of cosmic reionization that leave lasting relics in the post-reionization IGM, i.e. the so-called {\it memory of \ion{H}{I} reionization}. This effect leads to the breakdown of the temperature-density relation as a good approximation in the redshift range $2 < z < 4$ \citep{2018MNRAS.474.2173H,2019MNRAS.487.1047M,2020MNRAS.499.1640M}. As shown by \cite{2023MNRAS.520.4853M}, the reionization relics do change the broadband shape of the correlation function multipoles significantly, particularly at high redshifts. In contrast, the effect is diminished at lower redshift since the IGM has more time to relax the additional injected energy during the reionization process and to get closer to recovering the temperature-density power law. 

X-ray preheating of the IGM before reionization likely diminishes the imprint of reionization relics in the \lya forest at redshifts near the quasar luminosity peak ($z \sim 2$). However, at higher redshifts, it will strengthen the impact of reionization in the IGM \citep{2024MNRAS.529.3666M}. Despite uncertainties surrounding the sources of X-ray preheating, its lasting influence in the post-reionization IGM offers a new avenue to constrain the astrophysics that governs the X-ray sources during cosmic dawn. Furthermore, while the high-redshift \lya forest has often been shown to be a promising tool for constraining the nature of dark matter, particularly warm dark matter (\citealt{2020JCAP...04..038P, 2023PhRvD.108b3502V, 2024PhRvD.109d3511I}), this potential is somewhat tempered by the strong impact of the reionization relics in this regime \citep{2021MNRAS.502.2356G}.

The rest of this paper is organized as follows. We describe the data used throughout this work in \S\ref{sec:data}. In \S\ref{sec:model}, we introduce the ingredients necessary to model the correlations of the \lya forest. In particular, two distinct methodologies to account for the memory of reionization in the IGM are described in \S\ref{ssec:reio}. In \S\ref{sec:re}, we describe our findings and compare them with the original extended Baryonic Oscillation Spectroscopic Survey (eBOSS) Data Release 16 (DR16) analysis \citep{2020ApJ...901..153D}. Furthermore, we establish the impact of reionization relics in the \lya forest correlation functions and the inferred model parameters in \S\ref{ssec:re-mem}. \S\ref{sec:disc} discusses some caveats of our strategy. Finally, we summarize our findings in \S\ref{sec:con}. In Appendix \ref{app:yuk-par}, we introduce the parameter values of our Yukawa reionization model. We show the equivalency of our reionization approaches in Appendix \ref{app:pysr}, while Appendix \ref{app:yuk-free} presents an alternative reionization model allowed to span reionization timelines beyond those covered in our reionization simulations.  

Unless otherwise specified, a flat-$\rm{\Lambda CDM}$ cosmology from \cite{2016A&A...594A..13P} is adopted, with $\Omega _m=0.31457$, $\Omega _b=0.049$, and $h=0.6731$. Other relevant derived parameters are shown in \S\ref{sec:re} and Table 2 of \cite{2020ApJ...901..153D}.

\section{Data}
\label{sec:data}
Throughout this work, we utilize the publicly available\footnote{ \url{https://www.sdss4.org/dr16/algorithms/qso_catalog}} quasar catalog from the eBOSS DR16 \citep{2020ApJS..250....8L}. Here we briefly describe the catalog and measurements but refer interested readers to the main eBOSS DR 16 analysis for details \citep{2020ApJ...901..153D}.

This catalog contains more than 480,000 quasars within the redshift range $0.8 < z < 2.2$ and over 239,000 quasars for the \lya forest measurements. In comparison, the early data release from the Dark Energy Spectroscopic Instrument \citep[DESI;][]{2022AJ....164..207D} includes a total of 318,691 quasars between the early data release and two months of the main survey  \citep{2024MNRAS.528.6666R} while the DESI Year 1 \lya BAO analysis has 1,529,530 quasars in total with 709,565 at $z> 1.77$ \citep{2024arXiv240403001D}.

In this work, we analyze \lya absorption in two spectral regions. The ``$\rm Ly\alpha$'' region spans the rest-frame wavelength range $104 < \lambda_{\rm RF} < 120 \ \rm nm$, lying between the quasar \lya and $\rm Ly\beta -OVI$ emission peaks. The ``$\rm Ly\beta$'' region covers $92 < \lambda_{\rm RF} < 102 \  \rm nm$, extending from the $ \rm Ly\beta-OVI$ emission peak to the quasar rest-frame Lyman limit \citep{2020ApJ...901..153D}. The $\rm Ly\beta$ region is affected by absorption from higher-order Lyman lines, necessitating the separation of the forests for an accurate analysis.

With the eBOSS DR 16, one can construct the following correlation functions\footnote{There are two extra correlations that one could compute with this data, namely quasar auto-correlation and the auto-correlation of \lya fluctuations in the Ly$\beta$ region, i.e.\ Ly$\alpha$(Ly$\beta$) $\times$ Ly$\alpha$(Ly$\beta$). However, we follow the eBOSS analysis and focus on the four correlations with higher signal to noise.}:
\begin{itemize}
  \item $\rm Ly\alpha(\rm Ly\alpha)\times \rm Ly\alpha(\rm Ly\alpha)$
  \item $\rm Ly\alpha(\rm Ly\alpha)\times \rm Ly\alpha(\rm Ly\beta)$
  \item $\rm Ly\alpha(\rm Ly\alpha)\times \rm quasar$
  \item $\rm Ly\alpha(\rm Ly\beta)\times \rm quasar$
\end{itemize}
We refer to the \lya absorption in the $\rm Ly\alpha$ region as $\rm Ly\alpha(\rm Ly\alpha)$, while $\rm Ly\alpha(\rm Ly\beta)$ corresponds to \lya absorption in the $\rm Ly\beta$ region. 

In terms of observed wavelength, the eBOSS data used in this paper covers the range $\lambda \in [3600, 6000]$ \r{A}, where the upper limit is set by the focus on quasars with redshifts $z_q < 4$. The lower limit is caused by the high atmospheric absorption in the ultraviolet. 

Furthermore, damped Lyman-$\alpha$ absorbers (DLA) and broad absorption line (BAL) troughs were identified in the quasar spectra as described in \cite{2020ApJS..250....8L} following a similar procedure to that of \cite{2019ApJ...879...72G} for BALs and using a neural network for DLAs \citep{2018MNRAS.476.1151P}. 

The four correlation functions are computed using comoving separation bins ($r_{\parallel}, r_\perp$) adopting the $\Lambda$CDM cosmology of \cite{2016A&A...594A..13P} to convert angular and redshift separations into physical distances. This strategy is well-motivated, as the BAO scale remains relatively stable across redshifts, even when the fiducial cosmology deviates significantly from the true cosmology \citep{2024arXiv240403004C}. Although reionization can influence the fiducial cosmology \citep{2024arXiv240513680M}, its effects are unlikely to disrupt this stability. The auto-correlation functions are computed using comoving separation bins of length 4 $h^{-1}$Mpc along both parallel and perpendicular directions, resulting in a total of 2500 bins (covering $r_{\parallel}$, $r_\perp \in [0,200] \, h^{-1}$Mpc). In contrast, the cross-correlation functions are asymmetric under the permutation of the two tracers. Thus, the cross-correlations have $r_\parallel \in [-200, 200] \, h^{-1}$Mpc and utilize a total of 5000 bins. 

For further details regarding the computation of the correlation functions and covariance matrix from the data we refer the readers to \S3 of \cite{2020ApJ...901..153D}. Compared to recent analyses, such as \citet{2024arXiv240403001D}, the approach of \cite{2020ApJ...901..153D} for the covariance matrices computes each covariance of the four correlation functions in isolation, i.e. ignoring the cross-covariances between them. This can lead to $\sim 10\%$ underestimation of the uncertainties on the BAO parameters \citep{2024arXiv240403001D} and should also affect other non-BAO parameters. 

\begin{figure*}
    \centering
    \includegraphics[width=\linewidth]{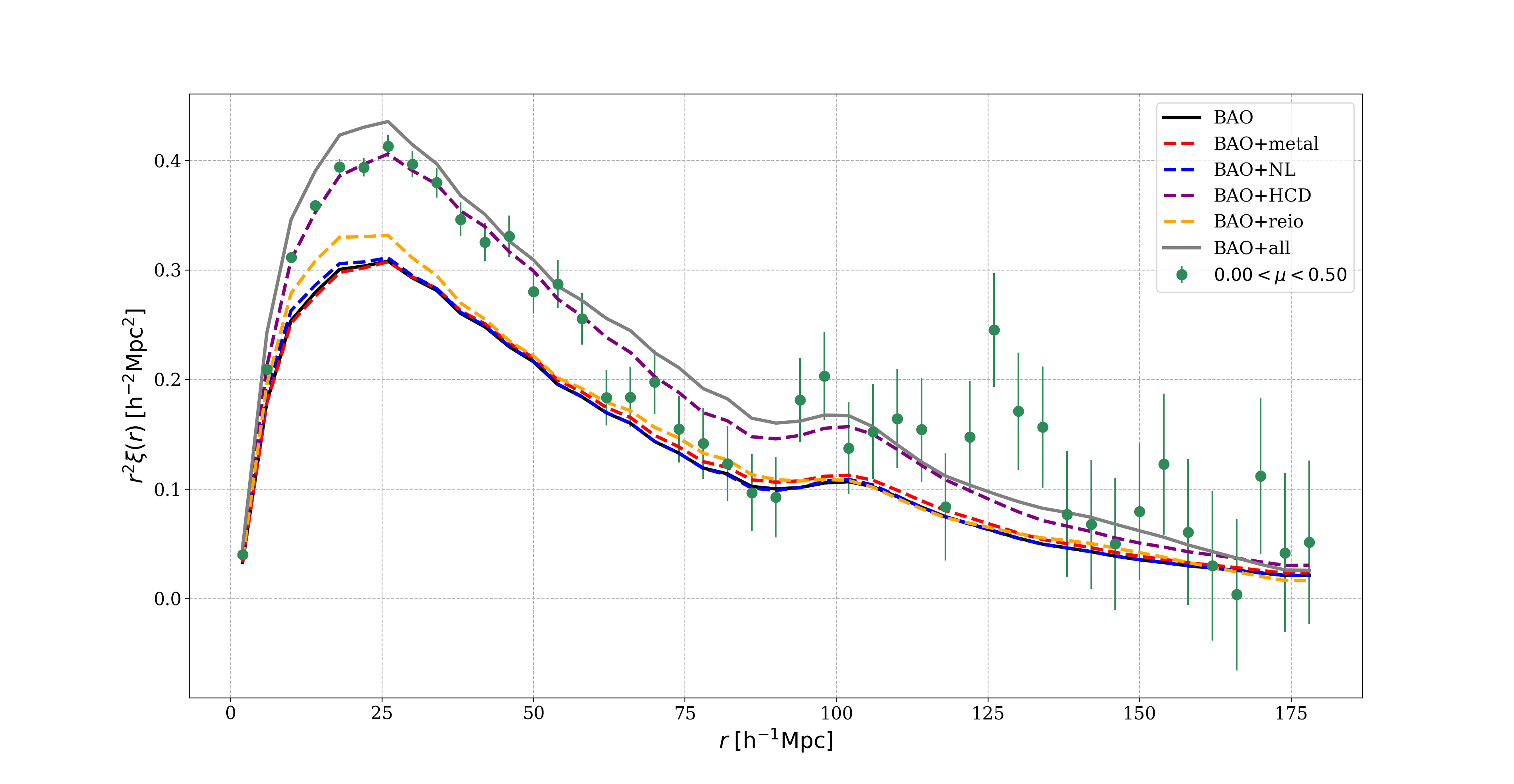}
    \caption{Wedge plot of the Ly$\alpha$(Ly$\alpha$) auto correlation function for $0.00<\mu<0.50$. The green points represent the eBOSS DR16 data. The black solid curve illustrates the auto-correlation function without any other broadband effects, highlighting the contribution of the BAO peak. Each colored dashed curve represents the result of adding one of the several broadband components of the model, including the Arinyo non-linear correction (dashed blue, \S\ref{ssec:nl}), high column density systems (dashed purple, \S\ref{ssec:hcd}), \ion{H}{I} reionization relics (dashed yellow, \S\ref{ssec:reio}), and metal contamination (dashed red, \S\ref{ssec:metal}). The gray solid curve shows the result of including them all together. The curves in this figure do not correspond to any best-fit because all the free parameters are fixed to make the comparison between different broadband components fair and clear. The reionization relics effect is represented by the PySR late reionization scenario, which gives the strongest deviations relative to the fiducial no reionization model (see \S\ref{ssec:reio} for a description of the reionization scenarios considered throughout this work).}
    \label{fig:baowedge}
\end{figure*}

Figure \ref{fig:baowedge} displays the measurement of the Ly$\alpha$ forest auto-correlation using Ly$\alpha$ pixels from the Ly$\alpha$ region, for orientations that are farther from the line of sight (LOS) with $0 < \mu < 0.5$, where $\mu$ is the cosine of the angle between the pixel pairs and the LOS. While the correlation function is measured over the full range $r_\parallel, \, r_\perp \in $ [0, 200] $h^{-1}$Mpc, the fits are restricted to separations $r \in [10, 180] \, h^{-1}$Mpc, with $ \mu \in [0,1]$ for auto-correlations and $\mu \in [-1,1]$ for cross-correlations. The fitting uses 1590 bins for each of the auto-correlations and 3180 bins for each of the cross-correlations. Hence, the combined analysis, encompassing two auto-correlation and two cross-correlations, involves a total of 9540 bins. 

\section{Model}
\label{sec:model}
Our base model largely follows the methodology of \cite{2023JCAP...11..045G} and \cite{2020ApJ...901..153D} (and references therein), taking full advantage of \texttt{picca} \citep{2021ascl.soft06018D}, a python package developed for IGM cosmological-correlation analyses. Furthermore, we closely follow the \texttt{picca} tutorial for reproducing the eBOSS DR16 analysis outside of NERSC\footnote{Available at \url{https://github.com/igmhub/picca/blob/master/tutorials/eboss_dr16/tutorial_instructions.ipynb}}. This section outlines the primary physical ingredients of the base model but we refer interested readers to the previously mentioned works. In essence, the model for the \lya forest and its cross-correlation with quasars needs the following ingredients: biases, including the redshift-space distortion parameters (RSD), and their redshift evolution (\S\ref{ssec:bias}), quasilinear power spectrum (separated into peak and smooth components), which is needed for the BAO extraction (\S\ref{ssec:pql}), nonlinear corrections (\S\ref{ssec:nl}), and a term to account for the binning of the data (\S\ref{ssec:g2}). Moreover, we include contamination due to high-column-density (HCD) systems (\S\ref{ssec:hcd}), quasar radiation effects and quasar redshift errors (\S\ref{ssec:qso}), metals (\S\ref{ssec:metal}), instrumental systematics (\S\ref{ssec:sky}), and continuum distortions (\S\ref{ssec:cont}). Beyond the \emph{conventional} modeling, we incorporate, for the first time, the long-term impact of inhomogeneous reionization on the \lya forest in the cosmological analysis of \lya forest data \citep[\S\ref{ssec:reio};][]{2019MNRAS.487.1047M}.

The base model for the auto-\lya and cross \lya $\times$ quasar power spectra, excluding the impact of inhomogeneous reionization, can be written as
\begin{eqnarray}
    \label{eq:base}
    P_{ab} (k, \mu_k,z) &=& b_a (z) b_b (z) (1 + \beta_a \mu_k^2)(1+\beta_b \mu_k^2)  \nonumber \\ 
    &&\times \ P_{\rm QL}(k,\mu_k,z) F_{\rm NL}(k, \mu_k) G(k,\mu_k) \, ,
\end{eqnarray}
where $b_a$ and $\beta_a$ are the bias and RSD parameter of tracer $a$, respectively, and the similar notation convention applies to the tracer $b$. Here $\mu_k = k_\parallel/k$ is the cosine of the angle between the mode ${\bf k}$ and the LOS, $P_{\rm QL}$ contains the matter power spectrum,  $F_{\rm NL}$ is the nonlinear correction, and $G$ is the grid term.

\subsection{Bias factors and RSD parameters}
\label{ssec:bias}
We measure the BAO parameters at an effective redshift $z_{\rm eff} = 2.334$ \citep{2020ApJ...901..153D}. To account for redshift evolution, we adopt the following parametrizations \citep{2006ApJS..163...80M,2019ApJ...878...47D}
\begin{eqnarray}
    b_{\rm F} (z) & = & b_{\rm F}(z_{\rm eff}) \left( \frac{1 + z}{1 + z_{\rm eff}} \right)^{2.9} \, , \label{eq:bf} \\ 
    b_{\rm q} (z) & = & 3.60 \left( \frac{1 + z}{1 + z_{\rm eff}}\right)^{1.44} \label{eq:bq} \, .
\end{eqnarray}
For consistency with previous \lya BAO analyses \citep{2020ApJ...901..153D,2023JCAP...11..045G}, we assume that the RSD parameters -- $\beta_{\rm F}$ and $\beta_{\rm q}$ -- do not evolve with redshift. 

In total, we introduce three free parameters to model the \lya forest and QSO RSD and bias factors. However, for analyses relying solely on cross-correlation data, we fix the QSO RSD parameter $\beta_{\rm q} = 0.2602$, as was done in the eBOSS DR16 analysis. This choice is justified by the limited constraining power of the dataset in such cases. Likewise, the choice of fixed QSO bias is justified since the cross-correlation is only sensitive to the product of the quasar and \lya biases. In contrast, the analysis of the combination of all correlation functions allows the QSO RSD parameter to be set as a free parameter due to the strong constraining power of the full dataset. 

\subsection{Quasi-linear matter power spectrum}
\label{ssec:pql}
In order to facilitate the extraction of the BAO parameters, we decompose the matter power spectrum into a peak component and a smooth component, incorporating nonlinear broadening corrections,
\begin{eqnarray}
    \label{eq:pql}
    P_{\rm m}^{\rm QL} (\boldsymbol{k}, z) = P_{\rm m}^{\rm peak} (k,z) \exp{\left( -\frac{k_\parallel^2 \Sigma^2_\parallel + k^2_\perp \Sigma^2_\perp}{2}\right)} + P^{\rm smooth}_{\rm m} (k,z) \, ,
\end{eqnarray}
where $\Sigma_{\parallel}= 6.42 \, h^{-1}$ Mpc and $\Sigma_\perp = 3.26 \, h^{-1}$ Mpc are the nonlinear broadening scales of the BAO peak \citep{2007ApJ...664..660E}. Note that only $P_{\rm m}^{\rm peak}$ contains BAO information.

The decomposition is achieved using the sideband technique detailed in \cite{2013JCAP...03..024K}. This method involves transforming the linear matter power spectrum  into correlation function through Fourier transformation and two sidebands are defined surrounding the BAO feature. A smooth function is fitted to connect these sidebands, effectively removing the BAO feature from the correlation function. The resulting smooth correlation function is then transformed back to $k$-space, yielding $P_{\rm m}^{\rm smooth}$. Consequently, the peak power spectrum is given by $P^{\rm peak}_{\rm m} = P_{\rm m}^{\rm L} - P_{\rm m}^{\rm smooth}$.

\subsection{Nonlinear correction}
\label{ssec:nl}
Even though the BAO feature is a large-scale phenomenon, accurately modeling systematics in the correlation functions requires sensitivity to smaller separations, scales below 80 $h^{-1}$Mpc, which are not strictly necessary to detect the BAO signal. The main reason for carefully addressing these smaller scales is that the observed transmitted flux fluctuations $\delta_{\rm F}$ can be distorted by the distortion matrix (see \S\ref{ssec:cont}), effectively propagating some of the small-scale nuisance into larger separations. 

The nonlinear correction for the auto-correlation is given by \citep{2015JCAP...12..017A}
\begin{eqnarray}
    \label{eq:fnl}
    \ln F^{\rm auto}_{\rm NL} (\boldsymbol{k},z_{\rm eff}) = q_1 \Delta^2_{\rm L}(k, z_{\rm eff}) \left[1 - \left(\frac{k}{k_v}\right)^{a_v} |\mu_k|^{b_v} \right] - \left(\frac{k}{k_p}\right)^2 \, ,
\end{eqnarray}
where $\Delta^2_{\rm L} \equiv k^3 P_{\rm L} / (2\pi^2)$ and with the remaining parameter values interpolated to $z_{\rm eff}$ using Table 7 of \cite{2015JCAP...12..017A}. The first term in Eq.~(\ref{eq:fnl}) corresponds to the expected isotropic nonlinear enhancement of the power spectrum, where $q_1$ is dimensionless and controls the importance of this nonlinear enhancement. The second term accounts for the suppression along the LOS due to peculiar velocities, the parameter $k_v$ is the characteristic wave number of the velocity broadening effect, while $a_v$ and $b_v$ are the power law exponents controlling the velocity broadening effect as a function of $k$ and $\mu$ respectively. The third term represents the isotropic suppression caused by gas pressure below the Jeans scale, where $k_p$ is related to the temperature-density relationship of the IGM and plays a role similar to Jeans scale. (It is not a strict physical Jeans scale, but an empirical parameter determined by fitting).

We emphasize that Eq.~(\ref{eq:fnl}) is evaluated at the effective redshift, which introduces a minor approximation. Specifically, this procedure neglects some small variation in the nonlinear parameters due to different $(r_\parallel, r_\perp)$-bins having mean redshifts that vary over the range $2.32 \lessapprox \overline{z} \lessapprox 2.39$ \citep{2020ApJ...901..153D}.

For the cross-correlation, we account for the effects of redshift errors and quasar nonlinear velocities \citep{2009MNRAS.393..297P}, which smear the signal along the LOS. We correct for these effects using
\begin{eqnarray}
    \label{eq:fnlc}
    F^{\rm cross}_{\rm NL} (k_\parallel) = \frac{1}{\sqrt{1 + k^2_\parallel \sigma^2_v}} \, ,
\end{eqnarray}
where $\sigma_v$ is a free parameter that must be marginalized over to extract BAO parameters. $\sigma_v$ encapsulates the smoothing induced by redshift uncertainties. Note that the auto-correlation function is also affected by statistical redshift errors \citep{2022MNRAS.516..421Y}. 

As shown in Figure \ref{fig:baowedge}, the impact of the nonlinear correction can be quite subdued compared to other broadband components of the model.

\subsection{Grid-binning term}
\label{ssec:g2}
To account for the effect of binning the correlation functions into a  $(r_\parallel, r_\perp)$-grid, we introduce a grid term. We assume a uniform distribution within each bin, which is a sufficiently accurate approximation \citep{2017A&A...603A..12B}. The grid term $G$ can be written as the product of two top-hat functions
\begin{eqnarray}
    \label{eq:G}
    G(\boldsymbol{k}) = {\rm sinc}\left(\frac{k_\parallel R_\parallel}{2}\right) {\rm sinc} \left( \frac{k_\perp R_\perp}{2} \right) \, ,
\end{eqnarray}
where $R_\parallel = R_\perp = 4 \ h^{-1}$ Mpc are the bin widths in the radial and transverse directions, respectively.

\subsection{Imprints from \ion{H}{I} reionization}
\label{ssec:reio}
Ultraviolet heating during reionization raises the temperature of the IGM to $\mathcal{O}(10^4 \ {\rm K})$; however, underdense gas within minivoids can also undergo shock heating and compression to mean density due to shocks propagating from surrounding denser regions \citep{2018MNRAS.474.2173H}. Consequently, the commonly assumed temperature-density relation, $T = T_0 \Delta^{\gamma-1}$ \citep{1997MNRAS.292...27H}, becomes bimodal. This new phase, referred to as the high-entropy mean-density (HEMD) mode, reaches a higher adiabat and traces the denser gas in its evolution. The relaxation of the excess energy deposited during reionization occurs over cosmological timescales, with the HEMD gas eventually converging back to the standard temperature-density relation by $z \sim 2$\footnote{Note that even in the absence of the mass-resolution needed to capture the HEMD gas, the temperature-density relation will still be affected by \ion{H}{I} reionization \citep[e.g., ][]{2018MNRAS.477.5501K} but it will recover much sooner ($z \sim 4$).}  \citep{2018MNRAS.474.2173H,2020MNRAS.499.1640M}. The ability to track the HEMD gas, and hence fluctuations to the temperature-density relation that persist to $z<4$, requires high-mass resolution, e.g. a gas particle mass of $1.22 \times 10^3 h^{-1} M_\odot$ \citep{2018MNRAS.474.2173H}, capable of going under the Jeans mass prior to the start of reionization (see also \citealt{2024MNRAS.533L.100C} for another high-mass resolution -- $M_{\rm gas} \approx 1 \times 10^2 h^{-1} M_\odot$ -- study of reionization's long-lasting impact to the temperature-density relation and the importance of capturing gas below the Jeans mass prior to the start of the reionization process).

Furthermore, reionization proceeds in an inhomogeneous manner, with denser regions ionizing earlier due to the proximity of ultraviolet radiation sources. Therefore, the reionization process does not occur simultaneously across all regions, leading to spatial variations in the strength and timing of reionization. These inhomogeneities imprint distinct signatures on cosmological observables of the post-reionization era \citep{2019MNRAS.486.4075O,2023MNRAS.525.6036L,2025MNRAS.536.1645M}, which couple with the spatial scale of ionized bubbles, typically ranging from $\sim 1 - 10 $ Mpc \citep{2019MNRAS.487.1047M}. 

These long-lasting relics of reionization are inherently sensitive to the astrophysical processes governing cosmic reionization, presenting a valuable and novel avenue for constraining the reionization timeline \citep{2017ApJ...847...63O,2021MNRAS.508.1262M}. Likewise, they provide insights into the era of cosmic dawn \citep{2024MNRAS.529.3666M}, particularly regarding X-ray preheating, offering the potential to deepen our understanding of the nature and sources of X-rays in the early Universe. Conversely, neglecting the impact of reionization relics can result in significant biases and an underestimation of uncertainties when inferring cosmological parameters from the \lya forest \citep{2023MNRAS.520.4853M}.

The memory of reionization in the \lya forest, i.e. the impact of inhomogeneous reionization in the post-reionization \lya forest, is given by \citep{2019MNRAS.487.1047M}
\begin{eqnarray}
    P_{\rm F} (\boldsymbol{k}, z) &=& b_{\rm F}^2(z) (1 + \beta_{\rm F} \mu_k^2)^2 P_{\rm m}(k,z) F_{\rm NL}(\boldsymbol{k},z) G(\boldsymbol{k}) \nonumber \\ \label{eq:auto-mem}
    && + \,2 b_{\rm F}(z) b_\Gamma(z) (1 + \beta_{\rm F} \mu_k^2) P_{\mathrm{m},\psi}(k,z) G(\boldsymbol{k}) \, ,
\end{eqnarray}
where we include the grid correction term $G$ and $b_\Gamma = \partial \ln \overline{F} / \partial \ln \tau_1$ is the radiation bias \citep{2015JCAP...12..017A,2018MNRAS.474.2173H}, i.e. the variation of the natural logarithm of the transmitted flux with respect to a uniform rescaling of the optical depth $\tau_1$. We interpolate the values provided by \cite{2019MNRAS.487.1047M} to obtain $b_\Gamma (z = 2.334) = 0.16$. The term $P_{\mathrm{m},\psi}$ corresponds to the cross-power spectrum between matter and transparency of the IGM. This cross-power spectrum captures the effect of reionization through its effect on the optical depth and the spatial distributions of ionized regions. A detailed derivation of this formalism is provided in \S2 of \cite{2019MNRAS.487.1047M}. Note that higher order terms in $\psi$ have been neglected and no nonlinear term has been included in the second term due to this choice\footnote{The absence of these higher order terms can lead to deviations, particularly at large $k$; however, the distortion matrix (\S\ref{ssec:cont}) and the comparison with real data in configuration space, may also be affected by this modeling choice at different scales due to mode mixing.}.

\begin{figure}
    \centering
    \includegraphics[width=\linewidth]{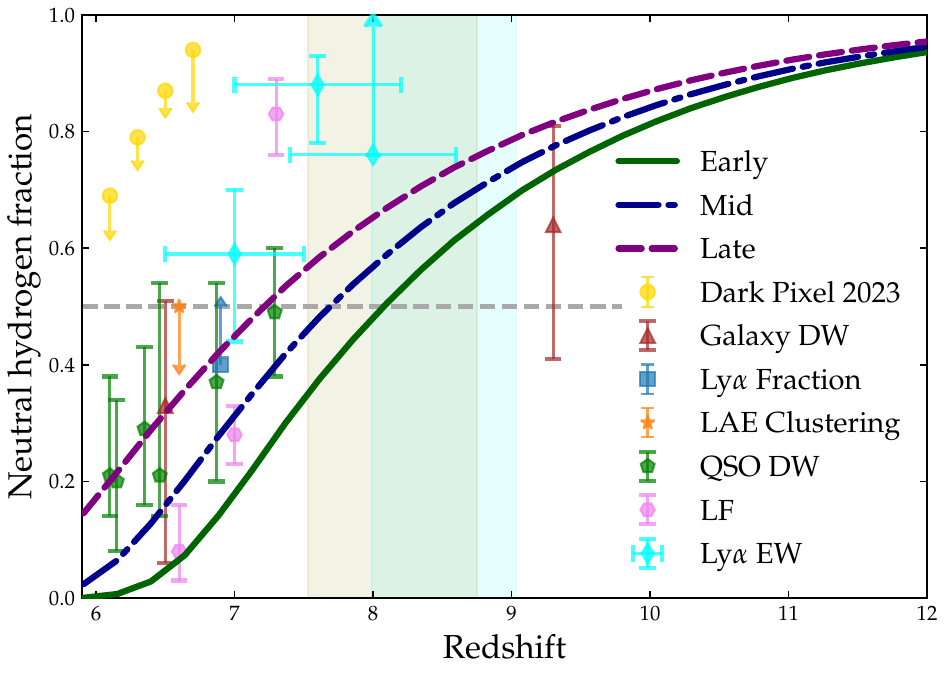}
    \caption{Evolution of the \ion{H}{I} fraction for the early (green solid), mid (blue dash-dotted), and late (purple dashed) reionization scenarios. For context, we include several astrophysical constraints such as: dark pixel \citep{2023ApJ...942...59J}, Ly$\alpha$ emission fraction \citep{2015MNRAS.446..566M}, clustering of Ly$\alpha$ emitters \citep{2015MNRAS.453.1843S}, Ly$\alpha$ equivalent width of Ly$\alpha$ emitters \citep{2018ApJ...856....2M,2019MNRAS.485.3947M,2019ApJ...878...12H}, the evolution of the galaxy Ly$\alpha$ luminosity function \citep{2021ApJ...919..120M}, galaxy damping wings \citep{2025arXiv250111702M}, and quasar damping wings \citep{2022MNRAS.512.5390G,2024MNRAS.530.3208G,2024A&A...688L..26S,2024ApJ...969..162D}. The gray dashed line corresponds to $x_{\rm HI} = 0.5$ while the olive and cyan colored regions indicate the inferred values for the midpoint of reionization obtained from CMB data by \citealt{2020A&A...635A..99P} and \citealt{2021MNRAS.507.1072D}, respectively.}
    \label{fig:xHI}
\end{figure}

Accurately modeling the effects of inhomogeneous reionization on the \lya forest within the redshift range $2 < z < 4$ poses significant computational challenges. Simulations must resolve the complex interactions between small-scale structure and reionization with sufficient mass resolution to capture processes occurring below the Jeans mass prior to the passage of an ionization front \citep{2020MNRAS.499.1640M}. Moreover, the patchy nature of reionization requires tracking spatial variations across large scales to achieve the necessary statistical power. Consequently, the simulations needed to capture these physical processes and to model the \lya forest demand a computationally expensive huge dynamical range. To circumvent these computational demands (which would make fitting real data excessively slow), we adopt the analytical approach introduced by \cite{2023MNRAS.520.4853M} to approximate the effects of reionization relics in the post-reionization IGM. This model employs a Yukawa-like template with three free, redshift-dependent parameters, and it is given by
\begin{eqnarray}
\label{eq:yukawa}
P_{\mathrm{m},\psi} (k,z) = -\frac{A_{\rm re}(z)}{(k/k_0)^{\beta_{\rm re}(z)}} e^{-\alpha_{\rm re}(z) (k/k_0)} \,\ \ \ \  [{\rm Mpc}^3] ,    
\end{eqnarray}
where $k_0 = 1 $ Mpc$^{-1}$, $A_{\rm re}$ corresponds to the amplitude of the cross-power spectrum, $\alpha_{\rm re}$ represents a screening coefficient, and $\beta_{\rm re}$ corrects for the large scale shape of the memory of reionization. These parameters are determined from simulation fits and interpolated to $z_{\rm eff}$. Table \ref{tab:yuk-par} in Appendix \ref{app:yuk-par} tabulates the values of the template parameters  -- at $z_{\rm eff}$ -- for three distinct fixed reionization scenarios presented in \citep{2020MNRAS.499.1640M}, we contextualize these early, mid, and late reionization scenarios in Figure \ref{fig:xHI}. The mid reionization scenario has a midpoint of reionization of $z_{\rm re} = 7.69$ ($\tau_{\rm reio} = 0.0547$) in good agreement with the midpoint of reionization and optical depth of  \cite{2020A&A...641A...6P}, while the late ($z_{\rm re} = 7.21$, $\tau_{\rm reio} = 0.0504)$ and early ($z_{\rm re} = 8.08$, $\tau_{\rm reio} = 0.0583$) reionization scenarios approximately span the $1\sigma$ range of the  Planck measurements.

As demonstrated in Figure 3 of \cite{2023MNRAS.520.4853M}, Eq.~(\ref{eq:yukawa}) performs remarkably well at large scales but its performance worsens for $k \gtrapprox 0.4 $ Mpc$^{-1}$. This limitation arises from the difficulty of the analytic template to fully capture the interplay between small-scale structure and reionization relics. Given the significant uncertainties in the reionization timeline \citep[e.g.][]{2021ApJ...919..120M} and the complexities in the epoch of reionization modeling \citep[see, for example,][]{2019MNRAS.484..933P}, the adopted Yukawa-like template provides sufficient robustness for the scope of this study. Notably, for a known reionization history, this template methodology has shown efficacy in reproducing the multipoles of the \lya auto-correlation function (see Figure 5 of \citealt{2023MNRAS.520.4853M}). However, to ensure that our results are not biased by the Yukawa template we will also use a separate analytic template based on the same simulation suite in \cite{2020MNRAS.499.1640M} but explicitly accounting for its redshift dependence. Moreover, to mitigate the declining performance at smaller scales from compromising the \lya model, we truncate the impact of reionization by setting it to zero for $k > 0.4$ Mpc$^{-1}$. This truncation is justified as reionization relics rapidly diminish with increasing wavenumber, as illustrated in Figure 3 of \cite{2020MNRAS.499.1640M}. 

Likewise, Eq.~(\ref{eq:yukawa}) is valid for $k_{\rm min} \approx 0.06$ Mpc$^{-1}$ due to the finite box size of the simulations used to derive this template. To model larger scales ($k < k_{\rm min}$), we adopt a linear biasing as follows
\begin{eqnarray}
    \label{eq:kmin}
    P_{\mathrm{m},\psi}(k < k_{\rm min},z_{\rm eff}) = \frac{P_{\mathrm{m},\psi} (k_{\rm min}, z_{\rm eff})}{P_{\rm m}^{\rm L}(k_{\rm min},z_{\rm eff})} P^{\rm L}_{\rm m} (k, z_{\rm eff}) \, .
\end{eqnarray}
This approach should provide sufficient accuracy for scales larger than the typical ionized bubble size. Nevertheless, we emphasize the importance of exploring larger simulation volumes in preparation for the scales DESI will probe with the 1D flux power spectrum \citep{2023MNRAS.526.5118R,2024MNRAS.528.6666R}. 

As an alternative to the analytical Yukawa-like template for describing the memory of reionization in the Lyman-$\alpha$ forest, and to facilitate its application to 1D \lya flux power spectrum measurements, we propose an additional expression valid for a Planck-like reionization. This new template is derived using {\sc PySR} \citep{2020arXiv200611287C}\footnote{\url{https://github.com/MilesCranmer/PySR}}, a Python package designed for high-performance Symbolic Regression, utilizing a {\sc Julia} backend. Symbolic regression, as implemented in {\sc PySR}, provides an efficient methodology for discovering interpretable mathematical expressions that best describe complex datasets \citep[see e.g. ][for more details regarding symbolic regression]{2020SciA....6.2631U}. For this work, the symbolic regression algorithm was trained on the same dataset used to derive the Yukawa-like template, specifically the \textit{bubble} models of \cite{2020MNRAS.499.1640M}. To enhance interpretability and physical relevance, we introduce a series of constraints during the training process. The exponential operator was preferentially weighted over binary operators, such as addition and subtraction. Furthermore, power-law terms ($x^y$) were limited in complexity, allowing $y$ to be either a constant or a single-variable factor. Exponential terms were restricted to arguments of limited complexity, capped at four instances of variables or variable-constant combinations -- e.g., $e^{x^4}$ and $e^{ax}$ are allowed, whereas nested forms like $e^{e^{x}}$ are excluded. The training data included wavenumber and redshift as the only independent variables. For further technical details on the derivation and implementation of the {\sc PySR} templates, we refer readers to \citet{igm_pysr}.

Under these constraints, and given the simulation data, we obtain the following expression for the memory of reionization in the \lya forest power spectrum
\begin{eqnarray}
    \label{eq:pysr-template}
    P_{\mathrm{m},\psi} (k,z) = -0.307 \left(\frac{11.9^z}{z^{5.86}}\right) \left(\frac{e^{-2.04 \, (k/k_0)}}{\left(k/k_0\right)^{2.04}}\right) \ \ \ \  [{\rm Mpc}^3] \, ,
\end{eqnarray}
with $k_0 = 1$ Mpc$^{-1}$.

Eq.~(\ref{eq:pysr-template}) preserves the Yukawa-like structure of the Yukawa-like template. While a comprehensive and quantitative evaluation of  Eq.~(\ref{eq:pysr-template}) --- the theme of a future companion paper --- is beyond the scope of this paper, Appendix \ref{app:pysr} offers a comparative analysis of its performance relative to the original Yukawa-like template for the purpose of this work. The {\sc PySR} template has the same domain of validity as Eq.~(\ref{eq:yukawa}), requiring the adoption of analogous strategies to handle the boundary conditions. Specifically, for $k > 0.4 $Mpc$^{-1}$, we set the contribution from reionization to zero due to its rapid decline at these scales. Similarly, for $k<0.06$ Mpc$^{-1}$, we employ the linear biasing model described in Eq.~(\ref{eq:kmin}) to address the limitations arising from the finite box size of the simulation data. 

The cross-correlation between \lya forest and quasars is also influenced by cosmic reionization. Besides its impact in the \lya forest, reionization can potentially influence the quasars directly through the imprints it leaves in the denser regions of the IGM during the post-reionization era \citep{2023MNRAS.525.6036L}. However, these effects primarily impact regions characterized by shallow potential wells during reionization, which are unlikely to represent the typical environments surrounding observed quasars. Furthermore, halos with initially shallow potential wells are expected to have recovered from the baryonic modulation induced by reionization by $z \approx 3.5 $. Consequently, the cross-power spectrum -- including reionization relics -- can be written as
\begin{eqnarray}
    \label{eq:cross-mem}
    P_{\rm cross} (\boldsymbol{k}, z) &=& b_{\rm F}(z) \,b_{\rm q}(z)  (1 + \beta_{\rm F} \mu_k^2) (1 + \beta_{\rm q} \mu_k^2) \nonumber \\
    && \times \, P_{\rm m}(k,z) F^{\rm cross}_{\rm NL} (k_\parallel) G (\boldsymbol{k}) \nonumber \\ 
    && + \, b_{\rm q}(z)\, b_\Gamma(z) (1 + \beta_{\rm q} \mu_k^2) P_{\mathrm{m},\psi} (k, z) G (\boldsymbol{k}) \, . \nonumber \\
\end{eqnarray}
The relics of reionization naturally depend on the timeline of cosmic reionization, as a more recent reionization would allow less time for the IGM to relax the additional energy injected during this epoch. To explore this dependency, we consider the three distinct reionization scenarios depicted in Figure \ref{fig:xHI}. The mid reionization scenario assumes a neutral hydrogen evolution consistent with \cite{2020A&A...641A...6P}, while the early and late scenarios are designated to be approximately consistent with Planck's $1\sigma$ errors on the inferred reionization timeline. Our terminology is that the Yukawa template with early reionization is referred to as ``Yukawa early'', whereas ``PySR late'' corresponds to the symbolic regression template with the late reionization scenario. Eq.~(\ref{eq:pysr-template}) corresponds to the PySR mid model, while the PySR early template has the form
\begin{eqnarray}
    \label{eq:pysr-early}
    \setstretch{1.3}
    \begin{cases}
    &x \equiv \frac{e^{-2(k/k_0)}}{(k/k_0)^2} \\
    &f(z) = (-0.19956973z+1.9220588)^z-2.8029568 \\
    &P_{\mathrm{m},\psi} (k,z) = f(z)\cdot x^{1.0633986} \ \ \ \  [{\rm Mpc}^3] \, .
    \end{cases}    
\end{eqnarray}
For late reionization, the PySR late model is given by
\begin{eqnarray}
    \label{eq:pysr-late}
    \setstretch{1.3}
    \begin{cases}
    &x = k/k_0 + 0.0037718469 \\
    &f(z) = \frac{z^{-4.6010156}}{0.13144803^z} \\
    &P_{\mathrm{m},\psi} (k,z) = f(z)\cdot \frac{e^{-2x}}{-2.2932026x^2} \ \ \ \  [{\rm Mpc}^3] \, ,
    \end{cases}    
\end{eqnarray}
where $k_0$ is still 1 Mpc$^{-1}$. 

In principle, one could use a generic template, say Eq.~(\ref{eq:yukawa}) with three free parameters and fit the data; however, Ly$\alpha$ BAO analyses are unlikely to robustly distinguish between arbitrary reionization models \citep[e.g., see Figure 4 of ][]{2023MNRAS.520.4853M}, and thus we opt to restrict the analysis in the body of this work to the fixed reionization scenarios presented in \cite{2020MNRAS.499.1640M} with the goal of finding which scenario yields a better fit to the data. Appendix \ref{app:yuk-par} introduces the values of $A_{\rm re}$, $\alpha_{\rm re}$, and $\beta_{\rm re}$ for the Yukawa early, mid, and late templates. 

All templates, including the Yukawa templates, are evaluated at the effective redshift $z_{\rm eff} = 2.334$ when entering the correlation function model.

Appendix \ref{app:pysr} presents a comparison of the two distinct methodologies employed to incorporate the memory of reionization into the \lya forest model. Our findings demonstrate that the templates yield results that are largely equivalent. Thus, for the sake of clarity and brevity, we focus solely on the {\sc PySR}-based results in the main text. However, we emphasize that the full analysis was conducted using both methodologies to ensure the robustness of the reionization modeling. Furthermore, we investigate the impact of more diverse reionization timelines in Appendix \ref{app:yuk-free}; where we explore a variation of the Yukawa template, introducing an additional free parameter by allowing its amplitude $A_{\rm re}$ to vary. This modified Yukawa template, with one free parameter, successfully recovers BAO parameters that are consistent with the results from the other templates including a preference for early reionization based on the best-fit amplitude. In addition, the recovered amplitude from the fit on the combination of all the correlations is in good agreement with the fixed value used in the Yukawa early template (see Table \ref{tab:freeA-yukawa(+)}), and hence with the PySR early model (Appendix \ref{app:pysr}).

We note that both the Yukawa and PySR templates do not include the effects of X-ray preheating prior to the onset of \ion{H}{I} reionization. The incorporation of X-ray heating during the cosmic dawn leads to a strengthening of the imprints from reionization at higher redshifts but also to a small suppression at lower redshifts ($z \sim 2$) -- depending on specific X-ray prescription \citep{2024MNRAS.529.3666M}.

Figure 1 indicates that besides the broadband nature of the impact of reionization on the correlation function, the main effect occurs for separations smaller than the BAO, particularly between $15 \lesssim r \lessapprox 100 \ h^{-1}$ Mpc.

\subsection{Absorption by HCD systems}
\label{ssec:hcd}
In the eBOSS context, high column density (HCD) systems are defined as neutral hydrogen absorbers with column densities exceeding $10^{17.2}$ cm$^{-2}$ \citep{2020ApJ...901..153D}. These systems mainly consist of Lyman limit systems and DLAs \citep{2022ApJS..259...28W}. An ideal \lya forest survey would perfectly identify and mask all HCD absorption regions to remove their effect on the correlation functions. Unfortunately, in practice, this task remains challenging. Residual HCDs that go undetected introduce systematic modifications to the correlation functions, particularly along the radial direction. On large scales, this effect manifests as a broadening of the signal that introduces a dependence on $k_\parallel$ for the \lya bias and RSD parameter as follows \citep{2012JCAP...07..028F}
\begin{eqnarray}
    \tilde{b}_{\rm F} &=&b_{\rm F} + b_{\rm HCD}F_{\rm HCD}(k_\parallel) \, , \\
    \tilde{b}_{\rm F}\, \tilde{\beta}_{\rm F} &=& b_{\rm F} \, \beta_{\rm F} + b_{\rm HCD} \,\beta_{\rm HCD} F_{\rm HCD}(k_\parallel) \, ,
\end{eqnarray}
where $b_{\rm HCD}$ and $\beta_{\rm HCD}$ are the bias and RSD parameters associated with HCD systems, respectively. In the fit of the auto-correlation functions, these parameters are treated as free parameters, while for the cross-correlations, they are fixed to $b_{\rm HCD} = -0.0501$ and $\beta_{\rm HCD} = 0.703$, respectively. Moreover, $F_{\rm HCD}$ models the suppression effect induced by HCD systems and is defined as \citep{2018MNRAS.476.3716R}
\begin{eqnarray}
    \label{eq:fhcd}
    \ln F_{\rm HCD}(k_\parallel) = - L_{\rm HCD} k_\parallel \, ,
\end{eqnarray}
where $L_{\rm HCD}$ characterizes the effective scale of the suppression. For consistency with the eBOSS DR16 analysis, $L_{\rm HCD}$ is fixed to $10 \, h^{-1}$Mpc\footnote{However, it is worth noting that the DESI Year 1 analysis allows $L_{\rm HCD}$ to vary, finding minimal impact on the BAO parameters \citep{2024arXiv240403001D}. Nonetheless, variations in $L_{\rm HCD}$ could potentially influence the inferred bias and RSD parameters due to their intrinsic degeneracies.}. 

As shown in Figure \ref{fig:baowedge}, the impact of HCD in the correlation function can be quite strong at all scales.

\subsection{Quasar proximity effect and quasar redshift errors}
\label{ssec:qso}
The radiation emitted by a quasar can have a strong influence on its surrounding neighborhood. Assuming isotropic emission, this transverse proximity effect can be expressed as \citep{2013JCAP...05..018F}
\begin{eqnarray}
    \label{eq:tp}
    \xi^{\rm TP} = \xi_0^{\rm TP} \left( \frac{1 \, h^{-1} \, \textup{Mpc}}{r}\right)^2 \exp\left(-\frac{r}{\lambda_{\rm UV}}\right) \, ,
\end{eqnarray}
where $\lambda_{\rm UV}$, representing the characteristic attenuation length, is set to $300 \, h^{-1}$Mpc \citep{2013ApJ...769..146R}, consistent with the eBOSS DR16 analysis. The amplitude $\xi^{\rm TP}_0$ is yet another free parameter of the model. However, in cases where only the cross-correlation information is used, $\xi^{\rm TP}_0$ is fixed to a value of 0.7386, again for consistency with the eBOSS DR16 analysis. 

Another caveat for the cross-correlation analysis is the impact of quasar redshift errors, which can systematically alter the inferred separation between quasars and \lya absorption regions \citep{2024arXiv240218009B}. To address these systematic quasar redshift errors, we introduce an additional free parameter $\Delta r_{\parallel, {\rm QSO}} \equiv r^{\rm True}_\parallel - r^{\rm Measured}_\parallel$, which accounts for potential biases in the measured redshifts. This correction mitigates asymmetries in the cross-correlation along positive and negative $r_\parallel$ caused by redshift errors \citep{2023JCAP...11..045G}. We remind the readers that Eq.~(\ref{eq:fnlc}), which models the smearing of the cross-correlations along the LOS, already incorporates another aspect of redshift errors by introducing the free parameter $\sigma_v$. This parameter accounts for the effects of redshift uncertainties on the velocity dispersion of quasars.

\subsection{Absorption by metals}
\label{ssec:metal}
Besides the absorption features from the \lya forest and HCD systems, the eBOSS data also captures absorption signatures from metals present in the IGM \citep{2022ApJ...935..121Y}. This includes the cross-correlation between \lya absorption and metal lines, as well as contributions from metal-metal correlations to the overall observed correlation functions.

We follow the methodology outlined in \cite{2020ApJ...901..153D} to tackle the impact of metals. Specifically, we define a power spectrum, $P_{mn}$, for each pair of absorber species $m$ and $n$ and process it similarly to the \lya power spectrum, with two notable modifications. First, the HCD contribution is excluded. Second, since redshifts are assigned under the assumption of \lya absorption, systematic deviations arise for metal absorbers due to the different transition lines involved. These deviations are addressed using a metal distortion matrix as established in \cite{2018JCAP...05..029B}. Specifically, we use Eq.~(35) in \cite{2020ApJ...901..153D}.

In line with the eBOSS DR16 analysis, we focus on metal contamination from four silicon transitions: \ion{Si}{II} (119 nm), \ion{Si}{II} (119.3 nm), \ion{Si}{II} (126 nm), and \ion{Si}{II} (120.7 nm). The RSD parameters for these transitions are fixed to $\beta_m = 0.50$, motivated by previous measurements of the cross-correlation between the \lya forest and DLAs \citep{2012JCAP...11..059F}, while the bias factors, e.g., $b_{\eta, {\rm SiII(119)}}$, are allowed to vary. In addition, we include only the auto-correlation of \ion{C}{IV}, with a fixed $\beta_{\ion{C}{IV}} = 0.27$, consistent with previous findings \citep{2018JCAP...05..029B}. The corresponding bias \ion{C}{IV} factor, $b_{\eta,{\rm CIV(eff)}}$, is a free parameter but it is kept fixed to $-4.8 \times 10^{-3}$ for cross-correlation analyses \citep{2020ApJ...901..153D}.

As illustrated in Figure \ref{fig:baowedge}, the broadband impact of metals tend to be more important near the BAO peak. 

We have opted not to include reionization terms in the metal lines power spectra, and instead use the standard Kaiser model. Naturally, we want to preserve the metal treatment of the official eBOSS DR16 but there is also a physical justification for this modeling choice. For the \ion{H}{I} reionization relics to survive up to $z_{\rm eff}$, the HEMD gas is necessary and at the low effective redshift of the BAO analysis HEMD gas is predominantly present in lower density regions ($\log_{10} \Delta \leq 0.5$) as shown in Figure 5 of \cite{2018MNRAS.474.2173H}. In contrast, metals in the IGM are more abundant in denser regions \citep{2003ApJ...596..768S,2022ApJ...935..121Y}. However, by $z \gtrsim 4$ even non-HEMD gas can exhibit imprints from \ion{H}{I} reionization \citep{2019MNRAS.486.4075O, 2019MNRAS.490.3177W, 2022MNRAS.509.6119M} in low and mild density regions. Furthermore, since metals are produced in galaxies and subsequently expelled into the IGM by winds and outflows, the baryonic modulation effect of \ion{H}{I} reionization in halos \citep{2022MNRAS.513..117L,2023MNRAS.525.6036L} may be important in this regime.

\subsection{Instrumental systematics}
\label{ssec:sky}
The eBOSS instrument has two spectrographs, each equipped with 500 fibers, of which 450 are allocated for science observations and approximately 40 are designated as sky fibers. These sky fibers play a critical role in monitoring atmospheric conditions and are essential for performing independent sky subtraction for each spectrograph. This subtraction mitigates the Poisson fluctuations present in the sky spectra, which can induce excess correlations in bins with $r_\parallel = 0$. This sky-based induced correlated noise is the dominant source of contamination caused by instrumental effects in the \lya forest observations \citep{2024arXiv240403003G}. 

Although this contamination initially manifests as an excess correlation at small parallel separations, continuum-fitting will propagate the distortion across all $r_\parallel$. To address this, \cite{2020ApJ...901..153D} proposed an empirical model to fit the distortion. This model introduces two free parameters and is given by
\begin{eqnarray}
    \label{eq:sky}
    \xi^{\rm sky} (r_\parallel, r_\perp) = \begin{cases}
        \frac{A_{\rm sky}}{\sigma_{\rm sky} \sqrt{2 \pi}} \exp{\left( -\frac{1}{2} \left( \frac{r_{\perp}}{\sigma_{\rm sky}}\right)^2\right)}, & \text{if $r_\parallel = 0$,} \\
        0, & \text{if $r_\parallel \neq 0$,}
    \end{cases}
\end{eqnarray}
where $A_{\rm sky}$ and $\sigma_{\rm sky}$ represent the amplitude and width of the induced correlation, respectively. Note that the contamination in the \lya region requires one set of $A_{\rm sky}$ and $\sigma_{\rm sky}$, while the Ly$\beta$ region needs a separate set to account for this instrumental effect.

\subsection{Continuum distortion}
\label{ssec:cont}
Fitting the quasar continua is a crucial step in determining the overall strength of absorption features in the \lya forest. The continuum fitting procedure employed for the eBOSS DR16 analysis involves modeling the continuum for each skewer using a linear fit, characterized by an amplitude and slope \citep[see e.g.,][]{2011JCAP...09..001S}. However, this method has the unintended consequence of introducing biases in both the mean and slope of flux fluctuations, which can distort the resulting correlation measurements.

To ensure a fair comparison between the model and the observed data, the physical model must be subjected to an analogous distortion. This is the responsibility of the distortion matrix $D$, which relates the true correlations to their distorted counterparts \citep{2017A&A...603A..12B}
\begin{eqnarray}
    \label{eq:dist}
    \hat{\xi}_{\rm distorted} = D \xi_{\rm true} \, .
\end{eqnarray}
Thus, the physical model of the correlations is then distorted by the distortion matrix prior to any comparison with the observed data. The specific forms of the distortion matrix for the auto-correlations and cross-correlations are detailed in Eqs.~(21) and (22) of \cite{2020ApJ...901..153D}.

Note that our correlation function model does not include broadband power-law correlations \citep{2017A&A...603A..12B}, as was the case for the main eBOSS DR16 analysis. These broadband corrections are useful to assess whether incomplete modeling affects the measurement of the BAO parameters. Thus, we will compare our results to several broadband tests done in previous Ly$\alpha$ correlation studies in \S\ref{sec:disc}.

In total, the model has 13 free parameters for the Ly$\alpha$(Ly$\alpha$) auto-correlation, 10 for the Ly$\alpha$(Ly$\alpha$) cross-correlation, and 19 for the combined analysis of all four correlation functions. Both the reionization model and the fiducial -- no reionization -- model have the same number of free parameters. The determination of the two BAO parameters requires marginalizing over the remaining nuisance parameters. As such, we introduce a Yukawa template with an extra degree of freedom in Appendix \ref{app:yuk-free} with the objective of inspecting the impact of the extra flexibility in the marginalization of nuisance parameters. Besides, this additional free parameter allows us to explore broader reionization histories than the three covered by the early, mid, and late versions. Interestingly, the recovered $A_{\rm re}$ values indicate a preference for early reionization, which agrees well with the main analysis (\S\ref{ssec:re-mem}).

\begin{figure*}
    \centering
    \includegraphics[width=\linewidth]{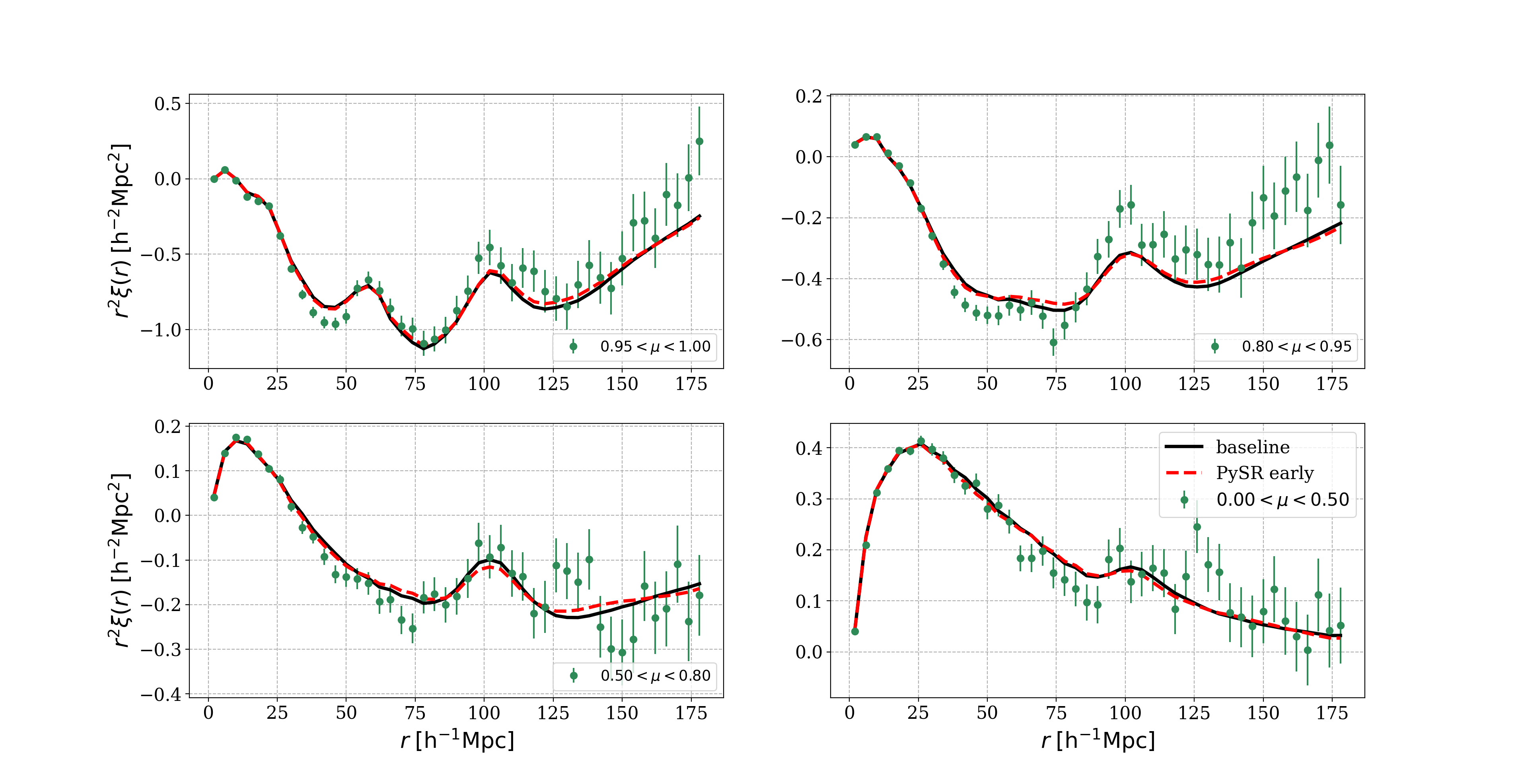}
    \caption{Wedges of the Ly$\alpha$(Ly$\alpha$) auto-correlation function of the \lya forest. The black solid curve corresponds to the best-fit model for the fiducial scenario, which assumes no reionization relics. The red dashed curve represents the early reionization scenario described by the PySR early template, our best performing model. Green points correspond to the measurements from eBOSS DR16. All models include metal contamination, which explains the bumps in the top-left panel at separations of 20, 60, and 111 $h^{-1}$Mpc, arising from correlations with \ion{Si}{II} and \ion{Si}{III} \citep[see \S\ref{ssec:metal} and][]{2023JCAP...11..045G}. }
    \label{fig:fid-lyalya}
\end{figure*}

\section{Results}
\label{sec:re}
To obtain the best fit from the eBOSS \lya and quasars correlation data, we follow the methodology outlined in the eBOSS DR16 tutorial available in \texttt{picca}. This approach maximizes the likelihood to determine the best-fit parameters, leveraging a Hessian matrix approach at the best-fit point to estimate uncertainties for most model parameters at the 1$\sigma$ confidence level. The use of the Hessian assumes symmetric errors and is particularly effective for Gaussian-like likelihoods. Since the main cosmological product of these analyses is the BAO parameters, one then employs the {\sc MINOS}\footnote{\url{https://github.com/scikit-hep/iminuit/blob/develop/src/minos.cpp}} algorithm from the {\sc iminuit}\footnote{\url{https://github.com/scikit-hep/iminuit}} minimization package to refine the determination of the $1\sigma$ confidence levels for the BAO parameters. This strategy is computationally inexpensive and well-suited for our purposes. For comparison, \cite{2020ApJ...901..153D} adopted a $\chi^2$ scan technique (detailed in their Appendix E) and derived the confidence levels for the BAO parameters using a suite of 1000 fastMC simulations of the correlation function. 

\subsection{No memory of reionization}
\label{sec:fid}
To ensure the reliability of our approach, we first re-analyze the eBOSS DR16 observations using our fiducial model, following the eBOSS DR16 \texttt{picca} tutorial. The primary objective of this sanity check was to validate our implementation by reproducing the best-fit values obtained by the eBOSS collaboration. For this initial analysis, we exclude the contribution from reionization relics and focus on fitting the fiducial model. 

The results of this re-analysis are presented in the first column of Table \ref{tab:uber}. A comparison with the results of \cite{2020ApJ...901..153D} (see their Table 6) reveals minor deviations. For most free parameters, such as the BAO parameters and the bias and RSD parameters of the \lya forest, the deviations are minimal, typically below $1\%$. This level of consistency demonstrates that our analysis successfully replicates the eBOSS pipeline as implemented in the \texttt{picca} tutorial. Small discrepancies, however, are expected. As noted in the tutorial, these discrepancies arise primarily due to the use of a slightly updated quasar catalog compared to the original catalog used in the eBOSS analysis. The differences are more pronounced for the bias parameters associated with metal contaminants, where deviations can reach approximately $20\%$. Nevertheless, the constraining power on metal parameters is relatively weak, and the differences between our results and those of \cite{2020ApJ...901..153D} remain within $1\sigma$. Therefore, these variations are deemed acceptable for our purposes. 

In Table \ref{tab:uber}, the bias parameter for the \ion{Si}{II} (119.3 nm) line in the Ly$\alpha$(Ly$\alpha$)$\times$ quasar cross-correlation has a positive value, whereas the biases for all other metal lines are negative. Similarly, a positive bias is inferred for the \ion{Si}{II} (119 nm) line in the Ly$\alpha$(Ly$\beta$)$\times$ quasar cross-correlation. This pattern of atypical metal bias parameters aligns with the findings of the eBOSS DR16 analysis, confirming that these results are not artifacts sourced by the {\texttt picca} tutorial. The occurrence of these positive bias signs can be attributed to statistical fluctuations, as the precision of measurements in the Ly$\beta$ region -- or in the cross-correlations with quasars -- is relatively limited. Consequently, the inferred biases may occasionally appear positive. However, none of the positive biases deviate by more than $2\sigma$, indicating that these values remain statistically consistent with expectations.

Figure \ref{fig:baowedge} shows the wedge plot for the  Ly$\alpha$(Ly$\alpha$) auto-correlation function in the range  $0.00 < \mu < 0.50$. This visualization is obtained by re-binning the original $(r_\parallel, r_\perp)$ rectangular grid into $(r,\mu)$ bins for visualization purposes. For clarity, different broadband components are shown with different colored curves. The black solid curve represents the correlation function without any other broadband effects, highlighting the contribution of the BAO peak. Both the observations (green points) and the model showcase the characteristic BAO bump near $r \approx 100 \, h^{-1}$Mpc.

A keen reader may notice that the probability ($p$-value) is higher, and the corresponding $\chi_{\rm min}^2$ is lower, for the Ly$\alpha$(Ly$\alpha$) auto-correlation than for the Ly$\alpha$(Ly$\alpha$) $\times$ quasar cross-correlation. Several factors contribute to this trend. The cross-correlation includes twice the number of bins as the auto-correlation, as the former lacks symmetry under the permutation of tracers. However, the cross-correlation provides less constraining power, leading to the fixing of certain non-BAO parameters, such as $b_{\eta, {\rm CIV (eff)}}$ and $\xi^{\rm TP}_0$, to the best-fit values reported in \cite{2020ApJ...901..153D}\footnote{The best-fit values reported in the eBOSS DR16 analysis show only minor discrepancies with those obtained from our combines analysis for the no-reionization model.}. As a result, the cross-correlation analysis involves 10 free parameters, while the auto-correlation uses 13.

When combining all four correlation functions, the probability decreases and the $\chi^2_{\rm min}$ increases further. This is due to the rise in degrees of freedom (the full model uses 19 free parameters in the combined analysis) and the inclusion of less constraining datasets such as Ly$\alpha$(Ly$\alpha$) $\times$ Ly$\alpha$(Ly$\beta$) or the Ly$\alpha$(Ly$\beta$) $\times$ quasar correlations. Hence, the uneven modeling and varying constraining power of the datasets make direct comparison of $p$-values across different correlations unwise. 

An interesting trend, also present in the recent DESI Year 1 \lya BAO analysis by \cite{2024arXiv240403001D}, is the significantly weaker RSD parameter for quasars compared to the \lya forest. Specifically, $\beta_{{\rm Ly}\alpha} = 1.6641$ for the forest, compared to $\beta_{\rm q} = 0.2593$ for quasars. These values are consistent with the DESI Year 1 results ($\beta_{{\rm Ly}\alpha} \sim 1.7, \, \beta_{\rm q} \sim 0.3$). 

\begin{figure*}
\begin{subfigure}{.49\textwidth}
    \centering
    \includegraphics[width=\linewidth]{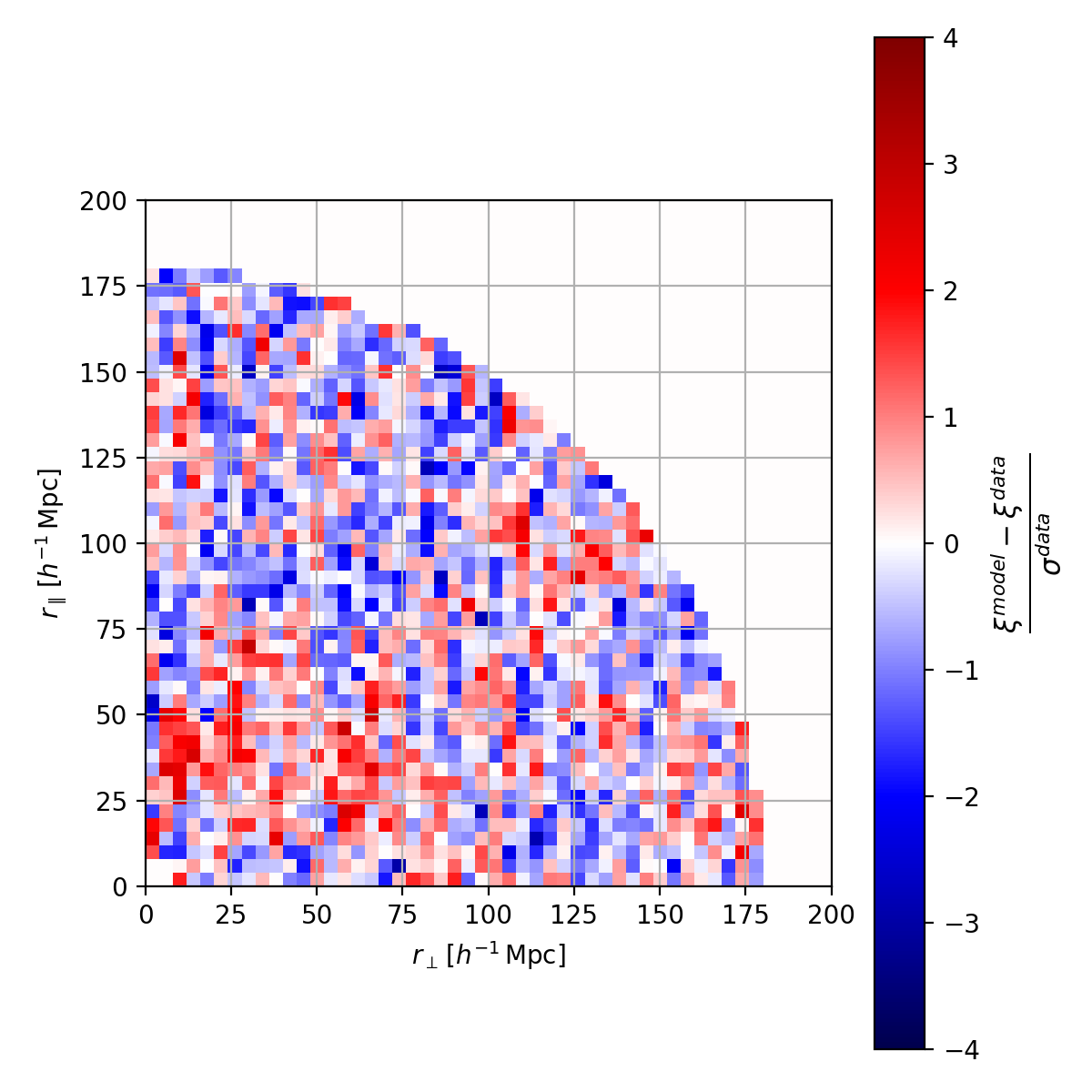}
    \caption{With reionization}
    \label{fig:dif_xi_withreio}
\end{subfigure}
\begin{subfigure}{.49\textwidth}
    \centering
    \includegraphics[width=\linewidth]{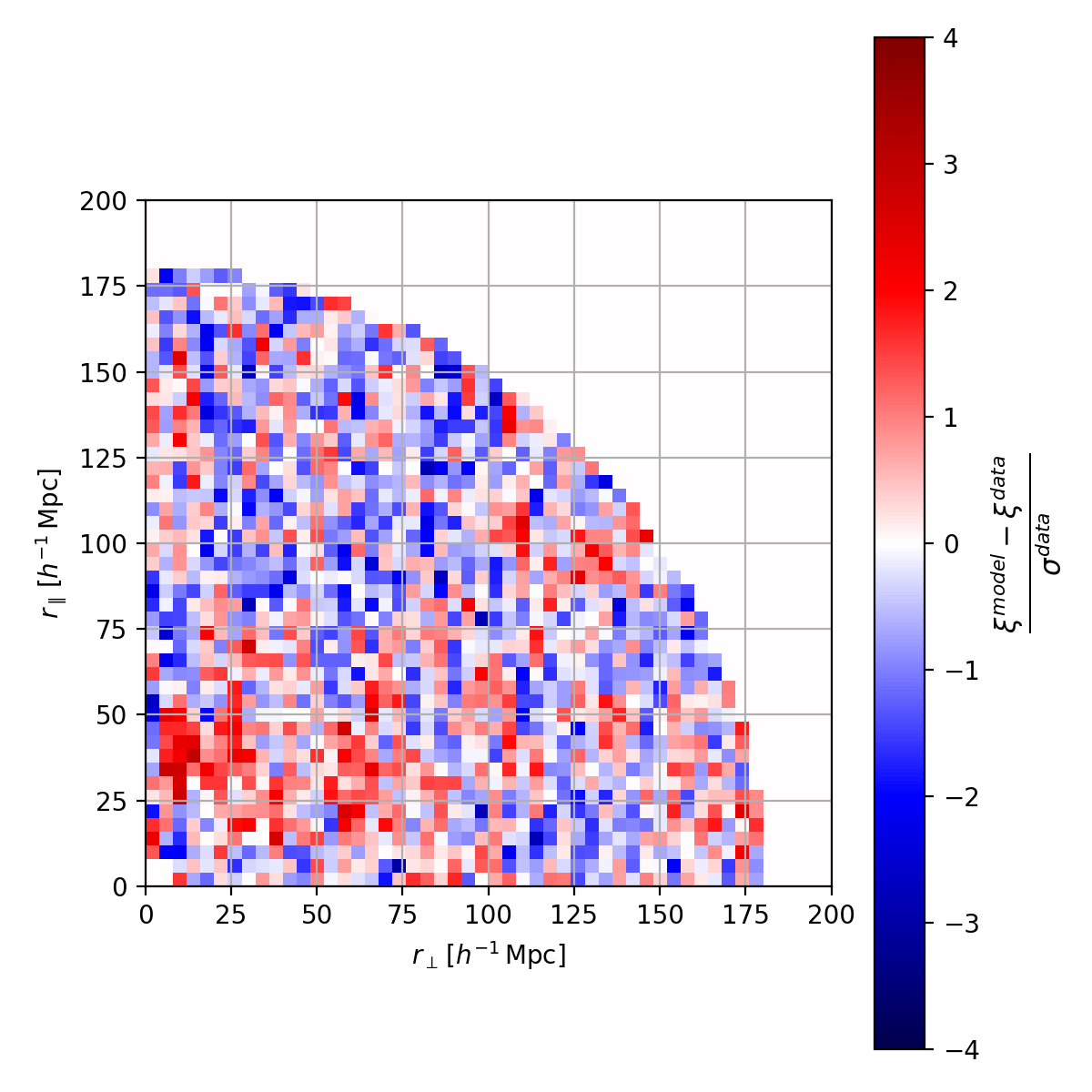}
    \caption{No reionization}
    \label{fig:dif_xi_noreio}
\end{subfigure}
\caption{Residuals of the 3D Ly$\alpha$(Ly$\alpha$) auto-correlation function. The residuals are defined as the difference between the model and the observed data, normalized by the observational uncertainties, i.e. $(\xi^{\rm model} - \xi^{\rm data}) / \sigma^{\rm data}$. In Figure \ref{fig:dif_xi_withreio}, the model incorporates reionization relics, as described by Eq.~(\ref{eq:auto-mem}), while Figure \ref{fig:dif_xi_noreio} represents the fiducial scenario, which excludes reionization effects. While differences between these models appear subtle, they are sufficiently significant to bias the non-BAO parameters in the fit. See Figure \ref{fig:xi_comp} for highlight of the difference between the two models.}
\end{figure*}

\begin{figure}
    \centering
    \includegraphics[width=\linewidth]{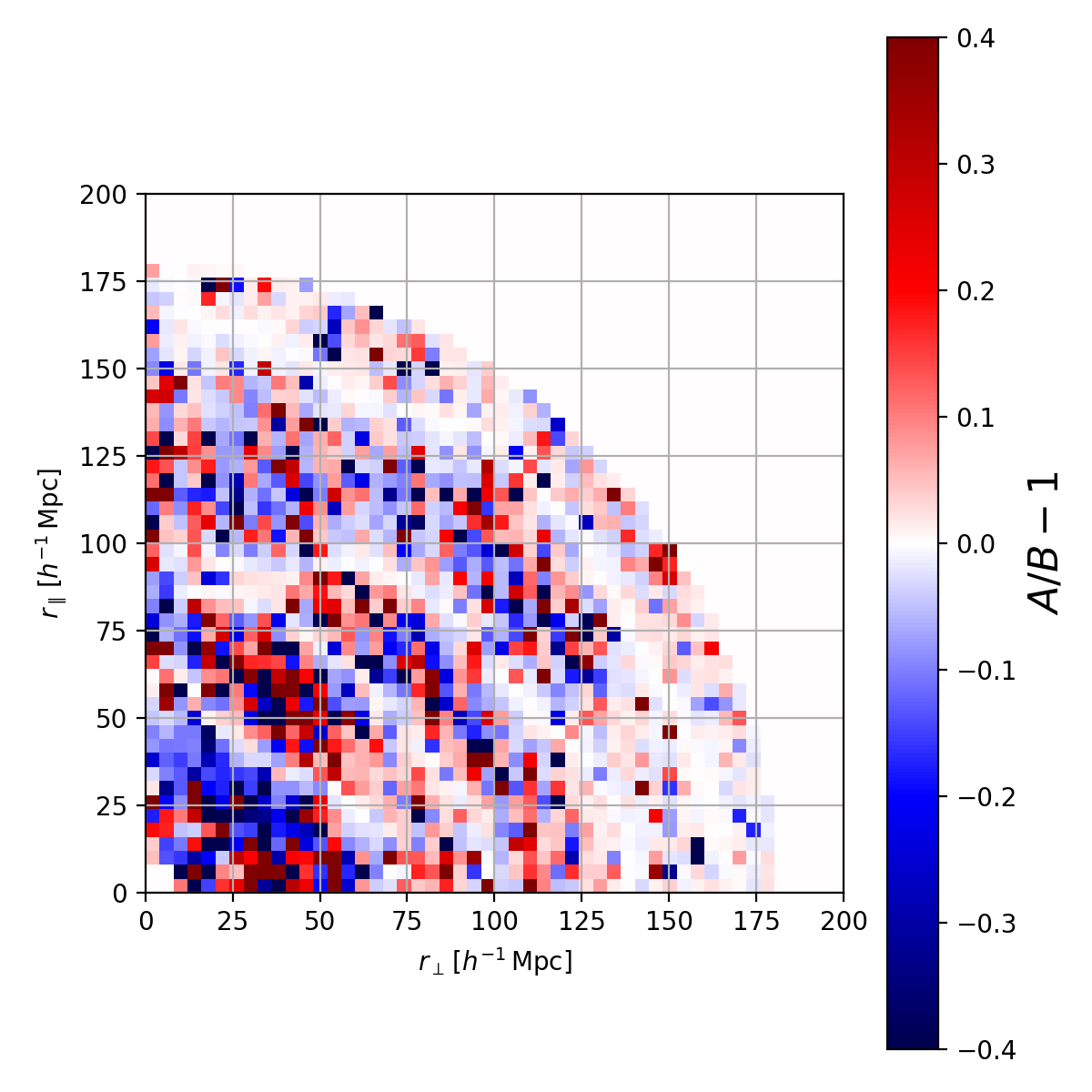}
    \caption{Comparison of Figure \ref{fig:dif_xi_withreio} and Figure \ref{fig:dif_xi_noreio}. For each pixel in Figure \ref{fig:dif_xi_withreio} (with reionization, labeled $A$) and Figure \ref{fig:dif_xi_noreio} (fiducial, labeled $B$), this plot displays the ratio $A/B$. To improve readability, we subtract 1 from the ratios, centering the values around zero. Around the typical BAO scale of 100 $h^{-1}$Mpc, an arc of lighter pixels is evident, indicating that the impact of reionization at the BAO peak is negligible.}
    \label{fig:xi_comp}
\end{figure}

\begin{table*}
    \centering
    \caption{Best-fit parameters for the auto-correlation function of the \lya forest (from the \lya region), the cross-correlation between the forest and quasars, and the combination of all the four correlations (including \lya absorption in the Ly$\beta$ region). We include a reference model that ignores reionization relics, alongside three variants that incorporate the impact of reionization relics in the post-reionization IGM, allowing for exploration of different reionization timelines. The gray-shaded entries identify the model that provides the best fit to the corresponding eBOSS DR16 measurements. Except for the $p$-value and minimum $\chi^2$, parameters without error bars are fixed in the fit. Unlike the original eBOSS DR16 analysis \citep{2020ApJ...901..153D}, which reports error bars on the BAO parameters derived from 1000 fastMC realizations, the error bars for all parameters in our table are computed using the {\sc iminuit} package.}
    \begin{tabular}{l|l|l|l|l}
    \hline
    \hline
    Parameter & No reionization & PySR early & PySR mid & PySR late\\
    \hline
    \multicolumn{5}{c}{\lya auto-correlation}\\
    \hline
    Prob. & 0.285 & \cellcolor{lightgray!70} 0.328 & 0.327 & 0.317\\
    $\chi^2_{\rm min}$ & 1608.49 & \cellcolor{lightgray!70} 1601.41 & 1601.58 & 1603.23\\
    $\alpha_\parallel$ & 1.0458 $\pm$ 0.0339 & 1.0256 $\pm$ 0.0372 & 1.0156 $\pm$ 0.0397 & 1.0085 $\pm$ 0.0405\\
    $\alpha_\perp$ & 0.9814 $\pm$ 0.0422 & 0.9913 $\pm$ 0.0487 & 1.0007 $\pm$ 0.0467 & 1.0069 $\pm$ 0.0452\\
    $\beta_{\rm{Ly}\alpha}$ & 1.6331 $\pm$ 0.0864 & 1.7221 $\pm$ 0.0943 & 1.7711 $\pm$ 0.0986 & 1.8180 $\pm$ 0.1028\\
    $b_{\rm{Ly\alpha}}$ & -0.1187 $\pm$ 0.0023 & -0.1131 $\pm$ 0.0022 & -0.1103 $\pm$ 0.0021 & -0.1079 $\pm$ 0.0021\\
    $b_{\eta,\rm{Ly\alpha}}$ & -0.1998 $\pm$ 0.0039 & -0.2007 $\pm$ 0.0039 & -0.2014 $\pm$ 0.0039 & -0.2021 $\pm$ 0.0039\\
    $10^3 b_{\eta, \rm{CIV(eff)}}$ & -5.4 $\pm$ 2.7 & -5.2 $\pm$ 2.6 & -5.2 $\pm$ 2.6 & -5.2 $\pm$ 2.6\\
    $10^3 b_{\eta, \rm{SiII(119)}}$ & -3.0 $\pm$ 0.5 & -2.9 $\pm$ 0.5 & -2.8 $\pm$ 0.5 & -2.8 $\pm$ 0.5\\ 
    $10^3 b_{\eta, \rm{SiII(119.3)}}$ & -1.7 $\pm$ 0.5 & -1.7 $\pm$ 0.5 & -1.7 $\pm$ 0.5 & -1.7 $\pm$ 0.5\\ 
    $10^3 b_{\eta, \rm{SiII(126)}}$ & -2.1 $\pm$ 0.6 & -2.1 $\pm$ 0.6 & -2.1 $\pm$ 0.6 & -2.1 $\pm$ 0.6\\ 
    $10^3 b_{\eta, \rm{SiIII(120.7)}}$ & -4.4 $\pm$ 0.5 & -4.8 $\pm$ 0.5 & -5.0 $\pm$ 0.5 & -5.2 $\pm$ 0.5\\
    $b_{\rm HCD}$ & -0.0518 $\pm$ 0.0045 & -0.0476 $\pm$ 0.0045 & -0.0467 $\pm$ 0.0045 & -0.0468 $\pm$ 0.0045 \\
    $\beta_{\rm HCD}$ & 0.6051 $\pm$ 0.0831 & 0.5985 $\pm$ 0.0839 & 0.5976 $\pm$ 0.0840 & 0.6003 $\pm$ 0.0840 \\
    $10^2 A_{\rm sky,auto}$ & 0.95 $\pm$ 0.6 & 0.94 $\pm$ 0.6 & 0.94 $\pm$ 0.6 & 0.94 $\pm$ 0.6 \\
    $\sigma_{\rm sky,auto}$ & 31.42 $\pm$ 1.68 & 31.78 $\pm$ 1.69 & 31.89 $\pm$ 1.69 & 31.93 $\pm$ 1.69 \\
    \hline
    \multicolumn{5}{c}{\lya $\times$ quasar cross-correlation}\\
    \hline
    Prob. & \cellcolor{lightgray!70} 0.200 & 0.184 & 0.175 & 0.169 \\
    $\chi^2_{\rm min}$ & \cellcolor{lightgray!70} 3236.92 & 3241.58 & 3244.22 & 3246.17 \\
    $\alpha_\parallel$ & 1.0617 $\pm$ 0.0321 & 1.0544 $\pm$ 0.0341 & 1.0509 $\pm$ 0.0349 & 1.0486 $\pm$ 0.0354\\
    $\alpha_\perp$ & 0.9289 $\pm$ 0.0381 & 0.9300 $\pm$ 0.0415 & 0.9331 $\pm$ 0.0431 & 0.9365 $\pm$ 0.0446\\
    $\beta_{\rm{Ly}\alpha}$ & 1.9311 $\pm$ 0.1428 & 2.1266 $\pm$ 0.1649 & 2.2282 $\pm$ 0.1761 & 2.3135 $\pm$ 0.1851\\
    $b_{\rm{Ly\alpha}}$ & -0.1127 $\pm$ 0.0052 & -0.1033 $\pm$ 0.0048 & -0.0995 $\pm$ 0.0047 & -0.0969 $\pm$ 0.0045\\
    $b_{\eta,\rm{Ly\alpha}}$ & -0.2242 $\pm$ 0.0103 & -0.2265 $\pm$ 0.0105 & -0.2285 $\pm$ 0.0107 & -0.2311 $\pm$ 0.0108\\
    $10^3 b_{\eta, \rm{CIV(eff)}}$ & -4.8 & -4.8 & -4.8 & -4.8\\
    $10^3 b_{\eta, \rm{SiII(119)}}$ & -4.7 $\pm$ 1.2 & -4.6 $\pm$ 1.2 & -4.6 $\pm$ 1.2 & -4.6 $\pm$ 1.2\\ 
    $10^3 b_{\eta, \rm{SiII(119.3)}}$ & 2.2 $\pm$ 1.1 & 2.2 $\pm$ 1.2 & 2.2 $\pm$ 1.2 & 2.2 $\pm$ 1.2\\ 
    $10^3 b_{\eta, \rm{SiII(126)}}$ & -1.9 $\pm$ 0.8 & -1.8 $\pm$ 0.8 & -1.8 $\pm$ 0.8 & -1.8 $\pm$ 0.8\\ 
    $10^3 b_{\eta, \rm{SiIII(120.7)}}$ & -1.0 $\pm$ 1.0 & -1.0 $\pm$ 1.0 & -1.0 $\pm$ 1.0 & -1.0 $\pm$ 1.0\\
    $b_{\rm HCD}$ & -0.0501 & -0.0501 & -0.0501 & -0.0501 \\
    $\beta_{\rm HCD}$ & 0.7031 & 0.7031 & 0.7031 & 0.7031 \\
    $\beta_{\rm QSO}$ & 0.2602 & 0.2602 & 0.2602 & 0.2602 \\
    $\xi_0^{\rm TP}$ & 0.7386 & 0.7386 & 0.7386 & 0.7386 \\
    $\Delta r_{\rm \parallel, QSO}(h^{-1} \rm{Mpc})$ & 0.2261 $\pm$ 0.1256 & 0.2244 $\pm$ 0.1258 & 0.2234 $\pm$ 0.1259 & 0.2220 $\pm$ 0.1260 \\
    $\sigma_{\rm \nu}$ & 7.6993 $\pm$ 0.4453 & 7.9460 $ \pm$ 0.4779 & 8.1010 $\pm$ 0.4942 & 8.2629 $\pm$ 0.5076 \\
    \hline
    \multicolumn{5}{c}{All combined}\\
    \hline
    Prob. & 0.184 & \cellcolor{lightgray!70} 0.195 & 0.188 & 0.176\\
    $\chi^2_{\rm min}$ & 9644.85 & \cellcolor{lightgray!70} 9639.20 & 9643.20 & 9649.33\\
    $\alpha_\parallel$ & 1.0454 $\pm$ 0.0218 & 1.0329 $\pm$ 0.0233 & 1.0267 $\pm$ 0.0239 & 1.0225 $\pm$ 0.0245\\
    $\alpha_\perp$ & 0.9558 $\pm$ 0.0281 & 0.9568 $\pm$ 0.0317 & 0.9621 $\pm$ 0.0326 & 0.9674 $\pm$ 0.0335\\
    $\beta_{\rm{Ly}\alpha}$ & 1.6641 $\pm$ 0.0707 & 1.7730 $\pm$ 0.0782 & 1.8312 $\pm$ 0.0826 & 1.8847 $\pm$ 0.0866\\
    $b_{\rm{Ly\alpha}}$ & -0.1172 $\pm$ 0.0019 & -0.1108 $\pm$ 0.0018 & -0.1077 $\pm$ 0.0017 & -0.1050 $\pm$ 0.0017\\
    $b_{\eta,\rm{Ly\alpha}}$ & -0.2010 $\pm$ 0.0032 & -0.2024 $\pm$ 0.0032 & -0.2032 $\pm$ 0.0032 & -0.2039 $\pm$ 0.0032\\
    $10^3 b_{\eta, \rm{CIV(eff)}}$ & -5.2 $\pm$ 2.6 & -4.9 $\pm$ 2.6 & -4.8 $\pm$ 2.6 & -4.8 $\pm$ 2.6\\
    $10^3 b_{\eta, \rm{SiII(119)}}$ & -2.7 $\pm$ 0.4 & -2.6 $\pm$ 0.4 & -2.5 $\pm$ 0.4 & -2.5 $\pm$ 0.4\\ 
    $10^3 b_{\eta, \rm{SiII(119.3)}}$ & -1.0 $\pm$ 0.4 & -1.0 $\pm$ 0.4 & -1.0 $\pm$ 0.4 & -1.0 $\pm$ 0.4\\ 
    $10^3 b_{\eta, \rm{SiII(126)}}$ & -2.2 $\pm$ 0.4 & -2.2 $\pm$ 0.4 & -2.2 $\pm$ 0.4 & -2.2 $\pm$ 0.4\\ 
    $10^3 b_{\eta, \rm{SiIII(120.7)}}$ & -3.6 $\pm$ 0.4 & -4.0 $\pm$ 0.4 & -4.2 $\pm$ 0.4 & -4.3 $\pm$ 0.4\\
    $b_{\rm HCD}$ & -0.0501 $\pm$ 0.0036 & -0.0468 $\pm$ 0.0036 & -0.0462 $\pm$ 0.0036 & -0.0466 $\pm$ 0.0036 \\
    $\beta_{\rm HCD}$ & 0.7058 $\pm$ 0.0797 & 0.6959 $\pm$ 0.0806 & 0.6957 $\pm$ 0.0807 & 0.7002 $\pm$ 0.0806 \\
    $\beta_{\rm QSO}$ & 0.2593 $\pm$ 0.0058 & 0.2610 $\pm$ 0.0059 & 0.2615 $\pm$ 0.0059 & 0.2615 $\pm$ 0.0059 \\
    $10^2 A_{\rm sky,auto}$ & 0.93 $\pm$ 0.06 & 0.93 $\pm$ 0.06 & 0.92 $\pm$ 0.06 & 0.92 $\pm$ 0.06 \\
    $10^2 A_{\rm sky,autoLyb}$ & 1.32 $\pm$ 0.09 & 1.31 $\pm$ 0.09 & 1.31 $\pm$ 0.09 & 1.31 $\pm$ 0.09 \\
    $\sigma_{\rm sky,auto}$ & 31.42 $\pm$ 1.69 & 31.84 $\pm$ 1.70 & 31.94 $\pm$ 1.70 & 31.99 $\pm$ 1.70 \\
    $\sigma_{\rm sky,autoLyb}$ & 34.23 $\pm$ 2.31 & 34.54 $\pm$ 2.32 & 34.61 $\pm$ 2.32 & 34.63 $\pm$ 2.32 \\
    $\xi_0^{\rm TP}$ & 0.7590 $\pm$ 0.0925 & 0.7323 $\pm$ 0.0916 & 0.7255 $\pm$ 0.0913 & 0.7252 $\pm$ 0.0912 \\
    $\Delta r_{\rm \parallel, QSO}(h^{-1} \rm{Mpc})$ & 0.1081 $\pm$ 0.1087 & 0.0868 $\pm$ 0.1088 & 0.0767 $\pm$ 0.1089 & 0.0683 $\pm$ 0.1089 \\
    $\sigma_{\rm \nu}$ & 6.8148 $\pm$ 0.2676 & 7.0160 $ \pm$ 0.2817 & 7.0953 $\pm$ 0.2876 & 7.1508 $\pm$ 0.2916 \\
    \hline
    \hline
    \end{tabular}
    \label{tab:uber}
\end{table*}

\subsection{Impact of reionization relics on the BAO analysis}
\label{ssec:re-mem}
After successfully reproducing the eBOSS DR16 analysis with the fiducial model, we incorporate the effects of reionization in the \lya forest, specifically by adding the second terms in Eqs.~(\ref{eq:auto-mem}) and (\ref{eq:cross-mem}) to the modeling. The impact of reionization depends inherently on its timing; however, for clarity, we first focus on incorporating a reionization model to evaluate its impact relative to the no-reionization fiducial scenario. We defer the detailed analysis of differences arising from distinct reionization histories to subsequent paragraphs. In Figure \ref{fig:baowedge}, we add reionization relics to the correlation function (depicted by the yellow dashed curve) while isolating the models from other broadband effects to emphasize the impact of reionization. Figure \ref{fig:baowedge} uses a model with late reionization -- the PySR late model described by Eq.~(\ref{eq:pysr-late}) -- because it provides the largest differences from the no-reionization scenario. To further enhance the visual contrast with the fiducial model, some parameters have been manually adjusted, and thus, the yellow dashed curve does not correspond to the best performing PySR late model. Even though reionization relics generally introduce additional broadband power to the \lya forest correlation function, the curve with reionization effects is not guaranteed to always be above the fiducial model. The power boost for this $\mu$-wedge is most easily appreciable at smaller separations ($ 25 \lessapprox r \lessapprox 80 \, h^{-1}$Mpc). 

It is worth noting that reionization relics contribute to the \lya forest correlation function in a distinct way compared to other effects introduced in \S\ref{sec:model}. The contribution of metal lines is primarily constrained by the peaks they generate near $r_\perp=0$, while DLAs also influence the power spectrum along the LOS. Although quasars are likely to emit light anisotropically \citep{2013JCAP...05..018F}, the conventional method described in \S\ref{ssec:pql} assumes quasar isotropic emission and thus only depends on $r$ with a strong decay for increasing $r$. In contrast, inhomogeneous reionization affects the power spectrum in both $k_\parallel$ and $k_\perp$. In addition, while DLAs are localized features in the transmitted flux, the effect of patchy reionization extends across the transmitted flux -- as showcased in Figure 7 of \cite{2023MNRAS.519.6162P}. Therefore, its effects cannot be identified individually at the spectral level. Furthermore, as mentioned in \S\ref{ssec:nl}, nonlinear growth and gas pressure suppression are isotropic effects, whereas reionization relics are an anisotropic effect, with the strongest imprint appearing at small wavenumbers due to the characteristic scale of ionized bubbles. Although the suppression due to peculiar velocities -- second term in Eq.~(\ref{eq:fnl}) -- is also anisotropic, it has a different behavior in both $k$ and $\mu$.

Furthermore, Figure \ref{fig:fid-lyalya} includes wedge plots for the full range of $\mu$, including metal contamination. The metals primarily contribute additional bumps in the correlation function along the LOS, as evident from the features at $r \sim$ 20, 60, and 111 $h^{-1}$Mpc in the top-left panel. Notably, the bottom-right panel of Figure \ref{fig:fid-lyalya} differs from Figure \ref{fig:baowedge}. This discrepancy is due to using a different reionization model in the former (PySR early,  which happens to provide the best fit to the data) and because Figure \ref{fig:baowedge} does not depict a best-fit model. Besides, in Figure \ref{fig:baowedge}, the broadband effects are shown separately. In general, the curve incorporating reionization in Figure \ref{fig:fid-lyalya} agrees well with the data points but shows some discrepancies with the fiducial model. While the reionization term introduces added power to the correlation function, the red curve (reionization) can appear higher or lower than the black curve (fiducial model) depending on the fit since the fitting process optimizes alignment with data points, balancing the inclusion of reionization relics. Moreover, as implied by Eq.~(\ref{eq:auto-mem}), the ratio of the reionization term to the conventional \lya term decreases with increasing $\mu$, thus diminishing the impact of reionization. As a result, the level of discrepancy between the two models varies across different wedges. For wedges with $\mu \approx 1$, the red dashed curve tends to align more closely with the black fiducial curve, particularly at small separations below the BAO scale, where the data also have small error bars. 

The BAO parameters $\alpha_{\rm \parallel}$ and $\alpha_{\rm \perp}$ have only minor deviations when incorporating the effects of reionization. This limited impact is expected, as the reionization term introduces relatively small perturbations to the correlation function near the BAO peak, as illustrated in Figure \ref{fig:baowedge} (see also Figure 4 of \citealt{2023MNRAS.520.4853M}). To visualize these subtle changes, we examine the residuals of the 3D \lya forest auto-correlation function. The scenario incorporating reionization relics -- described by the PySR early model -- is shown in Figure \ref{fig:dif_xi_withreio}, while the fiducial model is depicted in Figure \ref{fig:dif_xi_noreio}. Given the difficulty of discerning the difference by eye, we plot their ratio in Figure \ref{fig:xi_comp}. Near the typical BAO scale of 100 $h^{-1}$Mpc, there is an arc of lighter pixels, indicating minimal impact of reionization relics on the BAO peak location. Additional arcs, such as the one observed around 75 $h^{-1}$ Mpc, correspond to transition points between the reionization model giving a larger or smaller $\xi$ compared to the no-reionization model, which explains why there is a preference for blue points at $25 < r < 50 \ h^{-1}$Mpc (for example, see the bottom-right panel of Figure \ref{fig:fid-lyalya}). Note that some of these transitions are coincidentally located near the recovery from metal bumps, such as after the bump at $r = 60 \ h^{-1}$Mpc. Even though the location of these features are consistent across models, their amplitudes can vary due to the broadband effects of reionization. Furthermore, the Fourier transform of a Yukawa-like function yields a Lorentzian-type curve, which contributes to the stronger discrepancies -- in absolute value -- between models at smaller separations, as illustrated in Figure \ref{fig:xi_comp}.

To quantify the influence of reionization in this analysis, we tabulate the best-fit parameter values in Table \ref{tab:uber}. From top to bottom, the table lists results for $\rm Ly\alpha(\rm Ly\alpha)\times \rm Ly\alpha(\rm Ly\alpha)$, $\rm Ly\alpha(\rm Ly\alpha)\times \rm quasar$, and the combined correlation -- including $\rm Ly\alpha(\rm Ly\alpha)\times \rm Ly\alpha(\rm Ly\beta)$, $\rm Ly\alpha(\rm Ly\beta)\times \rm quasar$, and the two aforementioned correlation functions. Consistent with expectations, the BAO parameters exhibit only minor deviations upon including reionization relics. For instance, for the PySR early model, $\alpha_{\rm \parallel}$ deviates by $1.9\%$ and $\alpha_{\rm \perp}$ by $1.0\%$. In contrast, other cosmological parameters show more significant shifts. The \lya forest bias factor ($b_{\rm{Ly\alpha}}$) shifts by $4.7\%$, surpassing the $1\sigma$ range of the fiducial model. Likewise, the corresponding RSD parameter ($\beta_{\rm{Ly\alpha}}$) shifts by $5.4\%$. Metal line biases remain largely unchanged, which is expected because the reionization relics produce a broadband effect and is only included in the modeling of the \lya term. Other nuisance parameters, like $A_{\rm sky}$ and $\sigma_{\rm sky}$, also remain essentially unchanged. 

The results for the cross-correlation mirror those of the auto-correlation for the BAO parameters, even after incorporating the reionization term. However, significant variations are present, particularly in the \lya bias and RSD parameter, while other non-BAO parameters remain largely unaffected. A notable difference from the auto-correlation function is the larger deviation from unity between $\alpha_\parallel$ and $\alpha_\perp$ in the cross-correlation dataset, a trend consistent with what was observed for the fiducial model. Furthermore, $\beta_{{\rm Ly}\alpha}$ has a considerably larger value in the cross-correlation analysis than in the auto-correlation -- $\beta_{{\rm Ly}\alpha}^{\rm cross} = 1.9311 \pm 0.1428$ versus $\beta_{{\rm Ly}\alpha}^{\rm auto} = 1.6331 \pm 0.0864$. This divergence is further reinforced by the inclusion of reionization relics. These discrepancies between the two datasets likely arise from the lower constraining power of the cross-correlation and also due to the implicit subspace of parameter space that is selected by fixing some of the model parameters. 

\begin{figure*}
    \centering
    \includegraphics[width=\linewidth]{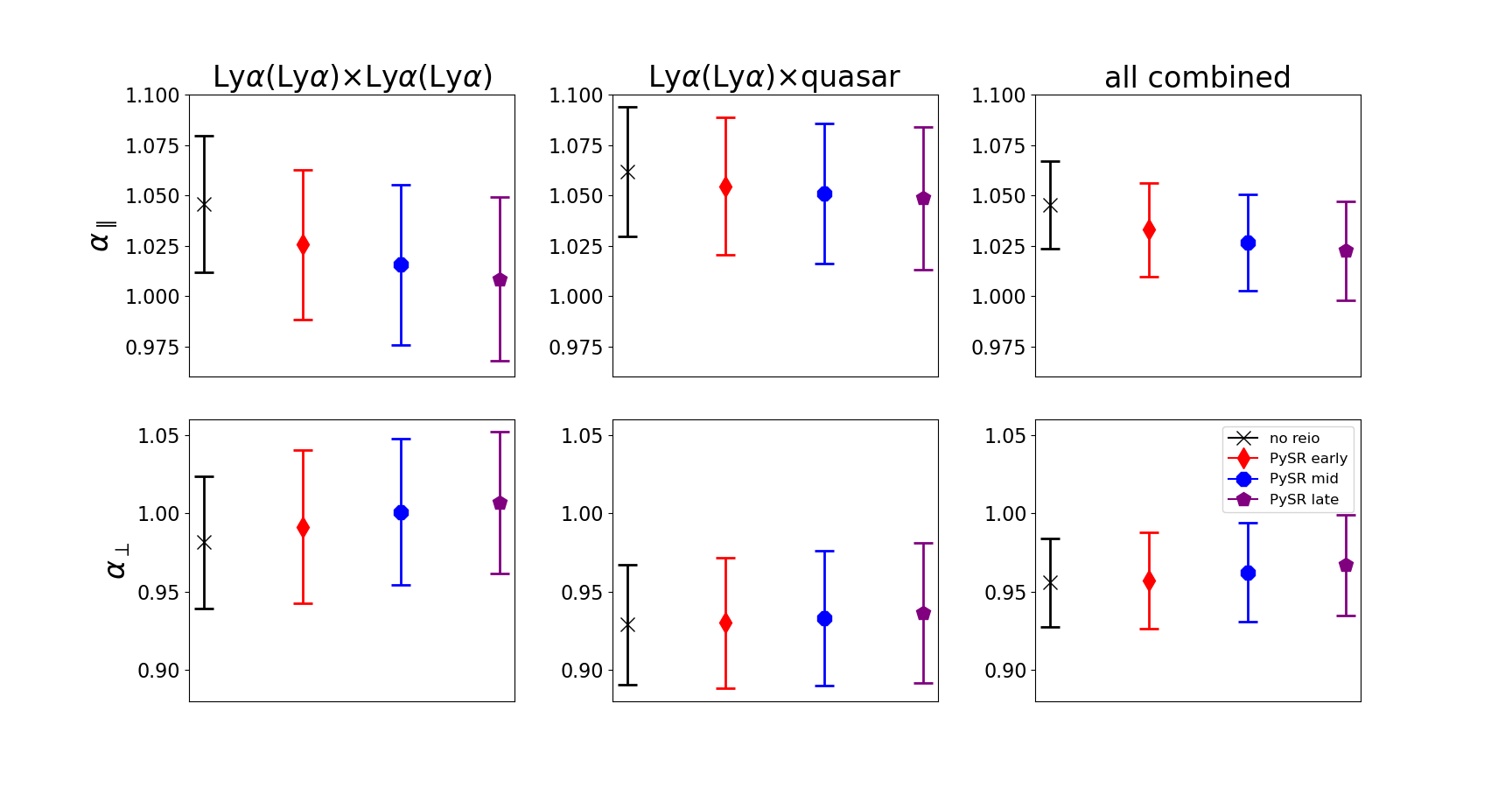}
    \caption{Comparison of constraints on the BAO parameters $\alpha_\parallel$ and $\alpha_\perp$ derived from the fiducial model (no impact of reionization) and various reionization scenarios modeled using the PySR template. Results are presented for the auto-correlation (left column), cross-correlation (center column), and the combined analysis (right column). The top row displays the inferred values of $\alpha_\parallel$, while the bottom row corresponds to $\alpha_\perp$. Error bars indicate the $1\sigma$ confidence, as detailed in Table \ref{tab:uber}. Across all reionization timelines, the inferred values remain consistent with those obtained from the no-reionization model. Note that the fixed reionization models do not account for uncertainties in the reionization timeline, which dominate the templates' error budget. For comparison, the template in Appendix \ref{app:yuk-free}, which marginalizes over some of this uncertainty, yields $\Delta \alpha_\parallel = 0.0441$ and $\Delta \alpha_\perp = 0.0483$ for the Ly$\alpha$ auto correlation.}
    \label{fig:error_alpha}
\end{figure*}

\begin{figure*}
    \centering
    \includegraphics[width=\linewidth]{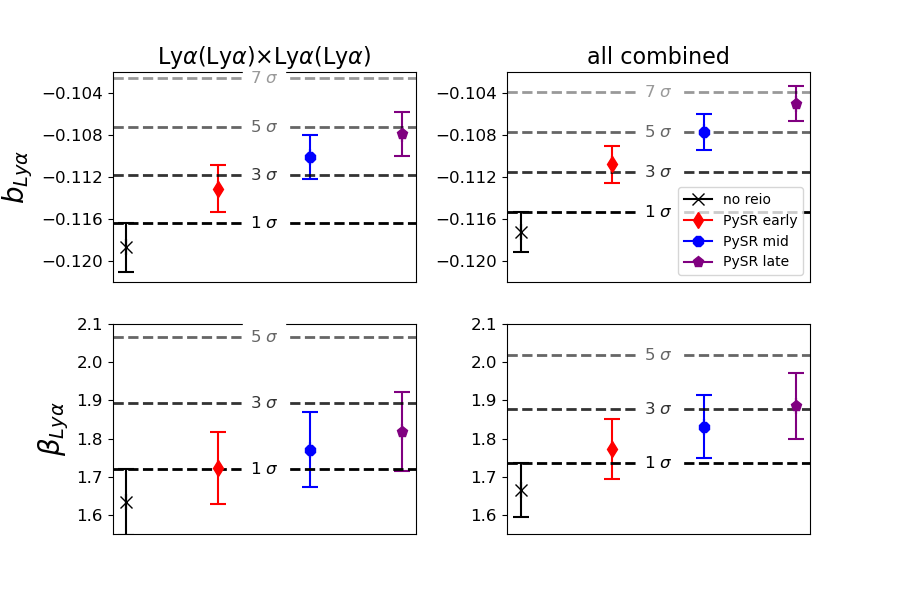}
    \caption{Similar to Figure \ref{fig:error_alpha} but focusing on the \lya bias and redshift space distortion parameter. The left and right columns correspond to the auto-correlation and the full correlation combination, respectively. The upper (lower) row presents the constraints on $b_{{\rm Ly}\alpha}$ ($\beta_{{\rm Ly}\alpha}$). The dashed lines describe different confidence levels based on the $1\sigma$ errors of the fiducial model. The fixed reionization models do not marginalize over the uncertainty of the reionization history, which dominates the templates' error budget. For reference, the template presented in Appendix \ref{app:yuk-free}, which incorporates this marginalization, leads to an increased error on the RSD parameter, $\Delta \beta_{\mathrm{Ly}\alpha} = 0.1100$, in the Ly$\alpha$ auto-correlation. In contrast, the error on the bias parameter has a similar value to that of the fixed models ($\Delta b_{\mathrm{Ly}\alpha} \approx 0.0022$).}
    \label{fig:error_biases}
\end{figure*}

We now change our focus to the influence of different reionization timelines, represented in our work by the early, mid, and late scenarios. These scenarios consistently yield $\alpha_{\rm \parallel}$ and $\alpha_{\rm \perp}$ values within $1\sigma$ agreement with the fiducial model across all correlations, as shown in Figure \ref{fig:error_alpha}. An interesting trend is the apparent convergence of $\alpha_{\rm \parallel}$ and $\alpha_{\rm \perp}$ toward unity with progressively later reionization models, although this trend appears to lack statistical or physical significance to our best knowledge. As mentioned above, the strongest effects of reionization are observed in $b_{\rm{Ly\alpha}}$ and $\beta_{\rm{Ly\alpha}}$ with a shift in the PySR early scenario of $5.4\%$ for the RSD parameter. The deviation increases with later reionization scenarios, as shown in Figure \ref{fig:error_biases}. From the fiducial to the PySR early, mid, and late models, the influence of reionization grows progressively, leading to an increase in both $\beta_{\rm{Ly\alpha}}$ and $b_{\rm{Ly\alpha}}$. Notably, in the PySR late model, $\beta_{\rm{Ly\alpha}}$ exceeds the $3\sigma$ range of the fiducial model, while $b_{\rm{Ly\alpha}}$ can reach shifts greater than $6\sigma$. These results underscore the substantial impact of reionization relics on the \lya forest. 

\begin{figure}
    \centering
    \includegraphics[width=\linewidth]{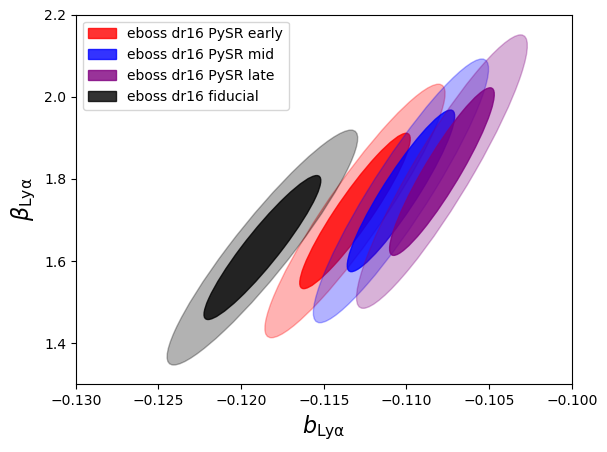}
    \caption{Contour plot of $\beta_{\rm Ly\alpha}$ and $b_{\rm Ly\alpha}$ for the fiducial and the PySR reionization models. The plot highlights a preference for larger RSD parameters and  smaller (in absolute value) biases in later reionization scenarios. As indicated in Eq.~(\ref{eq:auto-mem}), a smaller bias suppresses the overall \lya power while a larger RSD counteracts this suppression to some extent while simultaneously reducing the relative importance of the reionization term compared to the baseline \lya power spectrum.}
    \label{fig:contour-pysr-fid}
\end{figure}

We hypothesize that the observed effect on the biases is due to the minimizer attempting to balance the additional power introduced by reionization relics. According to Eq.~(\ref{eq:auto-mem}), the \lya contribution is modulated by $\beta_{\rm{Ly}\alpha}$ to the second power via the Kaiser term, while the reionization term depends on $\beta_{\rm{Ly}\alpha}$ to the first power instead. Consequently, as $\beta_{\rm{Ly}\alpha}$ increases, the relative contribution of the reionization perturbation diminishes. In scenarios with later reionization, the amplitude of the reionization term increases, prompting a corresponding increase in $\beta_{\rm{Ly}\alpha}$ to suppress the relative influence of the reionization term. An analog trend is evident for $b_{\rm{Ly\alpha}}$ in Eq.~(\ref{eq:auto-mem}), which explains why its value decreases (in absolute value) with the increasing impact of reionization relics, as can be seen in Figure \ref{fig:contour-pysr-fid}. Note that this interplay between the allowed freedom of the bias -- and RSD -- parameters and the reionization model is what ultimately leads to a better fit of the early reionization models, even when allowing for models that span different reionization histories (see Appendix \ref{app:yuk-free}) beyond the three fixed scenarios.

We note that our covariance matrices are underestimated due to the exclusion of cross-covariance terms among the different correlation functions in the eBOSS DR16 analysis, which has been shown to result in a 10$\%$ underestimation of uncertainties in BAO parameters \citep{2024arXiv240403004C}. This underestimation likely extends to other parameters but it is insufficient to account for the magnitude of the observed differences in bias and RSD parameters. However, non-BAO parameters are known to be sensitive to analysis choices \citep[see Figure 11 and surrounding text in][]{2024arXiv240403001D}. This variability could obscure the quantitative significance of the shifts in $b_{\rm{Ly\alpha}}$ and $\beta_{\rm{Ly}\alpha}$. Nonetheless, the qualitative preference for increased $b_{\rm{Ly\alpha}}$ and $\beta_{\rm{Ly}\alpha}$ in delayed reionization scenarios is likely robust and independent of analysis choices for the eBOSS DR16 catalog. 

Although it uses a different quasar catalog with an increased number of tracers, the DESI Year 1 \lya BAO analysis has not yet achieved a signal-to-noise ratio comparable to eBOSS DR16. Nevertheless, its increased density of tracers provides valuable insights into the range of non-BAO parameters. Specifically, DESI also exhibits a preference for larger \lya bias and RSD parameters ($b^{\rm DESI}_{{\rm Ly}\alpha} = -0.1078^{+0.0045}_{-0.0054}$ and $\beta_{{\rm Ly}\alpha}^{\rm DESI} = 1.743^{+0.074}_{-0.100}$) than the eBOSS DR16. 

While we observe significant deviations in the \lya bias and RSD parameters across different reionization models, the quasar RSD parameter ($\beta_{\rm q} \sim 0.26$) remains highly consistent in the combined analysis. However, the systematic trend of a greater RSD impact on the \lya forest compared to that on quasars persists across all models, in agreement with DESI and eBOSS findings.

Consistent with the findings from the eBOSS DR16 analysis, the reionization models also yield $\Delta r_\parallel$ values consistent with zero within the statistical uncertainties for the combined analysis. Hence, quasar redshifts remain unbiased. In addition, the reionization models recover values of $\sigma_v$ that align with the estimate of the fiducial model ($\sigma_v = 6.8148 \pm 0.2676$), and thus indicate that the Lorentzian profile, utilized in the nonlinear quasar modeling to capture the impact of random peculiar velocities and redshift errors (see \S\ref{ssec:nl}), remains effective. 

Our analysis incorporates four distinct models and multiple correlation functions, each with varying degrees of constraining power and thus different preferences for the model that best fits the data. The \lya auto-correlation has a clear preference for the PySR early model, as indicated by its highest $p$-value and lowest minimum $\chi^2$, as tabulated in Table \ref{tab:uber}. While this preference decreases progressively as the reionization timeline shifts from early to late, models with reionization relics consistently outperform the fiducial model. However, this trend does not hold for the $\rm{Ly} \alpha \times quasar$ cross-correlation, where the fiducial model provides a better fit compared to models that include reionization relics. This outcome can be partly attributed to the relatively lower quality of the data used in the cross-correlation analysis. Another contributing factor is that some of the model parameters are fixed, such as the quasar RSD parameter, the carbon bias, and the quasar proximity parameter. These fixed parameters restrict the fit to a constrained subspace of the parameter manifold, potentially biasing the computation against the inclusion of reionization relics. The fixed values, selected from \cite{2020ApJ...901..153D}, inherently exclude the impact of reionization and may slightly overestimate the fiducial cross-correlation value. To reinforce this last point, we multiply the fiducial cross-power spectrum by 0.9 and obtain a ``better'' fit, yielding a higher $p$-value and a lower minimum $\chi^2$. We interpret this as further evidence that the selected parameter space may overestimate the cross-correlation. 

When combining all correlations and allowing all model parameters to vary freely, a slight preference for some reionization over the fiducial models re-emerges. Our best performing model corresponds to the PySR early model achieving a $p$-value of 0.195 and $\chi_{\rm min}^2 = 9639.20$, compared to 0.184 and 9644.85, respectively, for the fiducial model. Once again, we caution against directly comparing the $p$-values and $\chi^2_{\rm min}$ values across different correlation functions to characterize the performance of the models. Such comparisons are problematic due to different number of free parameters and disparities in the quality of the data.

To contextualize the significance of the tantalizing evidence for reionization relics observed in the eBOSS DR16 correlations, it is important to recognize that other components of the model can yield comparable improvements in fit quality. For instance, in Table 5 of \cite{2017A&A...603A..12B}, the inclusion of fluctuations in the ionizing UV flux \citep{2014MNRAS.442..187G,2014PhRvD..89h3010P,2023MNRAS.520..948L} resulted in a decrease in the minimum chi-squared value by $\Delta \chi^2_{\rm min} = -4.9$ and an associated increase in probability of 0.027 when compared to a model that ignores UV clustering in the \lya auto-correlation\footnote{Note that the DESI Year 1 \lya BAO analysis did not find a preference for the inclusion of UV fluctuations \citep{2024arXiv240403001D}.}. For reference, our best-performing reionization model for the \lya auto-correlation achieves a $\Delta \chi^2_{\rm min} = -7.08$, which corresponds to ``strong evidence'' according to the standard usage \citep{kass1995bayes}\footnote{For the Akaike information criterion \citep[AIC,][]{1100705}, defined as $\mathrm{AIC} = \chi_{\rm min}^2 + 2n$, where $n$ the number of fitting parameters, strong evidence for the model with the smallest AIC is achieved when $6 \leq \Delta \mathrm{AIC} < 10$. In contrast, a range of $2 \leq \Delta \mathrm{AIC} < 6$ indicates ``positive evidence'' for our best performing reionization model.}, and a probability increase of 0.043 relative to the fiducial model. While a direct comparison between UV clustering and reionization models is not strictly fair due to differences in the data and the degrees of freedom, it is worth emphasizing that, as presented in Appendix \ref{app:yuk-free}, we explore a Yukawa template with a single free parameter -- matching the change in degrees of freedom associated with the UV clustering model in \cite{2017A&A...603A..12B}. This alternative template resulted in a similar statistical improvement, with $\Delta \chi^2_{\rm min} = -9$, also corresponding to strong evidence in the Jeffrey's scale, and an accompanying probability increase of 0.049, alongside a preference for early reionization. Note that the UV clustering effect originates deep in the post-reionization era as opposed to the reionization relics, which are seeded during the epoch of reionization. Their broadband impact is distinct since UV clustering can lead to a decrement in the power spectrum at $k\approx 3\times10^{-3}\, h$ Mpc$^{-1}$ as depicted in Figure 2 of \cite{2014MNRAS.442..187G}. 

As shown in Table \ref{tab:uber} and discussed above, our reionization models provide constraints on the BAO parameters $(\alpha_{\rm \parallel}, \alpha_{\rm \perp})$ comparable to those obtained with the fiducial model. To extract the BAO parameters, we adopt a peak-smooth decomposition -- see Eq.~(\ref{eq:pql}) -- which expresses the correlation function as
\begin{eqnarray}
    \label{eq:cor-dec}
    \xi(r_{\rm \parallel},r_{\rm \perp},\alpha_{\rm \parallel},\alpha_{\rm \perp})=\xi_{\rm sm}(r_{\rm \parallel},r_{\rm \perp})+\xi_{\rm peak}(\alpha_{\rm \parallel}r_{\rm \parallel},\alpha_{\rm \perp}r_{\rm \perp}) \, .
\end{eqnarray}
The BAO parameters are defined as
\begin{eqnarray}
    \label{eq:bao-definition}
    \alpha_{\rm \parallel}=\frac{[D_H(z_{\rm eff})/r_d]}{[D_H(z_{\rm eff})/r_d]_{\rm fid}} \  {\rm and} \ \ \  \alpha_{\rm \perp}=\frac{[D_M(z_{\rm eff})/r_d]}{[D_M(z_{\rm eff})/r_d]_{\rm fid}} \, ,
\end{eqnarray}
where $D_M(z)=(1+z)D_A(z)$ is the comoving angular diameter distance, $D_H(z)=c/H(z)$ is the Hubble distance, and $r_d$ is the sound horizon. The fiducial values for these ratios are calculated within the $\Lambda$CDM cosmology of \cite{2016A&A...594A..13P}:
\begin{eqnarray}
    \label{eq:bao-fid-value}
    \setstretch{1.3}
    \begin{cases}
    [D_H(z=2.334)/r_d]_{\rm fid}& = 8.6011 \\
    [D_M(z=2.334)/r_d]_{\rm fid}& = 39.2035
    \end{cases}
\end{eqnarray}
Substituting the fiducial values, the constraints on $\alpha_{\rm \parallel}$ and $\alpha_{\rm \perp}$ can be converted into constraints on the corresponding distance ratios. When combining all four correlation functions and using the fiducial model, the resulting distances are
\begin{eqnarray}
    \label{eq:bao-fid-all}
    \setstretch{1.3}
    \begin{cases}
    D_H(z=2.334)/r_d = 8.99^{+0.19}_{-0.19} \\
    D_M(z=2.334)/r_d = 37.5^{+1.1}_{-1.1}
    \end{cases}
\end{eqnarray}
Similarly, for the PySR early reionization model, we obtain
\begin{eqnarray}
    \label{eq:bao-reio-all}
    \setstretch{1.3}
    \begin{cases}
    D_H(z=2.334)/r_d = 8.88^{+0.20}_{-0.20} \\
    D_M(z=2.334)/r_d = 37.5^{+1.2}_{-1.2}
    \end{cases}
\end{eqnarray}
Since the BAO parameters have slightly non-Gaussian uncertainties, it is not straightforward to define appropriate confidence intervals. While \cite{2020ApJ...901..153D} used fastMC to generate 1000 realizations and map $(68.27\%,95.45\%)$ confidence levels to $\Delta\chi^2$ values, we opt for a different approach due to limited computational resources and environmental impact. Specifically, we use {\sc iminuit} to estimate the $1\sigma$ errors for the BAO parameters. When calculating the $1\sigma$ errors, the fitting algorithm strikes a balance between minimizing the residuals of the sum of squared differences and approximating a Gaussian fit. Note that this methodology has an over-reliance on the Gaussianity of the data for higher confidence levels.

The eBOSS DR16 data are better fitted by a model incorporating reionization relics, with an earlier reionization timeline featuring a midpoint at $z_{\rm re} = 8.08$ and optical depth of $\tau_{\rm reio} = 0.0583$. This value of midpoint is higher than the midpoint of reionization inferred indirectly from CMB data ($z \approx 7.7$) but remains within the $1\sigma$ uncertainty reported by \cite{2020A&A...641A...6P}. Moreover, the midpoint of reionization of our early model aligns well with the value inferred from the Planck high frequency instrument $z_{\rm re} = 8.14 \pm 0.61$ \citep{2020A&A...635A..99P} and that inferred using Planck's low multipole polarization and temperature data, $z_{\rm re} = 8.51 \pm 0.52$ \citep{2021MNRAS.507.1072D} as illustrated in Figure \ref{fig:xHI}. Likewise, they also agree with the allowed ranges for the moderate heating models of \cite{2017ApJ...847...63O} although their early model occurs earlier ($z_{\rm re} = 8.80$ and $\tau_{\rm reio} = 0.0698$). Furthermore, our findings are also in agreement with expectations from the thermal evolution of the IGM after reionization \citep{2016MNRAS.460.1885U}. Our results align with recent findings driven by data from the James Webb Space Telescope \citep[JWST;][]{2006SSRv..123..485G}. JWST's capabilities enable unprecedented observations of high-redshift quasars \citep[e.g.,][]{2023ApJ...950...68E} and high-redshift galaxies \citep[e.g.,][]{2023MNRAS.518.4755A}. JWST data suggest that structure formation may have progressed slightly faster than predicted by the concordance model \citep{2024MNRAS.535L..37M}, which could be interpreted as evidence for earlier reionization. For example, measurements of damping wings in high-redshift galaxies enabled by JWST indicate a preference for early reionization \citep{2025arXiv250111702M} while the detection of Ly$\alpha$ emission at $z = 13$ suggests that massive galaxies exist at such redshift, causing their surroundings to ionize at an early stage \citep{2025Natur.639..897W}. In contrast, late reionization is favored by measurements of the \lya effective optical depth \citep{2022MNRAS.514...55B}, ionizing photons mean free path \citep{2021ApJ...917L..37C}, CMB data using an universal shape for the timeline of reionization that captures the dependence of reionization with cosmology instead of a symmetric hyperbolic tangent \citep{2024arXiv240513680M}, and potentially measurements of the \lya mean free path at $3.2 \leq z \leq 4.6$ \citep{2025ApJ...981L..27G}. 

While our primary analysis was restricted to reionization models described in \cite{2023MNRAS.520.4853M} -- namely, the early and late reionization scenarios that span a range around Planck's $1\sigma$ uncertainty -- Appendix \ref{app:yuk-free} introduces a Yukawa model with a free amplitude, allowing the exploration of even earlier reionization scenarios. Remarkably, the best fit using the combination of all correlation functions corresponds to a reionization amplitude of $A_{\rm re} = 0.1900 \pm 0.0702$, which is remarkably close to the value associated with the Yukawa early reionization model ($A_{\rm re} = 0.2088$). 

\section{Discussion}
\label{sec:disc}
Our results indicate that eBOSS data overall favors models that incorporate the impact of reionization in the post-reionization IGM over conventional models that neglect this effect. Nonetheless, there remain some interesting avenues to consider regarding the modeling of reionization relics and their effects, particularly at higher redshifts $z \geq 3$.

The high-redshift \lya forest window into the IGM is increasingly recognized as a critical probe of cosmological structure formation. Alongside this growing importance is the necessity for better tools to derive accurate statistical insights from \lya forest spectra. For instance, \cite{2023ApJS..269....4S} recently introduced Quasar Factor Analysis (QFA), a new continuum fitting method based on unsupervised learning and latent factor analysis. This novel methodology achieves remarkable accuracy, with errors of less than one percent in the \lya forest one-dimensional power spectrum at $z = 3$. Such advancements are promising for studies leveraging the high-redshift \lya window, particularly for analyses accounting for reionization relics in the \lya forest or its cross-correlations \citep{2025MNRAS.536.1645M}. Moreover, the high-redshift \lya window provides a valuable opportunity to investigate small-scale effects before nonlinearities become dominant, offering an ideal environment to constrain dark matter candidates (see e.g. \citealt{2023MNRAS.519.6162P,2024arXiv241107970Z}; Zhang et al. {\it in prep.}). Given these ambitious cosmological programs at high redshift, an open question arises: how might improved continuum-fitting methods, specifically optimized for high-redshifts \lya skewers, influence the results presented in Table \ref{tab:uber}? We leave this exploration to future work. 

Likewise, alternative continuum-fitting methodologies \citep[e.g. ][]{2024arXiv240506743T,2023ApJS..269....4S,2024MNRAS.533.3312G} are also relevant due to their impact on the standard \lya forest analysis. The standard continuum-fitting procedure, which assumes a universal template for the quasar continuum $\overline{C} (\lambda_{\rm rf})$ \citep[e.g.,][]{2024MNRAS.528.6666R}, often modulated by a low-order polynomial, can sometimes yield unphysical results such as negative continuum estimates for specific spectra. For example, \cite{2024arXiv240403001D} reported that 3.6\% and 4.4\% of the forests in the Ly$\alpha$ and Ly$\beta$ regions, respectively, were discarded due to negative continuum.

Furthermore, in this work, we focused on the impact of inhomogeneous \ion{H}{I} reionization. While hydrogen constitutes the majority of the baryonic content of the Universe, a significant fraction is helium. The first reionization of helium occurs concurrently with hydrogen reionization; however, the reionization of \ion{He}{II} takes place later, around $z \sim 3$ \citep{2021MNRAS.506.4389G}, introducing further disruptions to the thermal state of the IGM that subsequently affect the \lya forest \citep{2015MNRAS.447.2503G}. Consequently, we emphasize that the impact of \ion{He}{II} reionization on \lya forest correlations is likely to alter the best-fit parameters and must also be investigated. Recent studies \citep[e.g.,][]{2023JCAP...10..037B,2024JCAP...07..029F} have dedicated efforts along these lines but lack the mass resolution necessary to fully resolve the high-entropy mean-density (HEMD) phase of the IGM evolution \citep{2018MNRAS.474.2173H}, which is crucial for modeling the long-lasting relics of \ion{H}{I} reionization \citep{2019MNRAS.487.1047M,2020MNRAS.499.1640M}. Achieving high mass resolution is essential for accurately tracking the response of gas to the passage of an ionization front, as the Jeans length increases with temperature. 

Even without being able to self-consistently resolve the HEMD gas, the quasar-driven \ion{He}{II} reionization is fundamentally different from the physics that governs the galaxy-driven \ion{H}{I} reionization. The latter is ionized via a two-photon population of ``long'' and ``short'' mean-free path photons \citep{2020MNRAS.496.4372U}, while the former only has short-mean-free path photons. The resulting ionization fluctuations in \ion{He}{II} reionization peak on $\sim 50$ Mpc scales \citep{2009ApJ...694..842M} while \ion{H}{I} fluctuations peak at larger scales. Although both can boost the large-scale power of the three-dimensional \lya forest power spectrum (e.g., see Figure 13 of \citealt{2017ApJ...841...87L} and Figure 5 of \citealt{2020MNRAS.499.1640M}), the way they affect the one-dimensional power spectrum is different. While \ion{H}{I} reionization at the relevant  ($z \leq 3.5$) primarily boosts the large-scale power of $P^{\rm 1D}_{\rm F}$, \ion{He}{II} reionization has a broader impact with deviations present up to quite small scales ($k \approx 3 \, h$ Mpc$^{-1}$) as depicted in Figure 10 of \cite{2020MNRAS.496.4372U}.

Regardless of the effects of \ion{He}{II} reionization in the IGM, our findings conclusively indicate that the eBOSS DR16 \lya correlations prefer an additional source of power. Whether this trend is entirely due to the HEMD phase of the temperature-density relation -- responsible for the long-lasting relics from \ion{H}{I} reionization -- or involves a combination of \ion{He}{II} reionization relics mixed in, remains an open question to be explored in future work. Besides, there is also the possibility that another unknown broadband component that mimics the characteristic shape of the H I reionization relics may be responsible for this behavior.

Although eBOSS data appear to favor the inclusion of reionization effects, this preference has only a minor impact on cosmology derived from BAO analyses, as evidenced by the small shifts  on $\alpha_\parallel$ and $\alpha_{\perp}$ (see Figure \ref{fig:error_alpha}). In contrast, the significant influence on bias and redshift-space distortion parameters underscores the strong effect on full-shape analyses of the \lya forest \citep{2018JCAP...01..003F,2023PhRvL.130s1003C,2023MNRAS.518.2567G}, with the potential to introduce significant biases. Future work will explore the signatures of reionization relics in measurements of the one-dimensional \lya forest power spectrum.

Nonlinear broadening of the BAO peak can substantially affect the estimated BAO errors \citep{2024arXiv240403001D}, motivating the need for more sophisticated and accurate nonlinear modeling approaches. In \S\ref{ssec:nl}, we discussed the standard nonlinear modeling applied in \lya BAO analyses; however, the treatment of the quasars remains relatively simplistic compared to that of the auto-correlation. \cite{2022JCAP...09..070G} and \cite{2025arXiv250104770C} introduced significant improvements to the nonlinear modeling of the cross-correlation. Despite these advancements, we chose not to adopt their formalism in this work to maintain consistency with the modeling choices employed in the eBOSS DR16 analysis.

Previous works have considered the addition of broadband polynomial corrections to the baseline analysis of the \lya correlation functions. The primary motivation of such tests is to demonstrate that the BAO parameters remain stable even in the presence of potential systematic effects unaccounted for in the modeling. These validation tests have relevant parallels to the memory of reionization, as their imprint manifests as a broadband effect.

The broadband terms are often introduced with informative priors on certain non-BAO parameters and are coupled with a restricted separation range to minimize their influence at small separations. (For example, \citealt{2017A&A...603A..12B} uses  $40 < r < 180 \, h^{-1}$Mpc). Compared to their baseline analyses, \cite{2017A&A...603A..12B} and \cite{2024arXiv240403001D} found slightly larger $\alpha_\parallel$ and smaller $\alpha_\perp$ when including broadband terms. In contrast, \cite{2020ApJ...901..153D} reported that the inclusion of broadband terms with physical priors resulted in a slight increase in both BAO parameters relative to the baseline model. Our results with reionization relics (see Figure \ref{fig:error_alpha}) reveal a distinct trend: the broadband effects of reionization lead to a small negative shift in $\alpha_\parallel$ and a positive shift in $\alpha_\perp$. 

These seemingly contradictory results are not necessarily indicative of tension. The implementation of the broadband polynomial correction adds 12 new free parameters per correlation function to the analysis. This addition is paired with a restricted range of separations and relies on priors to break degeneracies in parameters that require information from small separations. Conversely, the model presented in this work introduces no additional free parameters\footnote{In Appendix \ref{app:yuk-free}, we introduce a reionization model with one extra parameter, which recovers the trends identified in the main analysis.} and utilizes the same separation range as the fiducial analysis\footnote{A run of the PySR early model restricted to the range $40 < r < 180 \, h^{-1}$Mpc also recovers the same trend for the BAO parameters, i.e. a negative shift in $\alpha_\parallel$ and a positive in $\alpha_\perp$.}. Even a qualitative comparison by eye can already discern subtle differences between our model (Figure \ref{fig:fid-lyalya}) and the broadband corrections presented in Figure 18 of \cite{2020ApJ...901..153D}, Figure 14 in \cite{2017A&A...603A..12B}, and Figure 4 in \cite{2024arXiv240403001D}.

Similarly, \cite{2020ApJ...901..153D} reported that shifts in the BAO parameters due to the broadband corrections -- with physical priors -- were accompanied by a significant decrement in $\beta_{{\rm Ly}\alpha}$ and a strong reduction in the bias $b_{\eta,{\rm Ly}\alpha}$ (see their Table 11). When the priors were relaxed, however, the RSD parameter was found to increase relative to the baseline value, while the bias decreased more sharply. Notably, the no-prior broadband model also produced trends in the BAO parameters similar to those observed in this work, specifically a negative shift in $\alpha_\parallel$ and a positive shift in $\alpha_\perp$.  Our findings align more closely with the no-prior results. Table \ref{tab:uber} shows an increase in the RSD parameter, accompanied by a decrease in $b_{\eta,{\rm Ly}\alpha}$ with respect to the fiducial scenario\footnote{An analysis of the PySR early model restricted to $40 < r < 180 \, h^{-1}$Mpc  obtains a negative shift in the RSD parameter instead, in agreement with the physical prior broadband test used in \cite{2020ApJ...901..153D}, while still exhibiting a different trend for the BAO parameters.}. 

\section{Conclusions} 
\label{sec:con}
The \lya forest correlation functions have proven successful for probing the evolution of the Universe at redshifts that are typically inaccessible to galaxy surveys. In the era of stage IV spectroscopic surveys, DESI has already produced intriguing findings, including a potential preference for dynamic dark energy \citep{2024arXiv240403002D}, which could have significant repercussions for cosmological constraints on the sum of the neutrino masses \citep[e.g.,][]{2024arXiv240715640D}. With the enhanced statistical power provided by DESI observations and the anticipated capabilities of next-generation stage V spectroscopic instruments, such as the MUltiplexed Survey Telescope \citep[MUST;][]{2024arXiv241107970Z}, addressing potential biases from known contaminants in the \lya forest correlations is a critical task. 

This work builds extensively upon prior efforts that have meticulously identified various contaminants, systematics, and modeling challenges in \lya correlations \citep[][and references therein]{2020ApJ...901..153D,2017A&A...603A..12B,2023JCAP...11..045G}. Here, we have investigated the impact of an additional factor affecting \lya forest correlation functions, namely the impact of inhomogeneous reionization. Reionization leaves enduring imprints that survive for billions of years, well into the post-reionization era \citep[e.g.,][]{2019MNRAS.487.1047M}. These reionization relics induce a broadband enhancement in power within the correlation functions \citep{2023MNRAS.520.4853M}, and thus impact full-shape analysis significantly.

Here, we present the first analysis of the impact of reionization on \lya forest correlations using the publicly available eBOSS DR16. Our findings can be summarized as follows:
\begin{itemize} \setlength\itemsep{0.5em}
    \item The BAO parameters ($\alpha_\parallel$, $\alpha_\perp$) exhibit minimal sensitivity to the inclusion of reionization effects in the modeling, irrespective of the specific reionization model and scenario considered. The largest shifts are $\Delta \alpha_\parallel = -0.0373 $ and $\Delta \alpha_\perp = 0.0255 $, both arising from the PySR late model applied to the \lya auto-correlation function. These shifts remain well within the $1\sigma$ confidence level of the fiducial no-reionization model (see Figure \ref{fig:error_alpha}).
    \item Overall, the eBOSS DR16 \lya correlation data are better fitted by models that include the impact of reionization. This result represents the first tantalizing evidence of reionization relics persisting well into the post-reionization era. However, the preference is modest; for instance, an analysis encompassing all four correlation functions for the PySR early model yields $\chi^{2}_{\rm min} = 9639.20$ with a corresponding probability of 0.195, compared to $\chi^{2}_{\rm min} = 9644.85$ and a probability of 0.184 for the fiducial model without reionization effects.
    \item The inclusion of reionization effects into the modeling can have drastic influence on the \lya bias and RSD parameters, particularly for late reionization scenarios (see Figure \ref{fig:error_biases}). Notably, our PySR late reionization model leads to a positive shift in $\beta_{{\rm Ly}\alpha}$ exceeding $3\sigma$, accompanied by a positive shift in $b_{{\rm Ly}\alpha} \lessapprox 7\sigma$. 
    \item The eBOSS DR16 data favor an early reionization scenario, characterized by a reionization midpoint at $z \approx 8.1$ and an optical depth of $\tau_{\rm reio} \approx 0.0583$ \citep{2020MNRAS.499.1640M}, aligning well with the $1\sigma$ range reported by \cite{2020A&A...641A...6P} and in agreement with recent developments from JWST observations \citep[e.g., ][]{2025arXiv250111702M}. Note that this trend was also recovered in Appendix \ref{app:yuk-free}, which employs a varying $A_{\rm re}$ in the Yukawa template -- Eq.~(\ref{eq:yukawa}). Moreover, the value recovered for the fit on the combination of all the correlations $A_{\rm re} \approx 0.19$ aligns well with the fixed value of the Yukawa early model ($0.2088$). 
\end{itemize}

We conclude that the shape of the \lya correlations can be used to weakly constrain reionization scenarios. Besides, these effects must also be carefully incorporated into future shape-analysis efforts that aim to extend beyond BAO information \citep[e.g.][]{2018JCAP...01..003F,2023PhRvL.130s1003C,2023MNRAS.518.2567G}. Furthermore, we underscore the vital need for high-mass resolution\footnote{Note that a typical mass resolution of $M_{\rm gas} \sim \mathcal{O}(10^5) \, h^{-1} M_\odot$ \citep[e.g., Sherwood Relics,][]{2023MNRAS.519.6162P}, which is usually considered high-mass resolution, is at least two order of magnitudes away from the resolution used in simulations that do track gas below the Jeans mass prior to reionization \citep{2018MNRAS.474.2173H,2019MNRAS.487.1047M,2024MNRAS.533L.100C}.} in simulations to accurately capture the HEMD phase of the temperature-density evolution \citep{2018MNRAS.474.2173H}. Simply put, simulations that fail to resolve small-scale structures prior to the onset of reionization, i.e.\ before the significant increase in the Jeans mass caused by ionization fronts, will inherently underestimate the long-term survival of reionization relics. 

Likewise, independent of potential modeling improvements or limitations, our results clearly demonstrate that the eBOSS DR16 \lya forest correlation functions have a preference for an additional source of power.   

Finally, we highlight that the correlation function BAO analysis remains largely shielded from the impact of reionization. This robustness arises from two key factors: the BAO scale is significantly larger than the typical scale of reionization bubbles, and the analysis is conducted at an effective redshift of $z_{\rm eff} = 2.334$. At higher redshifts, the impact of reionization relics becomes stronger, as there is insufficient time for the IGM to relax into the usual temperature-density relation \citep{1997MNRAS.292...27H}. Given our findings, the redshift range probed directly by the one-dimensional \lya power spectrum, and the expected effects of reionization relics \citep{2021MNRAS.508.1262M}, future work will focus on conducting the first analysis of these measurements that explicitly account for the lasting impact of inhomogeneous reionization.

\section*{Acknowledgements}
We are sincerely grateful to Abby Bault, Andreu Font-Ribera, and other members of the DESI collaboration who contributed to the tutorial for reproducing the eBOSS DR16 analysis outside of NERSC. We are also thankful to James Rich and Paul Martini for their valuable insights. This work is supported by the National SKA Program of China (grant No. 2020SKA0110401) and the Major Key Project of PCL. We acknowledge the Tsinghua Astrophysics High-Performance Computing platform at Tsinghua University and PCL's Cloud Brain for providing computational and data storage resources that have contributed to the research results reported within this paper. This work made extensive use of the \hyperlink{https://ui.adsabs.harvard.edu}{NASA Astrophysics DataSystem} and the following open-source python libraries/packages: \texttt{matplotlib} \citep{2007CSE.....9...90H}, \texttt{numpy} \citep{2020Natur.585..357H}, {\sc PySR} \citep{pysr}, \texttt{scipy} \citep{2020NatMe..17..261V}, \texttt{iminuit}, and \texttt{picca} \citep{2021ascl.soft06018D}.

%%%%%%%%%%%%%%%%%%%%%%%%%%%%%%%%%%%%%%%%%%%%%%%%%%
\section*{Data Availability}
The data underlying this article will be shared on reasonable request to the corresponding authors.



\bibliographystyle{mnras}
\bibliography{re-eBOSS} % if your bibtex file is called example.bib

@ARTICLE{2022ApJS..259...28W,
       author = {{Wang}, Ben and {Zou}, Jiaqi and {Cai}, Zheng and {Prochaska}, J. Xavier and {Sun}, Zechang and {Ding}, Jiani and {Font-Ribera}, Andreu and {Gonzalez}, Alma and {Herrera-Alcantar}, Hiram K. and {Irsic}, Vid and {Lin}, Xiaojing and {Brooks}, David and {Chabanier}, Sol{\'e}ne and {Belsunce}, Roger de and {Palanque-Delabrouille}, Nathalie and {Tarle}, Gregory and {Zhou}, Zhimin},
        title = "{Deep Learning of Dark Energy Spectroscopic Instrument Mock Spectra to Find Damped Ly{\ensuremath{\alpha}} Systems}",
      journal = {\apjs},
         year = 2022,
        month = mar,
       volume = {259},
       number = {1},
          eid = {28},
        pages = {28},
          doi = {10.3847/1538-4365/ac4504},
       adsurl = {https://ui.adsabs.harvard.edu/abs/2022ApJS..259...28W}
}

@ARTICLE{2019ApJ...879...72G,
       author = {{Guo}, Zhiyuan and {Martini}, Paul},
        title = "{Classification of Broad Absorption Line Quasars with a Convolutional Neural Network}",
      journal = {\apj},
         year = 2019,
        month = jul,
       volume = {879},
       number = {2},
          eid = {72},
        pages = {72},
          doi = {10.3847/1538-4357/ab2590},
archivePrefix = {arXiv},
       eprint = {1901.04506},
 primaryClass = {astro-ph.GA},
       adsurl = {https://ui.adsabs.harvard.edu/abs/2019ApJ...879...72G}
}

@ARTICLE{2022MNRAS.511.3514E,
       author = {{Ennesser}, Lauren and {Martini}, Paul and {Font-Ribera}, Andreu and {P{\'e}rez-R{\`a}fols}, Ignasi},
        title = "{The impact and mitigation of broad-absorption-line quasars in Lyman {\ensuremath{\alpha}} forest correlations}",
      journal = {\mnras},
         year = 2022,
        month = apr,
       volume = {511},
       number = {3},
        pages = {3514-3523},
          doi = {10.1093/mnras/stac301},
archivePrefix = {arXiv},
       eprint = {2111.09439},
 primaryClass = {astro-ph.CO},
       adsurl = {https://ui.adsabs.harvard.edu/abs/2022MNRAS.511.3514E}
}

@ARTICLE{2016JCAP...02..051I,
       author = {{Ir{\v{s}}i{\v{c}}}, Vid and {Di Dio}, Enea and {Viel}, Matteo},
        title = "{Relativistic effects in Lyman-{\ensuremath{\alpha}} forest}",
      journal = {\jcap},
         year = 2016,
        month = feb,
       volume = {2016},
       number = {2},
          eid = {051},
        pages = {051},
          doi = {10.1088/1475-7516/2016/02/051},
archivePrefix = {arXiv},
       eprint = {1510.03436},
 primaryClass = {astro-ph.CO},
       adsurl = {https://ui.adsabs.harvard.edu/abs/2016JCAP...02..051I}
}

@ARTICLE{2017A&A...603A..12B,
       author = {{Bautista}, Julian E. and {Busca}, Nicol{\'a}s G. and {Guy}, Julien and {Rich}, James and {Blomqvist}, Michael and {du Mas des Bourboux}, H{\'e}lion and {Pieri}, Matthew M. and {Font-Ribera}, Andreu and {Bailey}, Stephen and {Delubac}, Timoth{\'e}e and {Kirkby}, David and {Le Goff}, Jean-Marc and {Margala}, Daniel and {Slosar}, An{\v{z}}e and {Vazquez}, Jose Alberto and {Brownstein}, Joel R. and {Dawson}, Kyle S. and {Eisenstein}, Daniel J. and {Miralda-Escud{\'e}}, Jordi and {Noterdaeme}, Pasquier and {Palanque-Delabrouille}, Nathalie and {P{\^a}ris}, Isabelle and {Petitjean}, Patrick and {Ross}, Nicholas P. and {Schneider}, Donald P. and {Weinberg}, David H. and {Y{\`e}che}, Christophe},
        title = "{Measurement of baryon acoustic oscillation correlations at z = 2.3 with SDSS DR12 Ly{\ensuremath{\alpha}}-Forests}",
      journal = {\aap},
         year = 2017,
        month = jun,
       volume = {603},
          eid = {A12},
        pages = {A12},
          doi = {10.1051/0004-6361/201730533},
archivePrefix = {arXiv},
       eprint = {1702.00176},
 primaryClass = {astro-ph.CO},
       adsurl = {https://ui.adsabs.harvard.edu/abs/2017A&A...603A..12B}
}

@ARTICLE{2014PhRvD..89h3010P,
       author = {{Pontzen}, Andrew},
        title = "{Scale-dependent bias in the baryonic-acoustic-oscillation-scale intergalactic neutral hydrogen}",
      journal = {\prd},
         year = 2014,
        month = apr,
       volume = {89},
       number = {8},
          eid = {083010},
        pages = {083010},
          doi = {10.1103/PhysRevD.89.083010},
archivePrefix = {arXiv},
       eprint = {1402.0506},
 primaryClass = {astro-ph.CO},
       adsurl = {https://ui.adsabs.harvard.edu/abs/2014PhRvD..89h3010P}
}

@ARTICLE{2014MNRAS.442..187G,
       author = {{Gontcho A Gontcho}, Satya and {Miralda-Escud{\'e}}, Jordi and {Busca}, Nicol{\'a}s G.},
        title = "{On the effect of the ionizing background on the Ly{\ensuremath{\alpha}} forest autocorrelation function}",
      journal = {\mnras},
         year = 2014,
        month = jul,
       volume = {442},
       number = {1},
        pages = {187-195},
          doi = {10.1093/mnras/stu860},
archivePrefix = {arXiv},
       eprint = {1404.7425},
 primaryClass = {astro-ph.CO},
       adsurl = {https://ui.adsabs.harvard.edu/abs/2014MNRAS.442..187G}
}

@ARTICLE{2018MNRAS.476.1151P,
       author = {{Parks}, David and {Prochaska}, J. Xavier and {Dong}, Shawfeng and {Cai}, Zheng},
        title = "{Deep learning of quasar spectra to discover and characterize damped Ly{\ensuremath{\alpha}} systems}",
      journal = {\mnras},
         year = 2018,
        month = may,
       volume = {476},
       number = {1},
        pages = {1151-1168},
          doi = {10.1093/mnras/sty196},
archivePrefix = {arXiv},
       eprint = {1709.04962},
 primaryClass = {astro-ph.GA},
       adsurl = {https://ui.adsabs.harvard.edu/abs/2018MNRAS.476.1151P}
}

@ARTICLE{2021MNRAS.507..704H,
       author = {{Ho}, Ming-Feng and {Bird}, Simeon and {Garnett}, Roman},
        title = "{Damped Lyman-{\ensuremath{\alpha}} absorbers from Sloan digital sky survey DR16Q with Gaussian processes}",
      journal = {\mnras},
         year = 2021,
        month = oct,
       volume = {507},
       number = {1},
        pages = {704-719},
          doi = {10.1093/mnras/stab2169},
archivePrefix = {arXiv},
       eprint = {2103.10964},
 primaryClass = {astro-ph.GA},
       adsurl = {https://ui.adsabs.harvard.edu/abs/2021MNRAS.507..704H}
}

@ARTICLE{2020JCAP...04..006L,
       author = {{Lepori}, Francesca and {Ir{\v{s}}i{\v{c}}}, Vid and {Di Dio}, Enea and {Viel}, Matteo},
        title = "{The impact of relativistic effects on the 3D Quasar-Lyman-{\ensuremath{\alpha}} cross-correlation}",
      journal = {\jcap},
         year = 2020,
        month = apr,
       volume = {2020},
       number = {4},
          eid = {006},
        pages = {006},
          doi = {10.1088/1475-7516/2020/04/006},
archivePrefix = {arXiv},
       eprint = {1910.06305},
 primaryClass = {astro-ph.CO},
       adsurl = {https://ui.adsabs.harvard.edu/abs/2020JCAP...04..006L}
}

@ARTICLE{2023MNRAS.520..948L,
       author = {{Long}, Heyang and {Hirata}, Christopher M.},
        title = "{Probing large-scale ionizing background fluctuation with Lyman {\ensuremath{\alpha}} forest and galaxy cross-correlation at z = 2.4}",
      journal = {\mnras},
         year = 2023,
        month = mar,
       volume = {520},
       number = {1},
        pages = {948-962},
          doi = {10.1093/mnras/stad184},
archivePrefix = {arXiv},
       eprint = {2209.07019},
 primaryClass = {astro-ph.CO},
       adsurl = {https://ui.adsabs.harvard.edu/abs/2023MNRAS.520..948L}
}

@ARTICLE{2019MNRAS.487.5346T,
       author = {{Tie}, Suk Sien and {Weinberg}, David H. and {Martini}, Paul and {Zhu}, Wei and {Peirani}, S{\'e}bastien and {Suarez}, Teresita and {Colombi}, St{\'e}phane},
        title = "{UV background fluctuations and three-point correlations in the large-scale clustering of the Lyman {\ensuremath{\alpha}} forest}",
      journal = {\mnras},
         year = 2019,
        month = aug,
       volume = {487},
       number = {4},
        pages = {5346-5362},
          doi = {10.1093/mnras/stz1632},
archivePrefix = {arXiv},
       eprint = {1905.02208},
 primaryClass = {astro-ph.CO},
       adsurl = {https://ui.adsabs.harvard.edu/abs/2019MNRAS.487.5346T}
}

@ARTICLE{2023ApJS..269....4S,
       author = {{Sun}, Zechang and {Ting}, Yuan-Sen and {Cai}, Zheng},
        title = "{Quasar Factor Analysis-An Unsupervised and Probabilistic Quasar Continuum Prediction Algorithm with Latent Factor Analysis}",
      journal = {\apjs},
         year = 2023,
        month = nov,
       volume = {269},
       number = {1},
          eid = {4},
        pages = {4},
          doi = {10.3847/1538-4365/acf2f1},
archivePrefix = {arXiv},
       eprint = {2211.11784},
 primaryClass = {astro-ph.CO},
       adsurl = {https://ui.adsabs.harvard.edu/abs/2023ApJS..269....4S}
}

@ARTICLE{2012AJ....143...51L,
       author = {{Lee}, Khee-Gan and {Suzuki}, Nao and {Spergel}, David N.},
        title = "{Mean-flux-regulated Principal Component Analysis Continuum Fitting of Sloan Digital Sky Survey Ly{\ensuremath{\alpha}} Forest Spectra}",
      journal = {\aj},
         year = 2012,
        month = feb,
       volume = {143},
       number = {2},
          eid = {51},
        pages = {51},
          doi = {10.1088/0004-6256/143/2/51},
archivePrefix = {arXiv},
       eprint = {1108.6080},
 primaryClass = {astro-ph.CO},
       adsurl = {https://ui.adsabs.harvard.edu/abs/2012AJ....143...51L}
}

@ARTICLE{2019MNRAS.487.1047M,
       author = {{Montero-Camacho}, Paulo and {Hirata}, Christopher M. and {Martini}, Paul and {Honscheid}, Klaus},
        title = "{Impact of inhomogeneous reionization on the Lyman-{\ensuremath{\alpha}} forest}",
      journal = {\mnras},
         year = 2019,
        month = jul,
       volume = {487},
       number = {1},
        pages = {1047-1056},
          doi = {10.1093/mnras/stz1388},
archivePrefix = {arXiv},
       eprint = {1902.02892},
 primaryClass = {astro-ph.CO},
       adsurl = {https://ui.adsabs.harvard.edu/abs/2019MNRAS.487.1047M}
}

@ARTICLE{2022MNRAS.509.6119M,
       author = {{Molaro}, Margherita and {Ir{\v{s}}i{\v{c}}}, Vid and {Bolton}, James S. and {Keating}, Laura C. and {Puchwein}, Ewald and {Gaikwad}, Prakash and {Haehnelt}, Martin G. and {Kulkarni}, Girish and {Viel}, Matteo},
        title = "{The effect of inhomogeneous reionization on the Lyman {\ensuremath{\alpha}} forest power spectrum at redshift z > 4: implications for thermal parameter recovery}",
      journal = {\mnras},
         year = 2022,
        month = feb,
       volume = {509},
       number = {4},
        pages = {6119-6137},
          doi = {10.1093/mnras/stab3416},
archivePrefix = {arXiv},
       eprint = {2109.06897},
 primaryClass = {astro-ph.CO},
       adsurl = {https://ui.adsabs.harvard.edu/abs/2022MNRAS.509.6119M}
}

@ARTICLE{2019MNRAS.490.3177W,
       author = {{Wu}, Xiaohan and {McQuinn}, Matthew and {Kannan}, Rahul and {D'Aloisio}, Anson and {Bird}, Simeon and {Marinacci}, Federico and {Dav{\'e}}, Romeel and {Hernquist}, Lars},
        title = "{Imprints of temperature fluctuations on the z {\ensuremath{\sim}} 5 Lyman-{\ensuremath{\alpha}} forest: a view from radiation-hydrodynamic simulations of reionization}",
      journal = {\mnras},
         year = 2019,
        month = dec,
       volume = {490},
       number = {3},
        pages = {3177-3195},
          doi = {10.1093/mnras/stz2807},
archivePrefix = {arXiv},
       eprint = {1907.04860},
 primaryClass = {astro-ph.CO},
       adsurl = {https://ui.adsabs.harvard.edu/abs/2019MNRAS.490.3177W}
}

@ARTICLE{2013MNRAS.435.3169C,
       author = {{Compostella}, Michele and {Cantalupo}, Sebastiano and {Porciani}, Cristiano},
        title = "{The imprint of inhomogeneous He II reionization on the H I and He II Ly{\ensuremath{\alpha}} forest}",
      journal = {\mnras},
         year = 2013,
        month = nov,
       volume = {435},
       number = {4},
        pages = {3169-3190},
          doi = {10.1093/mnras/stt1510},
archivePrefix = {arXiv},
       eprint = {1306.5745},
 primaryClass = {astro-ph.CO},
       adsurl = {https://ui.adsabs.harvard.edu/abs/2013MNRAS.435.3169C}
}

@ARTICLE{2020MNRAS.496.4372U,
       author = {{Upton Sanderbeck}, Phoebe and {Bird}, Simeon},
        title = "{Inhomogeneous He II reionization in hydrodynamic simulations}",
      journal = {\mnras},
         year = 2020,
        month = aug,
       volume = {496},
       number = {4},
        pages = {4372-4382},
          doi = {10.1093/mnras/staa1850},
archivePrefix = {arXiv},
       eprint = {2002.05733},
 primaryClass = {astro-ph.CO},
       adsurl = {https://ui.adsabs.harvard.edu/abs/2020MNRAS.496.4372U}
}

@ARTICLE{2017ApJ...841...87L,
       author = {{La Plante}, Paul and {Trac}, Hy and {Croft}, Rupert and {Cen}, Renyue},
        title = "{Helium Reionization Simulations. II. Signatures of Quasar Activity on the IGM}",
      journal = {\apj},
         year = 2017,
        month = jun,
       volume = {841},
       number = {2},
          eid = {87},
        pages = {87},
          doi = {10.3847/1538-4357/aa7136},
archivePrefix = {arXiv},
       eprint = {1610.02047},
 primaryClass = {astro-ph.CO},
       adsurl = {https://ui.adsabs.harvard.edu/abs/2017ApJ...841...87L}
}

@ARTICLE{2021MNRAS.506.5439C,
       author = {{Cuceu}, Andrei and {Font-Ribera}, Andreu and {Joachimi}, Benjamin and {Nadathur}, Seshadri},
        title = "{Cosmology beyond BAO from the 3D distribution of the Lyman-{\ensuremath{\alpha}} forest}",
      journal = {\mnras},
         year = 2021,
        month = oct,
       volume = {506},
       number = {4},
        pages = {5439-5450},
          doi = {10.1093/mnras/stab1999},
archivePrefix = {arXiv},
       eprint = {2103.14075},
 primaryClass = {astro-ph.CO},
       adsurl = {https://ui.adsabs.harvard.edu/abs/2021MNRAS.506.5439C}
}

@ARTICLE{2023MNRAS.518.2567G,
       author = {{Gerardi}, Francesca and {Cuceu}, Andrei and {Font-Ribera}, Andreu and {Joachimi}, Benjamin and {Lemos}, Pablo},
        title = "{Direct cosmological inference from three-dimensional correlations of the Lyman {\ensuremath{\alpha}} forest}",
      journal = {\mnras},
         year = 2023,
        month = jan,
       volume = {518},
       number = {2},
        pages = {2567-2573},
          doi = {10.1093/mnras/stac3257},
archivePrefix = {arXiv},
       eprint = {2209.11263},
 primaryClass = {astro-ph.CO},
       adsurl = {https://ui.adsabs.harvard.edu/abs/2023MNRAS.518.2567G}
}

@ARTICLE{2022arXiv220913942C,
       author = {{Cuceu}, Andrei and {Font-Ribera}, Andreu and {Nadathur}, Seshadri and {Joachimi}, Benjamin and {Martini}, Paul},
        title = "{New constraints on the expansion rate at redshift 2.3 from the Lyman-$\alpha$ forest}",
      journal = {arXiv e-prints},
         year = 2022,
        month = sep,
          eid = {arXiv:2209.13942},
        pages = {arXiv:2209.13942},
archivePrefix = {arXiv},
       eprint = {2209.13942},
 primaryClass = {astro-ph.CO},
       adsurl = {https://ui.adsabs.harvard.edu/abs/2022arXiv220913942C}
}

@ARTICLE{2022arXiv220912931C,
       author = {{Cuceu}, Andrei and {Font-Ribera}, Andreu and {Martini}, Paul and {Joachimi}, Benjamin and {Nadathur}, Seshadri and {Rich}, James and {Gonz{\'a}lez-Morales}, Alma X. and {du Mas des Bourboux}, H{\'e}lion and {Farr}, James},
        title = "{The Alcock-Paczy{\'n}ski effect from Lyman-$\alpha$ forest correlations: Analysis validation with synthetic data}",
      journal = {arXiv e-prints},
         year = 2022,
        month = sep,
          eid = {arXiv:2209.12931},
        pages = {arXiv:2209.12931},
archivePrefix = {arXiv},
       eprint = {2209.12931},
 primaryClass = {astro-ph.CO},
       adsurl = {https://ui.adsabs.harvard.edu/abs/2022arXiv220912931C}
}

@ARTICLE{2020A&A...641A...6P,
       author = {{Planck Collaboration} and {Aghanim}, N. and {Akrami}, Y. and {Ashdown}, M. and {Aumont}, J. and {Baccigalupi}, C. and {Ballardini}, M. and {Banday}, A.~J. and {Barreiro}, R.~B. and {Bartolo}, N. and {Basak}, S. and {Battye}, R. and {Benabed}, K. and {Bernard}, J. -P. and {Bersanelli}, M. and {Bielewicz}, P. and {Bock}, J.~J. and {Bond}, J.~R. and {Borrill}, J. and {Bouchet}, F.~R. and {Boulanger}, F. and {Bucher}, M. and {Burigana}, C. and {Butler}, R.~C. and {Calabrese}, E. and {Cardoso}, J. -F. and {Carron}, J. and {Challinor}, A. and {Chiang}, H.~C. and {Chluba}, J. and {Colombo}, L.~P.~L. and {Combet}, C. and {Contreras}, D. and {Crill}, B.~P. and {Cuttaia}, F. and {de Bernardis}, P. and {de Zotti}, G. and {Delabrouille}, J. and {Delouis}, J. -M. and {Di Valentino}, E. and {Diego}, J.~M. and {Dor{\'e}}, O. and {Douspis}, M. and {Ducout}, A. and {Dupac}, X. and {Dusini}, S. and {Efstathiou}, G. and {Elsner}, F. and {En{\ss}lin}, T.~A. and {Eriksen}, H.~K. and {Fantaye}, Y. and {Farhang}, M. and {Fergusson}, J. and {Fernandez-Cobos}, R. and {Finelli}, F. and {Forastieri}, F. and {Frailis}, M. and {Fraisse}, A.~A. and {Franceschi}, E. and {Frolov}, A. and {Galeotta}, S. and {Galli}, S. and {Ganga}, K. and {G{\'e}nova-Santos}, R.~T. and {Gerbino}, M. and {Ghosh}, T. and {Gonz{\'a}lez-Nuevo}, J. and {G{\'o}rski}, K.~M. and {Gratton}, S. and {Gruppuso}, A. and {Gudmundsson}, J.~E. and {Hamann}, J. and {Handley}, W. and {Hansen}, F.~K. and {Herranz}, D. and {Hildebrandt}, S.~R. and {Hivon}, E. and {Huang}, Z. and {Jaffe}, A.~H. and {Jones}, W.~C. and {Karakci}, A. and {Keih{\"a}nen}, E. and {Keskitalo}, R. and {Kiiveri}, K. and {Kim}, J. and {Kisner}, T.~S. and {Knox}, L. and {Krachmalnicoff}, N. and {Kunz}, M. and {Kurki-Suonio}, H. and {Lagache}, G. and {Lamarre}, J. -M. and {Lasenby}, A. and {Lattanzi}, M. and {Lawrence}, C.~R. and {Le Jeune}, M. and {Lemos}, P. and {Lesgourgues}, J. and {Levrier}, F. and {Lewis}, A. and {Liguori}, M. and {Lilje}, P.~B. and {Lilley}, M. and {Lindholm}, V. and {L{\'o}pez-Caniego}, M. and {Lubin}, P.~M. and {Ma}, Y. -Z. and {Mac{\'\i}as-P{\'e}rez}, J.~F. and {Maggio}, G. and {Maino}, D. and {Mandolesi}, N. and {Mangilli}, A. and {Marcos-Caballero}, A. and {Maris}, M. and {Martin}, P.~G. and {Martinelli}, M. and {Mart{\'\i}nez-Gonz{\'a}lez}, E. and {Matarrese}, S. and {Mauri}, N. and {McEwen}, J.~D. and {Meinhold}, P.~R. and {Melchiorri}, A. and {Mennella}, A. and {Migliaccio}, M. and {Millea}, M. and {Mitra}, S. and {Miville-Desch{\^e}nes}, M. -A. and {Molinari}, D. and {Montier}, L. and {Morgante}, G. and {Moss}, A. and {Natoli}, P. and {N{\o}rgaard-Nielsen}, H.~U. and {Pagano}, L. and {Paoletti}, D. and {Partridge}, B. and {Patanchon}, G. and {Peiris}, H.~V. and {Perrotta}, F. and {Pettorino}, V. and {Piacentini}, F. and {Polastri}, L. and {Polenta}, G. and {Puget}, J. -L. and {Rachen}, J.~P. and {Reinecke}, M. and {Remazeilles}, M. and {Renzi}, A. and {Rocha}, G. and {Rosset}, C. and {Roudier}, G. and {Rubi{\~n}o-Mart{\'\i}n}, J.~A. and {Ruiz-Granados}, B. and {Salvati}, L. and {Sandri}, M. and {Savelainen}, M. and {Scott}, D. and {Shellard}, E.~P.~S. and {Sirignano}, C. and {Sirri}, G. and {Spencer}, L.~D. and {Sunyaev}, R. and {Suur-Uski}, A. -S. and {Tauber}, J.~A. and {Tavagnacco}, D. and {Tenti}, M. and {Toffolatti}, L. and {Tomasi}, M. and {Trombetti}, T. and {Valenziano}, L. and {Valiviita}, J. and {Van Tent}, B. and {Vibert}, L. and {Vielva}, P. and {Villa}, F. and {Vittorio}, N. and {Wandelt}, B.~D. and {Wehus}, I.~K. and {White}, M. and {White}, S.~D.~M. and {Zacchei}, A. and {Zonca}, A.},
        title = "{Planck 2018 results. VI. Cosmological parameters}",
      journal = {\aap},
         year = 2020,
        month = sep,
       volume = {641},
          eid = {A6},
        pages = {A6},
          doi = {10.1051/0004-6361/201833910},
archivePrefix = {arXiv},
       eprint = {1807.06209},
 primaryClass = {astro-ph.CO},
       adsurl = {https://ui.adsabs.harvard.edu/abs/2020A&A...641A...6P}
}

@ARTICLE{2019JCAP...07..017C,
       author = {{Chabanier}, Sol{\`e}ne and {Palanque-Delabrouille}, Nathalie and
         {Y{\`e}che}, Christophe and {Le Goff}, Jean-Marc and {Armengaud}, Eric and
         {Bautista}, Julian and {Blomqvist}, Michael and {Busca}, Nicolas and
         {Dawson}, Kyle and {Etourneau}, Thomas and {Font-Ribera}, Andreu and
         {Lee}, Youngbae and {du Mas des Bourboux}, H{\'e}lion and
         {Pieri}, Matthew and {Rich}, James and {Rossi}, Graziano and
         {Schneider}, Donald and {Slosar}, An{\v{z}}e},
        title = "{The one-dimensional power spectrum from the SDSS DR14 Ly{\ensuremath{\alpha}} forests}",
      journal = {\jcap},
         year = 2019,
        month = jul,
       volume = {2019},
       number = {7},
          eid = {017},
        pages = {017},
          doi = {10.1088/1475-7516/2019/07/017},
archivePrefix = {arXiv},
       eprint = {1812.03554},
 primaryClass = {astro-ph.CO},
       adsurl = {https://ui.adsabs.harvard.edu/abs/2019JCAP...07..017C}
}

@ARTICLE{1998ARA&A..36..267R,
       author = {{Rauch}, Michael},
        title = "{The Lyman Alpha Forest in the Spectra of QSOs}",
      journal = {\araa},
         year = 1998,
        month = jan,
       volume = {36},
        pages = {267-316},
          doi = {10.1146/annurev.astro.36.1.267},
archivePrefix = {arXiv},
       eprint = {astro-ph/9806286},
 primaryClass = {astro-ph},
       adsurl = {https://ui.adsabs.harvard.edu/abs/1998ARA&A..36..267R}
}

@ARTICLE{2000ApJ...543....1M,
       author = {{McDonald}, Patrick and {Miralda-Escud{\'e}}, Jordi and {Rauch}, Michael and {Sargent}, Wallace L.~W. and {Barlow}, Tom A. and {Cen}, Renyue and {Ostriker}, Jeremiah P.},
        title = "{The Observed Probability Distribution Function, Power Spectrum, and Correlation Function of the Transmitted Flux in the Ly{\ensuremath{\alpha}} Forest}",
      journal = {\apj},
         year = 2000,
        month = nov,
       volume = {543},
       number = {1},
        pages = {1-23},
          doi = {10.1086/317079},
archivePrefix = {arXiv},
       eprint = {astro-ph/9911196},
 primaryClass = {astro-ph},
       adsurl = {https://ui.adsabs.harvard.edu/abs/2000ApJ...543....1M}
}

@ARTICLE{2004MNRAS.354..684V,
       author = {{Viel}, Matteo and {Haehnelt}, Martin G. and {Springel}, Volker},
        title = "{Inferring the dark matter power spectrum from the Lyman {\ensuremath{\alpha}} forest in high-resolution QSO absorption spectra}",
      journal = {\mnras},
         year = 2004,
        month = nov,
       volume = {354},
       number = {3},
        pages = {684-694},
          doi = {10.1111/j.1365-2966.2004.08224.x},
archivePrefix = {arXiv},
       eprint = {astro-ph/0404600},
 primaryClass = {astro-ph},
       adsurl = {https://ui.adsabs.harvard.edu/abs/2004MNRAS.354..684V}
}

@ARTICLE{2002ApJ...581...20C,
       author = {{Croft}, Rupert A.~C. and {Weinberg}, David H. and {Bolte}, Mike and {Burles}, Scott and {Hernquist}, Lars and {Katz}, Neal and {Kirkman}, David and {Tytler}, David},
        title = "{Toward a Precise Measurement of Matter Clustering: Ly{\ensuremath{\alpha}} Forest Data at Redshifts 2-4}",
      journal = {\apj},
         year = 2002,
        month = dec,
       volume = {581},
       number = {1},
        pages = {20-52},
          doi = {10.1086/344099},
archivePrefix = {arXiv},
       eprint = {astro-ph/0012324},
 primaryClass = {astro-ph},
       adsurl = {https://ui.adsabs.harvard.edu/abs/2002ApJ...581...20C}
}

@ARTICLE{1997MNRAS.292...27H,
       author = {{Hui}, Lam and {Gnedin}, Nickolay Y.},
        title = "{Equation of state of the photoionized intergalactic medium}",
      journal = {\mnras},
         year = 1997,
        month = nov,
       volume = {292},
       number = {1},
        pages = {27-42},
          doi = {10.1093/mnras/292.1.27},
archivePrefix = {arXiv},
       eprint = {astro-ph/9612232},
 primaryClass = {astro-ph},
       adsurl = {https://ui.adsabs.harvard.edu/abs/1997MNRAS.292...27H}
}

@ARTICLE{2022AJ....164..207A,
       author = {{Abareshi}, B. and {Aguilar}, J. and {Ahlen}, S. and {Alam}, Shadab and {Alexander}, David M. and {Alfarsy}, R. and {Allen}, L. and {Allende Prieto}, C. and {Alves}, O. and {Ameel}, J. and {Armengaud}, E. and {Asorey}, J. and {Aviles}, Alejandro and {Bailey}, S. and {Balaguera-Antol{\'\i}nez}, A. and {Ballester}, O. and {Baltay}, C. and {Bault}, A. and {Beltran}, S.~F. and {Benavides}, B. and {BenZvi}, S. and {Berti}, A. and {Besuner}, R. and {Beutler}, Florian and {Bianchi}, D. and {Blake}, C. and {Blanc}, P. and {Blum}, R. and {Bolton}, A. and {Bose}, S. and {Bramall}, D. and {Brieden}, S. and {Brodzeller}, A. and {Brooks}, D. and {Brownewell}, C. and {Buckley-Geer}, E. and {Cahn}, R.~N. and {Cai}, Z. and {Canning}, R. and {Capasso}, R. and {Carnero Rosell}, A. and {Carton}, P. and {Casas}, R. and {Castander}, F.~J. and {Cervantes-Cota}, J.~L. and {Chabanier}, S. and {Chaussidon}, E. and {Chuang}, C. and {Circosta}, C. and {Cole}, S. and {Cooper}, A.~P. and {da Costa}, L. and {Cousinou}, M. -C. and {Cuceu}, A. and {Davis}, T.~M. and {Dawson}, K. and {de la Cruz-Noriega}, R. and {de la Macorra}, A. and {de Mattia}, A. and {Della Costa}, J. and {Demmer}, P. and {Derwent}, M. and {Dey}, A. and {Dey}, B. and {Dhungana}, G. and {Ding}, Z. and {Dobson}, C. and {Doel}, P. and {Donald-McCann}, J. and {Donaldson}, J. and {Douglass}, K. and {Duan}, Y. and {Dunlop}, P. and {Edelstein}, J. and {Eftekharzadeh}, S. and {Eisenstein}, D.~J. and {Enriquez-Vargas}, M. and {Escoffier}, S. and {Evatt}, M. and {Fagrelius}, P. and {Fan}, X. and {Fanning}, K. and {Fawcett}, V.~A. and {Ferraro}, S. and {Ereza}, J. and {Flaugher}, B. and {Font-Ribera}, A. and {Forero-Romero}, J.~E. and {Frenk}, C.~S. and {Fromenteau}, S. and {G{\"a}nsicke}, B.~T. and {Garcia-Quintero}, C. and {Garrison}, L. and {Gazta{\~n}aga}, E. and {Gerardi}, F. and {Gil-Mar{\'\i}n}, H. and {Gontcho}, S. Gontcho A. and {Gonzalez-Morales}, Alma X. and {Gonzalez-de-Rivera}, G. and {Gonzalez-Perez}, V. and {Gordon}, C. and {Graur}, O. and {Green}, D. and {Grove}, C. and {Gruen}, D. and {Gutierrez}, G. and {Guy}, J. and {Hahn}, C. and {Harris}, S. and {Herrera}, D. and {Herrera-Alcantar}, Hiram K. and {Honscheid}, K. and {Howlett}, C. and {Huterer}, D. and {Ir{\v{s}}i{\v{c}}}, V. and {Ishak}, M. and {Jelinsky}, P. and {Jiang}, L. and {Jimenez}, J. and {Jing}, Y.~P. and {Joyce}, R. and {Jullo}, E. and {Juneau}, S. and {Kara{\c{c}}ayl{\i}}, N.~G. and {Karamanis}, M. and {Karcher}, A. and {Karim}, T. and {Kehoe}, R. and {Kent}, S. and {Kirkby}, D. and {Kisner}, T. and {Kitaura}, F. and {Koposov}, S.~E. and {Kov{\'a}cs}, A. and {Kremin}, A. and {Krolewski}, Alex and {L'Huillier}, B. and {Lahav}, O. and {Lambert}, A. and {Lamman}, C. and {Lan}, Ting-Wen and {Landriau}, M. and {Lane}, S. and {Lang}, D. and {Lange}, J.~U. and {Lasker}, J. and {Guillou}, L. Le and {Leauthaud}, A. and {Le Van Suu}, A. and {Levi}, Michael E. and {Li}, T.~S. and {Magneville}, C. and {Manera}, M. and {Manser}, Christopher J. and {Marshall}, B. and {Martini}, Paul and {McCollam}, W. and {McDonald}, P. and {Meisner}, Aaron M. and {Mena-Fern{\'a}ndez}, J. and {Meneses-Rizo}, J. and {Mezcua}, M. and {Miller}, T. and {Miquel}, R. and {Montero-Camacho}, P. and {Moon}, J. and {Moustakas}, J. and {Mueller}, E. and {Mu{\~n}oz-Guti{\'e}rrez}, Andrea and {Myers}, Adam D. and {Nadathur}, S. and {Najita}, J. and {Napolitano}, L. and {Neilsen}, E. and {Newman}, Jeffrey A. and {Nie}, J.~D. and {Ning}, Y. and {Niz}, G. and {Norberg}, P. and {Noriega}, Hern{\'a}n E. and {O'Brien}, T. and {Obuljen}, A. and {Palanque-Delabrouille}, N. and {Palmese}, A. and {Zhiwei}, P. and {Pappalardo}, D. and {Peng}, X. and {Percival}, W.~J. and {Perruchot}, S. and {Pogge}, R. and {Poppett}, C. and {Porredon}, A. and {Prada}, F. and {Prochaska}, J. and {Pucha}, R. and {P{\'e}rez-Fern{\'a}ndez}, A. and {P{\'e}rez-R{\`a}fols}, I. and {Rabinowitz}, D. and {Raichoor}, A. and {Ramirez-Solano}, S. and {Ram{\'\i}rez-P{\'e}rez}, C{\'e}sar and {Ravoux}, C. and {Reil}, K. and {Rezaie}, M. and {Rocher}, A. and {Rockosi}, C. and {Roe}, N.~A. and {Roodman}, A. and {Ross}, A.~J. and {Rossi}, G. and {Ruggeri}, R. and {Ruhlmann-Kleider}, V. and {Sabiu}, C.~G. and {Safonova}, S. and {Said}, K. and {Saintonge}, A. and {Catonga}, Javier Salas and {Samushia}, L. and {Sanchez}, E. and {Saulder}, C. and {Schaan}, E. and {Schlafly}, E. and {Schlegel}, D. and {Schmoll}, J. and {Scholte}, D. and {Schubnell}, M. and {Secroun}, A. and {Seo}, H. and {Serrano}, S. and {Sharples}, Ray M. and {Sholl}, Michael J. and {Silber}, Joseph Harry and {Silva}, D.~R. and {Sirk}, M. and {Siudek}, M. and {Smith}, A. and {Sprayberry}, D. and {Staten}, R. and {Stupak}, B. and {Tan}, T. and {Tarl{\'e}}, Gregory and {Tie}, Suk Sien and {Tojeiro}, R. and {Ure{\~n}a-L{\'o}pez}, L.~A. and {Valdes}, F. and {Valenzuela}, O. and {Valluri}, M. and {Vargas-Maga{\~n}a}, M. and {Verde}, L. and {Walther}, M. and {Wang}, B. and {Wang}, M.~S. and {Weaver}, B.~A. and {Weaverdyck}, C. and {Wechsler}, R. and {Wilson}, Michael J. and {Yang}, J. and {Yu}, Y. and {Yuan}, S. and {Y{\`e}che}, Christophe and {Zhang}, H. and {Zhang}, K. and {Zhao}, Cheng and {Zhou}, Rongpu and {Zhou}, Zhimin and {Zou}, H. and {Zou}, J. and {Zou}, S. and {Zu}, Y.},
        title = "{Overview of the Instrumentation for the Dark Energy Spectroscopic Instrument}",
      journal = {\aj},
         year = 2022,
        month = nov,
       volume = {164},
       number = {5},
          eid = {207},
        pages = {207},
          doi = {10.3847/1538-3881/ac882b},
archivePrefix = {arXiv},
       eprint = {2205.10939},
 primaryClass = {astro-ph.IM},
       adsurl = {https://ui.adsabs.harvard.edu/abs/2022AJ....164..207A}
}

@ARTICLE{2014JCAP...05..027F,
       author = {{Font-Ribera}, Andreu and {Kirkby}, David and {Busca}, Nicolas and {Miralda-Escud{\'e}}, Jordi and {Ross}, Nicholas P. and {Slosar}, An{\v{z}}e and {Rich}, James and {Aubourg}, {\'E}ric and {Bailey}, Stephen and {Bhardwaj}, Vaishali and {Bautista}, Julian and {Beutler}, Florian and {Bizyaev}, Dmitry and {Blomqvist}, Michael and {Brewington}, Howard and {Brinkmann}, Jon and {Brownstein}, Joel R. and {Carithers}, Bill and {Dawson}, Kyle S. and {Delubac}, Timoth{\'e}e and {Ebelke}, Garrett and {Eisenstein}, Daniel J. and {Ge}, Jian and {Kinemuchi}, Karen and {Lee}, Khee-Gan and {Malanushenko}, Viktor and {Malanushenko}, Elena and {Marchante}, Moses and {Margala}, Daniel and {Muna}, Demitri and {Myers}, Adam D. and {Noterdaeme}, Pasquier and {Oravetz}, Daniel and {Palanque-Delabrouille}, Nathalie and {P{\^a}ris}, Isabelle and {Petitjean}, Patrick and {Pieri}, Matthew M. and {Rossi}, Graziano and {Schneider}, Donald P. and {Simmons}, Audrey and {Viel}, Matteo and {Yeche}, Christophe and {York}, Donald G.},
        title = "{Quasar-Lyman {\ensuremath{\alpha}} forest cross-correlation from BOSS DR11: Baryon Acoustic Oscillations}",
      journal = {\jcap},
         year = 2014,
        month = may,
       volume = {2014},
       number = {5},
          eid = {027},
        pages = {027},
          doi = {10.1088/1475-7516/2014/05/027},
archivePrefix = {arXiv},
       eprint = {1311.1767},
 primaryClass = {astro-ph.CO},
       adsurl = {https://ui.adsabs.harvard.edu/abs/2014JCAP...05..027F}
}

@ARTICLE{1979Natur.281..358A,
       author = {{Alcock}, C. and {Paczynski}, B.},
        title = "{An evolution free test for non-zero cosmological constant}",
      journal = {\nat},
         year = 1979,
        month = oct,
       volume = {281},
        pages = {358},
          doi = {10.1038/281358a0},
       adsurl = {https://ui.adsabs.harvard.edu/abs/1979Natur.281..358A}
}

@ARTICLE{2020PhRvD.102b3515G,
       author = {{Givans}, Jahmour J. and {Hirata}, Christopher M.},
        title = "{Redshift-space streaming velocity effects on the Lyman-{\ensuremath{\alpha}} forest baryon acoustic oscillation scale}",
      journal = {\prd},
         year = 2020,
        month = jul,
       volume = {102},
       number = {2},
          eid = {023515},
        pages = {023515},
          doi = {10.1103/PhysRevD.102.023515},
archivePrefix = {arXiv},
       eprint = {2002.12296},
 primaryClass = {astro-ph.CO},
       adsurl = {https://ui.adsabs.harvard.edu/abs/2020PhRvD.102b3515G}
}

@ARTICLE{2018MNRAS.474.2173H,
       author = {{Hirata}, Christopher M.},
        title = "{Small-scale structure and the Lyman-{\ensuremath{\alpha}} forest baryon acoustic oscillation feature}",
      journal = {\mnras},
         year = 2018,
        month = feb,
       volume = {474},
       number = {2},
        pages = {2173-2193},
          doi = {10.1093/mnras/stx2854},
archivePrefix = {arXiv},
       eprint = {1707.03358},
 primaryClass = {astro-ph.CO},
       adsurl = {https://ui.adsabs.harvard.edu/abs/2018MNRAS.474.2173H}
}

@ARTICLE{2020MNRAS.499.1640M,
       author = {{Montero-Camacho}, Paulo and {Mao}, Yi},
        title = "{Ly-={\ensuremath{\alpha}} forest power spectrum as an emerging window into the epoch of reionization and cosmic dawn}",
      journal = {\mnras},
         year = 2020,
        month = dec,
       volume = {499},
       number = {2},
        pages = {1640-1651},
          doi = {10.1093/mnras/staa2918},
archivePrefix = {arXiv},
       eprint = {2003.10077},
 primaryClass = {astro-ph.CO},
       adsurl = {https://ui.adsabs.harvard.edu/abs/2020MNRAS.499.1640M}
}

@ARTICLE{2023MNRAS.520.4853M,
       author = {{Montero-Camacho}, Paulo and {Liu}, Yuchen and {Mao}, Yi},
        title = "{Separating the memory of reionisation from cosmology in the Ly{\ensuremath{\alpha}} forest power spectrum at the post-reionisation era}",
      journal = {\mnras},
         year = 2023,
        month = apr,
       volume = {520},
       number = {4},
        pages = {4853-4866},
          doi = {10.1093/mnras/stad437},
archivePrefix = {arXiv},
       eprint = {2207.09005},
 primaryClass = {astro-ph.CO},
       adsurl = {https://ui.adsabs.harvard.edu/abs/2023MNRAS.520.4853M}
}

@ARTICLE{2020ApJ...901..153D,
       author = {{du Mas des Bourboux}, H{\'e}lion and {Rich}, James and {Font-Ribera}, Andreu and {de Sainte Agathe}, Victoria and {Farr}, James and {Etourneau}, Thomas and {Le Goff}, Jean-Marc and {Cuceu}, Andrei and {Balland}, Christophe and {Bautista}, Julian E. and {Blomqvist}, Michael and {Brinkmann}, Jonathan and {Brownstein}, Joel R. and {Chabanier}, Sol{\`e}ne and {Chaussidon}, Edmond and {Dawson}, Kyle and {Gonz{\'a}lez-Morales}, Alma X. and {Guy}, Julien and {Lyke}, Brad W. and {de la Macorra}, Axel and {Mueller}, Eva-Maria and {Myers}, Adam D. and {Nitschelm}, Christian and {Mu{\~n}oz Guti{\'e}rrez}, Andrea and {Palanque-Delabrouille}, Nathalie and {Parker}, James and {Percival}, Will J. and {P{\'e}rez-R{\`a}fols}, Ignasi and {Petitjean}, Patrick and {Pieri}, Matthew M. and {Ravoux}, Corentin and {Rossi}, Graziano and {Schneider}, Donald P. and {Seo}, Hee-Jong and {Slosar}, An{\v{z}}e and {Stermer}, Julianna and {Vivek}, M. and {Y{\`e}che}, Christophe and {Youles}, Samantha},
        title = "{The Completed SDSS-IV Extended Baryon Oscillation Spectroscopic Survey: Baryon Acoustic Oscillations with Ly{\ensuremath{\alpha}} Forests}",
      journal = {\apj},
         year = 2020,
        month = oct,
       volume = {901},
       number = {2},
          eid = {153},
        pages = {153},
          doi = {10.3847/1538-4357/abb085},
archivePrefix = {arXiv},
       eprint = {2007.08995},
 primaryClass = {astro-ph.CO},
       adsurl = {https://ui.adsabs.harvard.edu/abs/2020ApJ...901..153D}
}

@ARTICLE{2020JCAP...04..038P,
       author = {{Palanque-Delabrouille}, Nathalie and {Y{\`e}che}, Christophe and {Sch{\"o}neberg}, Nils and {Lesgourgues}, Julien and {Walther}, Michael and {Chabanier}, Sol{\`e}ne and {Armengaud}, Eric},
        title = "{Hints, neutrino bounds, and WDM constraints from SDSS DR14 Lyman-{\ensuremath{\alpha}} and Planck full-survey data}",
      journal = {\jcap},
         year = 2020,
        month = apr,
       volume = {2020},
       number = {4},
          eid = {038},
        pages = {038},
          doi = {10.1088/1475-7516/2020/04/038},
archivePrefix = {arXiv},
       eprint = {1911.09073},
 primaryClass = {astro-ph.CO},
       adsurl = {https://ui.adsabs.harvard.edu/abs/2020JCAP...04..038P}
}

@ARTICLE{2021MNRAS.502.2356G,
       author = {{Garzilli}, Antonella and {Magalich}, Andrii and {Ruchayskiy}, Oleg and {Boyarsky}, Alexey},
        title = "{How to constrain warm dark matter with the Lyman-{\ensuremath{\alpha}} forest}",
      journal = {\mnras},
         year = 2021,
        month = apr,
       volume = {502},
       number = {2},
        pages = {2356-2363},
          doi = {10.1093/mnras/stab192},
archivePrefix = {arXiv},
       eprint = {1912.09397},
 primaryClass = {astro-ph.CO},
       adsurl = {https://ui.adsabs.harvard.edu/abs/2021MNRAS.502.2356G}
}

@software{pysr,
  author       = {Miles Cranmer},
  title        = {PySR: Fast \& Parallelized Symbolic Regression in Python/Julia},
  month        = sep,
  year         = 2020,
  publisher    = {Zenodo},
  doi          = {10.5281/zenodo.4041459},
  url          = {http://doi.org/10.5281/zenodo.4041459}
}

@ARTICLE{2020arXiv200611287C,
       author = {{Cranmer}, Miles and {Sanchez-Gonzalez}, Alvaro and {Battaglia}, Peter and {Xu}, Rui and {Cranmer}, Kyle and {Spergel}, David and {Ho}, Shirley},
        title = "{Discovering Symbolic Models from Deep Learning with Inductive Biases}",
      journal = {arXiv e-prints},
         year = 2020,
        month = jun,
          eid = {arXiv:2006.11287},
        pages = {arXiv:2006.11287},
          doi = {10.48550/arXiv.2006.11287},
archivePrefix = {arXiv},
       eprint = {2006.11287},
 primaryClass = {cs.LG},
       adsurl = {https://ui.adsabs.harvard.edu/abs/2020arXiv200611287C}
}

@ARTICLE{2020SciA....6.2631U,
       author = {{Udrescu}, Silviu-Marian and {Tegmark}, Max},
        title = "{AI Feynman: A physics-inspired method for symbolic regression}",
      journal = {Science Advances},
         year = 2020,
        month = apr,
       volume = {6},
       number = {16},
        pages = {eaay2631},
          doi = {10.1126/sciadv.aay2631},
archivePrefix = {arXiv},
       eprint = {1905.11481},
 primaryClass = {physics.comp-ph},
       adsurl = {https://ui.adsabs.harvard.edu/abs/2020SciA....6.2631U}
}

@ARTICLE{2021ApJ...919..120M,
       author = {{Morales}, Alexa M. and {Mason}, Charlotte A. and {Bruton}, Sean and {Gronke}, Max and {Haardt}, Francesco and {Scarlata}, Claudia},
        title = "{The Evolution of the Lyman-alpha Luminosity Function during Reionization}",
      journal = {\apj},
         year = 2021,
        month = oct,
       volume = {919},
       number = {2},
          eid = {120},
        pages = {120},
          doi = {10.3847/1538-4357/ac1104},
archivePrefix = {arXiv},
       eprint = {2101.01205},
 primaryClass = {astro-ph.GA},
       adsurl = {https://ui.adsabs.harvard.edu/abs/2021ApJ...919..120M}
}

@ARTICLE{2019MNRAS.484..933P,
       author = {{Park}, Jaehong and {Mesinger}, Andrei and {Greig}, Bradley and
         {Gillet}, Nicolas},
        title = "{Inferring the astrophysics of reionization and cosmic dawn from galaxy luminosity functions and the 21-cm signal}",
      journal = {\mnras},
         year = 2019,
        month = mar,
       volume = {484},
       number = {1},
        pages = {933-949},
          doi = {10.1093/mnras/stz032},
archivePrefix = {arXiv},
       eprint = {1809.08995},
 primaryClass = {astro-ph.GA},
       adsurl = {https://ui.adsabs.harvard.edu/abs/2019MNRAS.484..933P}
}

@ARTICLE{2020Natur.585..357H,
       author = {{Harris}, Charles R. and {Millman}, K. Jarrod and {van der Walt}, St{\'e}fan J. and {Gommers}, Ralf and {Virtanen}, Pauli and {Cournapeau}, David and {Wieser}, Eric and {Taylor}, Julian and {Berg}, Sebastian and {Smith}, Nathaniel J. and {Kern}, Robert and {Picus}, Matti and {Hoyer}, Stephan and {van Kerkwijk}, Marten H. and {Brett}, Matthew and {Haldane}, Allan and {del R{\'\i}o}, Jaime Fern{\'a}ndez and {Wiebe}, Mark and {Peterson}, Pearu and {G{\'e}rard-Marchant}, Pierre and {Sheppard}, Kevin and {Reddy}, Tyler and {Weckesser}, Warren and {Abbasi}, Hameer and {Gohlke}, Christoph and {Oliphant}, Travis E.},
        title = "{Array programming with NumPy}",
      journal = {\nat},
         year = 2020,
        month = sep,
       volume = {585},
       number = {7825},
        pages = {357-362},
          doi = {10.1038/s41586-020-2649-2},
archivePrefix = {arXiv},
       eprint = {2006.10256},
 primaryClass = {cs.MS},
       adsurl = {https://ui.adsabs.harvard.edu/abs/2020Natur.585..357H}
}

@ARTICLE{2007CSE.....9...90H,
       author = {{Hunter}, John D.},
        title = "{Matplotlib: A 2D Graphics Environment}",
      journal = {Computing in Science and Engineering},
         year = 2007,
        month = may,
       volume = {9},
       number = {3},
        pages = {90-95},
          doi = {10.1109/MCSE.2007.55},
       adsurl = {https://ui.adsabs.harvard.edu/abs/2007CSE.....9...90H}
}

@ARTICLE{2020NatMe..17..261V,
       author = {{Virtanen}, Pauli and {Gommers}, Ralf and {Oliphant}, Travis E. and {Haberland}, Matt and {Reddy}, Tyler and {Cournapeau}, David and {Burovski}, Evgeni and {Peterson}, Pearu and {Weckesser}, Warren and {Bright}, Jonathan and {van der Walt}, St{\'e}fan J. and {Brett}, Matthew and {Wilson}, Joshua and {Millman}, K. Jarrod and {Mayorov}, Nikolay and {Nelson}, Andrew R.~J. and {Jones}, Eric and {Kern}, Robert and {Larson}, Eric and {Carey}, C.~J. and {Polat}, {\.I}lhan and {Feng}, Yu and {Moore}, Eric W. and {VanderPlas}, Jake and {Laxalde}, Denis and {Perktold}, Josef and {Cimrman}, Robert and {Henriksen}, Ian and {Quintero}, E.~A. and {Harris}, Charles R. and {Archibald}, Anne M. and {Ribeiro}, Ant{\^o}nio H. and {Pedregosa}, Fabian and {van Mulbregt}, Paul and {SciPy 1. 0 Contributors}},
        title = "{SciPy 1.0: fundamental algorithms for scientific computing in Python}",
      journal = {Nature Methods},
         year = 2020,
        month = feb,
       volume = {17},
        pages = {261-272},
          doi = {10.1038/s41592-019-0686-2},
archivePrefix = {arXiv},
       eprint = {1907.10121},
 primaryClass = {cs.MS},
       adsurl = {https://ui.adsabs.harvard.edu/abs/2020NatMe..17..261V}
}

@ARTICLE{2023arXiv230405855A,
       author = {{{\'A}ngela Garc{\'\i}a}, Luz and {Martini}, Paul and {Gonzalez-Morales}, Alma X. and {Font-Ribera}, Andreu and {Herrera-Alcantar}, Hiram K. and {Aguilar}, Jessica Nicole and {Ahlen}, Steve and {Brooks}, David and {de la Macorra}, Axel and {Doel}, Peter and {Forero-Romero}, Jaime E. and {Guy}, Julien and {Kisner}, Theodore and {Landriau}, Martin and {Miquel}, Ramon and {Moustakas}, John and {Nie}, Jundan and {Poppett}, Claire and {Tarl{\'e}}, Gregory and {Zhou}, Zhimin},
        title = "{Analysis of the impact of broad absorption lines on quasar redshift measurements with synthetic observations}",
      journal = {arXiv e-prints},
         year = 2023,
        month = apr,
          eid = {arXiv:2304.05855},
        pages = {arXiv:2304.05855},
          doi = {10.48550/arXiv.2304.05855},
archivePrefix = {arXiv},
       eprint = {2304.05855},
 primaryClass = {astro-ph.CO},
       adsurl = {https://ui.adsabs.harvard.edu/abs/2023arXiv230405855A}
}

@ARTICLE{2015JCAP...12..017A,
       author = {{Arinyo-i-Prats}, Andreu and {Miralda-Escud{\'e}}, Jordi and {Viel}, Matteo and {Cen}, Renyue},
        title = "{The non-linear power spectrum of the Lyman alpha forest}",
      journal = {\jcap},
         year = 2015,
        month = dec,
       volume = {2015},
       number = {12},
          eid = {017},
        pages = {017},
          doi = {10.1088/1475-7516/2015/12/017},
archivePrefix = {arXiv},
       eprint = {1506.04519},
 primaryClass = {astro-ph.CO},
       adsurl = {https://ui.adsabs.harvard.edu/abs/2015JCAP...12..017A}
}

@ARTICLE{2018MNRAS.476.3716R,
       author = {{Rogers}, Keir K. and {Bird}, Simeon and {Peiris}, Hiranya V. and {Pontzen}, Andrew and {Font-Ribera}, Andreu and {Leistedt}, Boris},
        title = "{Correlations in the three-dimensional Lyman-alpha forest contaminated by high column density absorbers}",
      journal = {\mnras},
         year = 2018,
        month = may,
       volume = {476},
       number = {3},
        pages = {3716-3728},
          doi = {10.1093/mnras/sty603},
archivePrefix = {arXiv},
       eprint = {1711.06275},
 primaryClass = {astro-ph.CO},
       adsurl = {https://ui.adsabs.harvard.edu/abs/2018MNRAS.476.3716R}
}

@ARTICLE{2023JCAP...11..045G,
       author = {{Gordon}, C. and {Cuceu}, A. and {Chaves-Montero}, J. and {Font-Ribera}, A. and {Gonz{\'a}lez-Morales}, A.~X. and {Aguilar}, J. and {Ahlen}, S. and {Armengaud}, E. and {Bailey}, S. and {Bault}, A. and {Brodzeller}, A. and {Brooks}, D. and {Claybaugh}, T. and {de la Cruz}, R. and {Dawson}, K. and {Doel}, P. and {Forero-Romero}, J.~E. and {Gontcho}, S. Gontcho A. and {Guy}, J. and {Herrera-Alcantar}, H.~K. and {Ir{\v{s}}i{\v{c}}}, V. and {Kara{\c{c}}ayl{\i}}, N.~G. and {Kirkby}, D. and {Landriau}, M. and {Le Guillou}, L. and {Levi}, M.~E. and {de la Macorra}, A. and {Manera}, M. and {Martini}, P. and {Meisner}, A. and {Miquel}, R. and {Montero-Camacho}, P. and {Mu{\~n}oz-Guti{\'e}rrez}, A. and {Napolitano}, L. and {Nie}, J. and {Niz}, G. and {Palanque-Delabrouille}, N. and {Percival}, W.~J. and {Pieri}, M. and {Poppett}, C. and {Prada}, F. and {P{\'e}rez-R{\`a}fols}, I. and {Ram{\'\i}rez-P{\'e}rez}, C. and {Ravoux}, C. and {Rezaie}, M. and {Ross}, A.~J. and {Rossi}, G. and {Sanchez}, E. and {Schlegel}, D. and {Schubnell}, M. and {Seo}, H. and {Sinigaglia}, F. and {Tan}, T. and {Tarl{\'e}}, G. and {Walther}, M. and {Weaver}, B.~A. and {Y{\`e}che}, C. and {Zhou}, Z. and {Zou}, H.},
        title = "{3D correlations in the Lyman-{\ensuremath{\alpha}} forest from early DESI data}",
      journal = {\jcap},
         year = 2023,
        month = nov,
       volume = {2023},
       number = {11},
          eid = {045},
        pages = {045},
          doi = {10.1088/1475-7516/2023/11/045},
archivePrefix = {arXiv},
       eprint = {2308.10950},
 primaryClass = {astro-ph.CO},
       adsurl = {https://ui.adsabs.harvard.edu/abs/2023JCAP...11..045G}
}

@ARTICLE{2006ApJS..163...80M,
       author = {{McDonald}, Patrick and {Seljak}, Uro{\v{s}} and {Burles}, Scott and {Schlegel}, David J. and {Weinberg}, David H. and {Cen}, Renyue and {Shih}, David and {Schaye}, Joop and {Schneider}, Donald P. and {Bahcall}, Neta A. and {Briggs}, John W. and {Brinkmann}, J. and {Brunner}, Robert J. and {Fukugita}, Masataka and {Gunn}, James E. and {Ivezi{\'c}}, {\v{Z}}eljko and {Kent}, Stephen and {Lupton}, Robert H. and {Vanden Berk}, Daniel E.},
        title = "{The Ly{\ensuremath{\alpha}} Forest Power Spectrum from the Sloan Digital Sky Survey}",
      journal = {\apjs},
         year = 2006,
        month = mar,
       volume = {163},
       number = {1},
        pages = {80-109},
          doi = {10.1086/444361},
archivePrefix = {arXiv},
       eprint = {astro-ph/0405013},
 primaryClass = {astro-ph},
       adsurl = {https://ui.adsabs.harvard.edu/abs/2006ApJS..163...80M}
}

@ARTICLE{2019ApJ...878...47D,
       author = {{du Mas des Bourboux}, H{\'e}lion and {Dawson}, Kyle S. and {Busca}, Nicol{\'a}s G. and {Blomqvist}, Michael and {de Sainte Agathe}, Victoria and {Balland}, Christophe and {Bautista}, Julian E. and {Guy}, Julien and {Kamble}, Vikrant and {Myers}, Adam D. and {P{\'e}rez-R{\`a}fols}, Ignasi and {Pieri}, Matthew M. and {Rich}, James and {Schneider}, Donald P. and {Slosar}, An{\v{z}}e},
        title = "{The Extended Baryon Oscillation Spectroscopic Survey: Measuring the Cross-correlation between the Mg II Flux Transmission Field and Quasars and Galaxies at z = 0.59}",
      journal = {\apj},
         year = 2019,
        month = jun,
       volume = {878},
       number = {1},
          eid = {47},
        pages = {47},
          doi = {10.3847/1538-4357/ab1d49},
archivePrefix = {arXiv},
       eprint = {1901.01950},
 primaryClass = {astro-ph.CO},
       adsurl = {https://ui.adsabs.harvard.edu/abs/2019ApJ...878...47D}
}

@ARTICLE{2024MNRAS.529.3666M,
       author = {{Montero-Camacho}, Paulo and {Zhang}, Yao and {Mao}, Yi},
        title = "{The long-lasting effect of X-ray pre-heating in the post-reionization intergalactic medium}",
      journal = {\mnras},
         year = 2024,
        month = apr,
       volume = {529},
       number = {4},
        pages = {3666-3683},
          doi = {10.1093/mnras/stae751},
archivePrefix = {arXiv},
       eprint = {2307.10598},
 primaryClass = {astro-ph.CO},
       adsurl = {https://ui.adsabs.harvard.edu/abs/2024MNRAS.529.3666M}
}

@ARTICLE{2016MNRAS.456...47M,
       author = {{McQuinn}, Matthew and {Upton Sanderbeck}, Phoebe R.},
        title = "{On the intergalactic temperature-density relation}",
      journal = {\mnras},
         year = 2016,
        month = feb,
       volume = {456},
       number = {1},
        pages = {47-54},
          doi = {10.1093/mnras/stv2675},
archivePrefix = {arXiv},
       eprint = {1505.07875},
 primaryClass = {astro-ph.CO},
       adsurl = {https://ui.adsabs.harvard.edu/abs/2016MNRAS.456...47M}
}

@ARTICLE{2021MNRAS.508.1262M,
       author = {{Montero-Camacho}, Paulo and {Mao}, Yi},
        title = "{Extracting the astrophysics of reionization from the Ly{\ensuremath{\alpha}} forest power spectrum: a first forecast}",
      journal = {\mnras},
         year = 2021,
        month = nov,
       volume = {508},
       number = {1},
        pages = {1262-1279},
          doi = {10.1093/mnras/stab2569},
archivePrefix = {arXiv},
       eprint = {2106.14492},
 primaryClass = {astro-ph.CO},
       adsurl = {https://ui.adsabs.harvard.edu/abs/2021MNRAS.508.1262M}
}

@ARTICLE{2007ApJ...664..660E,
       author = {{Eisenstein}, Daniel J. and {Seo}, Hee-Jong and {White}, Martin},
        title = "{On the Robustness of the Acoustic Scale in the Low-Redshift Clustering of Matter}",
      journal = {\apj},
         year = 2007,
        month = aug,
       volume = {664},
       number = {2},
        pages = {660-674},
          doi = {10.1086/518755},
archivePrefix = {arXiv},
       eprint = {astro-ph/0604361},
 primaryClass = {astro-ph},
       adsurl = {https://ui.adsabs.harvard.edu/abs/2007ApJ...664..660E}
}

@ARTICLE{2013JCAP...03..024K,
       author = {{Kirkby}, David and {Margala}, Daniel and {Slosar}, An{\v{z}}e and {Bailey}, Stephen and {Busca}, Nicol{\'a}s G. and {Delubac}, Timoth{\'e}e and {Rich}, James and {Bautista}, Julian E. and {Blomqvist}, Michael and {Brownstein}, Joel R. and {Carithers}, Bill and {Croft}, Rupert A.~C. and {Dawson}, Kyle S. and {Font-Ribera}, Andreu and {Miralda-Escud{\'e}}, Jordi and {Myers}, Adam D. and {Nichol}, Robert C. and {Palanque-Delabrouille}, Nathalie and {P{\^a}ris}, Isabelle and {Petitjean}, Patrick and {Rossi}, Graziano and {Schlegel}, David J. and {Schneider}, Donald P. and {Viel}, Matteo and {Weinberg}, David H. and {Y{\`e}che}, Christophe},
        title = "{Fitting methods for baryon acoustic oscillations in the Lyman-{\ensuremath{\alpha}} forest fluctuations in BOSS data release 9}",
      journal = {\jcap},
         year = 2013,
        month = mar,
       volume = {2013},
       number = {3},
          eid = {024},
        pages = {024},
          doi = {10.1088/1475-7516/2013/03/024},
archivePrefix = {arXiv},
       eprint = {1301.3456},
 primaryClass = {astro-ph.CO},
       adsurl = {https://ui.adsabs.harvard.edu/abs/2013JCAP...03..024K}
}

@ARTICLE{2009MNRAS.393..297P,
       author = {{Percival}, Will J. and {White}, Martin},
        title = "{Testing cosmological structure formation using redshift-space distortions}",
      journal = {\mnras},
         year = 2009,
        month = feb,
       volume = {393},
       number = {1},
        pages = {297-308},
          doi = {10.1111/j.1365-2966.2008.14211.x},
archivePrefix = {arXiv},
       eprint = {0808.0003},
 primaryClass = {astro-ph},
       adsurl = {https://ui.adsabs.harvard.edu/abs/2009MNRAS.393..297P}
}

@ARTICLE{2022MNRAS.516..421Y,
       author = {{Youles}, Samantha and {Bautista}, Julian E. and {Font-Ribera}, Andreu and {Bacon}, David and {Rich}, James and {Brooks}, David and {Davis}, Tamara M. and {Dawson}, Kyle and {de la Macorra}, Axel and {Dhungana}, Govinda and {Doel}, Peter and {Fanning}, Kevin and {Gazta{\~n}aga}, Enrique and {Gontcho A Gontcho}, Satya and {Gonzalez-Morales}, Alma X. and {Guy}, Julien and {Honscheid}, Klaus and {Ir{\v{s}}i{\v{c}}}, Vid and {Kehoe}, Robert and {Kirkby}, David and {Kisner}, Theodore and {Landriau}, Martin and {Le Guillou}, Laurent and {Levi}, Michael E. and {Martini}, Paul and {Mu{\~n}oz-Guti{\'e}rrez}, Andrea and {Palanque-Delabrouille}, Nathalie and {P{\'e}rez-R{\`a}fols}, Ignasi and {Poppett}, Claire and {Ram{\'\i}rez-P{\'e}rez}, C{\'e}sar and {Schubnell}, Michael and {Tarl{\'e}}, Gregory and {Walther}, Michael},
        title = "{The effect of quasar redshift errors on Lyman-{\ensuremath{\alpha}} forest correlation functions}",
      journal = {\mnras},
         year = 2022,
        month = oct,
       volume = {516},
       number = {1},
        pages = {421-433},
          doi = {10.1093/mnras/stac2102},
archivePrefix = {arXiv},
       eprint = {2205.06648},
 primaryClass = {astro-ph.CO},
       adsurl = {https://ui.adsabs.harvard.edu/abs/2022MNRAS.516..421Y}
}

@ARTICLE{2023MNRAS.526.5118R,
       author = {{Ravoux}, Corentin and {Abdul Karim}, Marie Lynn and {Armengaud}, Eric and {Walther}, Michael and {Kara{\c{c}}ayl{\i}}, Naim G{\"o}ksel and {Martini}, Paul and {Guy}, Julien and {Aguilar}, Jessica Nicole and {Ahlen}, Steven and {Bailey}, Stephen and {Bautista}, Julian and {Beltran}, Sergio Felipe and {Brooks}, David and {Cabayol-Garcia}, Laura and {Chabanier}, Sol{\`e}ne and {Chaussidon}, Edmond and {Chaves-Montero}, Jon{\'a}s and {Dawson}, Kyle and {de la Cruz}, Rodrigo and {de la Macorra}, Axel and {Doel}, Peter and {Fanning}, Kevin and {Font-Ribera}, Andreu and {Forero-Romero}, Jaime and {Gontcho A Gontcho}, Satya and {Gonzalez-Morales}, Alma X. and {Gordon}, Calum and {Herrera-Alcantar}, Hiram K. and {Honscheid}, Klaus and {Ir{\v{s}}i{\v{c}}}, Vid and {Ishak}, Mustapha and {Kehoe}, Robert and {Kisner}, Theodore and {Kremin}, Anthony and {Landriau}, Martin and {Le Guillou}, Laurent and {Levi}, Michael and {Luki{\'c}}, Zarija and {Magneville}, Christophe and {Meisner}, Aaron and {Miquel}, Ramon and {Moustakas}, John and {Mueller}, Eva-Maria and {Mu{\~n}oz-Guti{\'e}rrez}, Andrea and {Napolitano}, Lucas and {Nie}, Jundan and {Niz}, Gustavo and {Palanque-Delabrouille}, Nathalie and {Percival}, Will and {P{\'e}rez-R{\`a}fols}, Ignasi and {Pieri}, Matthew and {Poppett}, Claire and {Prada}, Francisco and {Ram{\'\i}rez P{\'e}rez}, C{\'e}sar and {Rossi}, Graziano and {Sanchez}, Eusebio and {Schlegel}, David and {Schubnell}, Michael and {Seo}, Hee-Jong and {Sinigaglia}, Francesco and {Tan}, Ting and {Tarl{\'e}}, Gregory and {Wang}, Ben and {Weaver}, Benjamin and {Y{\`e}che}, Christophe and {Zhou}, Zhimin},
        title = "{The Dark Energy Spectroscopic Instrument: one-dimensional power spectrum from first Ly {\ensuremath{\alpha}} forest samples with Fast Fourier Transform}",
      journal = {\mnras},
         year = 2023,
        month = dec,
       volume = {526},
       number = {4},
        pages = {5118-5140},
          doi = {10.1093/mnras/stad3008},
archivePrefix = {arXiv},
       eprint = {2306.06311},
 primaryClass = {astro-ph.CO},
       adsurl = {https://ui.adsabs.harvard.edu/abs/2023MNRAS.526.5118R}
}

@ARTICLE{2022arXiv220307491V,
       author = {{Valluri}, Monica and {Chabanier}, Solene and {Irsic}, Vid and {Armengaud}, Eric and {Walther}, Michael and {Rockosi}, Connie and {Sanchez-Conde}, Miguel A. and {Beraldo e Silva}, Leandro and {Cooper}, Andrew P. and {Darragh-Ford}, Elise and {Dawson}, Kyle and {Deason}, Alis J. and {Ferraro}, Simone and {Forero-Romero}, Jaime E. and {Garzilli}, Antonella and {Li}, Ting and {Lukic}, Zarija and {Manser}, Christopher J. and {Palanque-Delabrouille}, Nathalie and {Ravoux}, Corentin and {Tan}, Ting and {Wang}, Wenting and {Wechsler}, Risa and {Carrillo}, Andreia and {Dey}, Arjun and {Koposov}, Sergey E. and {Mao}, Yao-Yuan and {Montero-Camacho}, Paulo and {Patel}, Ekta and {Rossi}, Graziano and {Urena-Lopez}, L. Arturo and {Valenzuela}, Octavio},
        title = "{Snowmass2021 Cosmic Frontier White Paper: Prospects for obtaining Dark Matter Constraints with DESI}",
      journal = {arXiv e-prints},
         year = 2022,
        month = mar,
          eid = {arXiv:2203.07491},
        pages = {arXiv:2203.07491},
          doi = {10.48550/arXiv.2203.07491},
archivePrefix = {arXiv},
       eprint = {2203.07491},
 primaryClass = {astro-ph.CO},
       adsurl = {https://ui.adsabs.harvard.edu/abs/2022arXiv220307491V}
}

@ARTICLE{2023arXiv230606308D,
       author = {{DESI Collaboration} and {Adame}, A.~G. and {Aguilar}, J. and {Ahlen}, S. and {Alam}, S. and {Aldering}, G. and {Alexander}, D.~M. and {Alfarsy}, R. and {Allende Prieto}, C. and {Alvarez}, M. and {Alves}, O. and {Anand}, A. and {Andrade-Oliveira}, F. and {Armengaud}, E. and {Asorey}, J. and {Avila}, S. and {Aviles}, A. and {Bailey}, S. and {Balaguera-Antol{\'\i}nez}, A. and {Ballester}, O. and {Baltay}, C. and {Bault}, A. and {Bautista}, J. and {Behera}, J. and {Beltran}, S.~F. and {BenZvi}, S. and {Beraldo e Silva}, L. and {Bermejo-Climent}, J.~R. and {Berti}, A. and {Besuner}, R. and {Beutler}, F. and {Bianchi}, D. and {Blake}, C. and {Blum}, R. and {Bolton}, A.~S. and {Brieden}, S. and {Brodzeller}, A. and {Brooks}, D. and {Brown}, Z. and {Buckley-Geer}, E. and {Burtin}, E. and {Cabayol-Garcia}, L. and {Cai}, Z. and {Canning}, R. and {Cardiel-Sas}, L. and {Carnero Rosell}, A. and {Castander}, F.~J. and {Cervantes-Cota}, J.~L. and {Chabanier}, S. and {Chaussidon}, E. and {Chaves-Montero}, J. and {Chen}, S. and {Chuang}, C. and {Claybaugh}, T. and {Cole}, S. and {Cooper}, A.~P. and {Cuceu}, A. and {Davis}, T.~M. and {Dawson}, K. and {de Belsunce}, R. and {de la Cruz}, R. and {de la Macorra}, A. and {de Mattia}, A. and {Demina}, R. and {Demirbozan}, U. and {DeRose}, J. and {Dey}, A. and {Dey}, B. and {Dhungana}, G. and {Ding}, J. and {Ding}, Z. and {Doel}, P. and {Doshi}, R. and {Douglass}, K. and {Edge}, A. and {Eftekharzadeh}, S. and {Eisenstein}, D.~J. and {Elliott}, A. and {Escoffier}, S. and {Fagrelius}, P. and {Fan}, X. and {Fanning}, K. and {Fawcett}, V.~A. and {Ferraro}, S. and {Ereza}, J. and {Flaugher}, B. and {Font-Ribera}, A. and {Forero-S{\'a}nchez}, D. and {Forero-Romero}, J.~E. and {Frenk}, C.~S. and {G{\"a}nsicke}, B.~T. and {Garc{\'\i}a}, L. {\'A}. and {Garc{\'\i}a-Bellido}, J. and {Garcia-Quintero}, C. and {Garrison}, L.~H. and {Gil-Mar{\'\i}n}, H. and {Golden-Marx}, J. and {Gontcho}, S. Gontcho A and {Gonzalez-Morales}, A.~X. and {Gonzalez-Perez}, V. and {Gordon}, C. and {Graur}, O. and {Green}, D. and {Gruen}, D. and {Guy}, J. and {Hadzhiyska}, B. and {Hahn}, C. and {Han}, J.~J. and {Hanif}, M.~M. S and {Herrera-Alcantar}, H.~K. and {Honscheid}, K. and {Hou}, J. and {Howlett}, C. and {Huterer}, D. and {Ir{\v{s}}i{\v{c}}}, V. and {Ishak}, M. and {Jacques}, A. and {Jana}, A. and {Jiang}, L. and {Jimenez}, J. and {Jing}, Y.~P. and {Joudaki}, S. and {Jullo}, E. and {Juneau}, S. and {Kizhuprakkat}, N. and {Kara{\c{c}}ayl{\i}}, N.~G. and {Karim}, T. and {Kehoe}, R. and {Kent}, S. and {Khederlarian}, A. and {Kim}, S. and {Kirkby}, D. and {Kisner}, T. and {Kitaura}, F. and {Kneib}, J. and {Koposov}, S.~E. and {Kov{\'a}cs}, A. and {Kremin}, A. and {Krolewski}, A. and {L'Huillier}, B. and {Lambert}, A. and {Lamman}, C. and {Lan}, T. -W. and {Landriau}, M. and {Lang}, D. and {Lange}, J.~U. and {Lasker}, J. and {Le Guillou}, L. and {Leauthaud}, A. and {Levi}, M.~E. and {Li}, T.~S. and {Linder}, E. and {Lyons}, A. and {Magneville}, C. and {Manera}, M. and {Manser}, C.~J. and {Margala}, D. and {Martini}, P. and {McDonald}, P. and {Medina}, G.~E. and {Medina-Varela}, L. and {Meisner}, A. and {Mena-Fern{\'a}ndez}, J. and {Meneses-Rizo}, J. and {Mezcua}, M. and {Miquel}, R. and {Montero-Camacho}, P. and {Moon}, J. and {Moore}, S. and {Moustakas}, J. and {Mueller}, E. and {Mundet}, J. and {Mu{\~n}oz-Guti{\'e}rrez}, A. and {Myers}, A.~D. and {Nadathur}, S. and {Napolitano}, L. and {Neveux}, R. and {Newman}, J.~A. and {Nie}, J. and {Nikutta}, R. and {Niz}, G. and {Norberg}, P. and {Noriega}, H.~E. and {Paillas}, E. and {Palanque-Delabrouille}, N. and {Palmese}, A. and {Zhiwei}, P. and {Parkinson}, D. and {Penmetsa}, S. and {Percival}, W.~J. and {P{\'e}rez-Fern{\'a}ndez}, A. and {P{\'e}rez-R{\`a}fols}, I. and {Pieri}, M. and {Poppett}, C. and {Porredon}, A. and {Pothier}, S. and {Prada}, F. and {Pucha}, R. and {Raichoor}, A. and {Ram{\'\i}rez-P{\'e}rez}, C. and {Ramirez-Solano}, S. and {Rashkovetskyi}, M. and {Ravoux}, C. and {Rocher}, A. and {Rockosi}, C. and {Ross}, A.~J. and {Rossi}, G. and {Ruggeri}, R. and {Ruhlmann-Kleider}, V. and {Sabiu}, C.~G. and {Said}, K. and {Saintonge}, A. and {Samushia}, L. and {Sanchez}, E. and {Saulder}, C. and {Schaan}, E. and {Schlafly}, E.~F. and {Schlegel}, D. and {Scholte}, D. and {Schubnell}, M. and {Seo}, H. and {Shafieloo}, A. and {Sharples}, R. and {Sheu}, W. and {Silber}, J. and {Sinigaglia}, F. and {Siudek}, M. and {Slepian}, Z. and {Smith}, A. and {Sprayberry}, D. and {Stephey}, L. and {Su{\'a}rez-P{\'e}rez}, J. and {Sun}, Z. and {Tan}, T. and {Tarl{\'e}}, G. and {Tojeiro}, R. and {Ure{\~n}a-L{\'o}pez}, L.~A. and {Vaisakh}, R. and {Valcin}, D. and {Valdes}, F. and {Valluri}, M. and {Vargas-Maga{\~n}a}, M. and {Variu}, A. and {Verde}, L. and {Walther}, M. and {Wang}, B. and {Wang}, M.~S. and {Weaver}, B.~A. and {Weaverdyck}, N. and {Wechsler}, R.~H. and {White}, M. and {Xie}, Y. and {Yang}, J. and {Y{\`e}che}, C. and {Yu}, J. and {Yuan}, S. and {Zhang}, H. and {Zhang}, Z. and {Zhao}, C. and {Zheng}, Z. and {Zhou}, R. and {Zhou}, Z. and {Zou}, H. and {Zou}, S. and {Zu}, Y.},
        title = "{The Early Data Release of the Dark Energy Spectroscopic Instrument}",
      journal = {arXiv e-prints},
         year = 2023,
        month = jun,
          eid = {arXiv:2306.06308},
        pages = {arXiv:2306.06308},
          doi = {10.48550/arXiv.2306.06308},
archivePrefix = {arXiv},
       eprint = {2306.06308},
 primaryClass = {astro-ph.CO},
       adsurl = {https://ui.adsabs.harvard.edu/abs/2023arXiv230606308D}
}

@ARTICLE{2023arXiv230606307D,
       author = {{DESI Collaboration} and {Adame}, A.~G. and {Aguilar}, J. and {Ahlen}, S. and {Alam}, S. and {Aldering}, G. and {Alexander}, D.~M. and {Alfarsy}, R. and {Allende Prieto}, C. and {Alvarez}, M. and {Alves}, O. and {Anand}, A. and {Andrade-Oliveira}, F. and {Armengaud}, E. and {Asorey}, J. and {Avila}, S. and {Aviles}, A. and {Bailey}, S. and {Balaguera-Antol{\'\i}nez}, A. and {Ballester}, O. and {Baltay}, C. and {Bault}, A. and {Bautista}, J. and {Behera}, J. and {Beltran}, S.~F. and {BenZvi}, S. and {Beraldo e Silva}, L. and {Bermejo-Climent}, J.~R. and {Berti}, A. and {Besuner}, R. and {Beutler}, F. and {Bianchi}, D. and {Blake}, C. and {Blum}, R. and {Bolton}, A.~S. and {Brieden}, S. and {Brodzeller}, A. and {Brooks}, D. and {Brown}, Z. and {Buckley-Geer}, E. and {Burtin}, E. and {Cabayol-Garcia}, L. and {Cai}, Z. and {Canning}, R. and {Cardiel-Sas}, L. and {Carnero Rosell}, A. and {Castander}, F.~J. and {Cervantes-Cota}, J.~L. and {Chabanier}, S. and {Chaussidon}, E. and {Chaves-Montero}, J. and {Chen}, S. and {Chuang}, C. and {Claybaugh}, T. and {Cole}, S. and {Cooper}, A.~P. and {Cuceu}, A. and {Davis}, T.~M. and {Dawson}, K. and {de Belsunce}, R. and {de la Cruz}, R. and {de la Macorra}, A. and {de Mattia}, A. and {Demina}, R. and {Demirbozan}, U. and {DeRose}, J. and {Dey}, A. and {Dey}, B. and {Dhungana}, G. and {Ding}, J. and {Ding}, Z. and {Doel}, P. and {Doshi}, R. and {Douglass}, K. and {Edge}, A. and {Eftekharzadeh}, S. and {Eisenstein}, D.~J. and {Elliott}, A. and {Escoffier}, S. and {Fagrelius}, P. and {Fan}, X. and {Fanning}, K. and {Fawcett}, V.~A. and {Ferraro}, S. and {Ereza}, J. and {Flaugher}, B. and {Font-Ribera}, A. and {Forero-S{\'a}nchez}, D. and {Forero-Romero}, J.~E. and {Frenk}, C.~S. and {G{\"a}nsicke}, B.~T. and {Garc{\'\i}a}, L. {\'A}. and {Garc{\'\i}a-Bellido}, J. and {Garcia-Quintero}, C. and {Garrison}, L.~H. and {Gil-Mar{\'\i}n}, H. and {Golden-Marx}, J. and {Gontcho}, S. Gontcho A and {Gonzalez-Morales}, A.~X. and {Gonzalez-Perez}, V. and {Gordon}, C. and {Graur}, O. and {Green}, D. and {Gruen}, D. and {Guy}, J. and {Hadzhiyska}, B. and {Hahn}, C. and {Han}, J.~J. and {Hanif}, M.~M. S and {Herrera-Alcantar}, H.~K. and {Honscheid}, K. and {Hou}, J. and {Howlett}, C. and {Huterer}, D. and {Ir{\v{s}}i{\v{c}}}, V. and {Ishak}, M. and {Jana}, A. and {Jiang}, L. and {Jimenez}, J. and {Jing}, Y.~P. and {Joudaki}, S. and {Jullo}, E. and {Juneau}, S. and {Kizhuprakkat}, N. and {Kara{\c{c}}ayl{\i}}, N.~G. and {Karim}, T. and {Kehoe}, R. and {Kent}, S. and {Khederlarian}, A. and {Kim}, S. and {Kirkby}, D. and {Kisner}, T. and {Kitaura}, F. and {Kneib}, J. and {Koposov}, S.~E. and {Kov{\'a}cs}, A. and {Kremin}, A. and {Krolewski}, A. and {L'Huillier}, B. and {Lambert}, A. and {Lamman}, C. and {Lan}, T. -W. and {Landriau}, M. and {Lang}, D. and {Lange}, J.~U. and {Lasker}, J. and {Le Guillou}, L. and {Leauthaud}, A. and {Levi}, M.~E. and {Li}, T.~S. and {Linder}, E. and {Lyons}, A. and {Magneville}, C. and {Manera}, M. and {Manser}, C.~J. and {Margala}, D. and {Martini}, P. and {McDonald}, P. and {Medina}, G.~E. and {Medina-Varela}, L. and {Meisner}, A. and {Mena-Fern{\'a}ndez}, J. and {Meneses-Rizo}, J. and {Mezcua}, M. and {Miquel}, R. and {Montero-Camacho}, P. and {Moon}, J. and {Moore}, S. and {Moustakas}, J. and {Mueller}, E. and {Mundet}, J. and {Mu{\~n}oz-Guti{\'e}rrez}, A. and {Myers}, A.~D. and {Nadathur}, S. and {Napolitano}, L. and {Neveux}, R. and {Newman}, J.~A. and {Nie}, J. and {Niz}, G. and {Norberg}, P. and {Noriega}, H.~E. and {Paillas}, E. and {Palanque-Delabrouille}, N. and {Palmese}, A. and {Zhiwei}, P. and {Parkinson}, D. and {Penmetsa}, S. and {Percival}, W.~J. and {P{\'e}rez-Fern{\'a}ndez}, A. and {P{\'e}rez-R{\`a}fols}, I. and {Pieri}, M. and {Poppett}, C. and {Porredon}, A. and {Prada}, F. and {Pucha}, R. and {Raichoor}, A. and {Ram{\'\i}rez-P{\'e}rez}, C. and {Ramirez-Solano}, S. and {Rashkovetskyi}, M. and {Ravoux}, C. and {Rocher}, A. and {Rockosi}, C. and {Ross}, A.~J. and {Rossi}, G. and {Ruggeri}, R. and {Ruhlmann-Kleider}, V. and {Sabiu}, C.~G. and {Said}, K. and {Saintonge}, A. and {Samushia}, L. and {Sanchez}, E. and {Saulder}, C. and {Schaan}, E. and {Schlafly}, E.~F. and {Schlegel}, D. and {Scholte}, D. and {Schubnell}, M. and {Seo}, H. and {Shafieloo}, A. and {Sharples}, R. and {Sheu}, W. and {Silber}, J. and {Sinigaglia}, F. and {Siudek}, M. and {Slepian}, Z. and {Smith}, A. and {Sprayberry}, D. and {Stephey}, L. and {Su{\'a}rez-P{\'e}rez}, J. and {Sun}, Z. and {Tan}, T. and {Tarl{\'e}}, G. and {Tojeiro}, R. and {Ure{\~n}a-L{\'o}pez}, L.~A. and {Vaisakh}, R. and {Valcin}, D. and {Valdes}, F. and {Valluri}, M. and {Vargas-Maga{\~n}a}, M. and {Variu}, A. and {Verde}, L. and {Walther}, M. and {Wang}, B. and {Wang}, M.~S. and {Weaver}, B.~A. and {Weaverdyck}, N. and {Wechsler}, R.~H. and {White}, M. and {Xie}, Y. and {Yang}, J. and {Y{\`e}che}, C. and {Yu}, J. and {Yuan}, S. and {Zhang}, H. and {Zhang}, Z. and {Zhao}, C. and {Zheng}, Z. and {Zhou}, R. and {Zhou}, Z. and {Zou}, H. and {Zou}, S. and {Zu}, Y.},
        title = "{Validation of the Scientific Program for the Dark Energy Spectroscopic Instrument}",
      journal = {arXiv e-prints},
         year = 2023,
        month = jun,
          eid = {arXiv:2306.06307},
        pages = {arXiv:2306.06307},
          doi = {10.48550/arXiv.2306.06307},
archivePrefix = {arXiv},
       eprint = {2306.06307},
 primaryClass = {astro-ph.CO},
       adsurl = {https://ui.adsabs.harvard.edu/abs/2023arXiv230606307D}
}

@ARTICLE{2023arXiv230903434F,
       author = {{Filbert}, S. and {Martini}, P. and {Seebaluck}, K. and {Ennesser}, L. and {Alexander}, D.~M. and {Bault}, A. and {Brodzeller}, A. and {Herrera-Alcantar}, H.~K. and {Montero-Camacho}, P. and {P{\'e}rez-R{\`a}fols}, I. and {Ram{\'\i}rez-P{\'e}rez}, C. and {Ravoux}, C. and {Tan}, T. and {Aguilar}, J. and {Ahlen}, S. and {Bailey}, S. and {Brooks}, D. and {Claybaugh}, T. and {Dawson}, K. and {de la Macorra}, A. and {Doel}, P. and {Fanning}, K. and {Font-Ribera}, A. and {Forero-Romero}, J.~E. and {Gontcho}, S. Gontcho A and {Guy}, J. and {Kirkby}, D. and {Kremin}, A. and {Magneville}, C. and {Manera}, M. and {Meisner}, A. and {Miquel}, R. and {Moustakas}, J. and {Nie}, J. and {Percival}, W.~J. and {Prada}, F. and {Rezaie}, M. and {Rossi}, G. and {Sanchez}, E. and {Schubnell}, M. and {Seo}, H. and {Tarl{\'e}}, G. and {Weaver}, B.~A. and {Zhou}, Z.},
        title = "{Broad Absorption Line Quasars in the Dark Energy Spectroscopic Instrument Early Data Release}",
      journal = {arXiv e-prints},
         year = 2023,
        month = sep,
          eid = {arXiv:2309.03434},
        pages = {arXiv:2309.03434},
          doi = {10.48550/arXiv.2309.03434},
archivePrefix = {arXiv},
       eprint = {2309.03434},
 primaryClass = {astro-ph.CO},
       adsurl = {https://ui.adsabs.harvard.edu/abs/2023arXiv230903434F}
}

@ARTICLE{2012JCAP...07..028F,
       author = {{Font-Ribera}, Andreu and {Miralda-Escud{\'e}}, Jordi},
        title = "{The effect of high column density systems on the measurement of the Lyman-{\ensuremath{\alpha}} forest correlation function}",
      journal = {\jcap},
         year = 2012,
        month = jul,
       volume = {2012},
       number = {7},
          eid = {028},
        pages = {028},
          doi = {10.1088/1475-7516/2012/07/028},
archivePrefix = {arXiv},
       eprint = {1205.2018},
 primaryClass = {astro-ph.CO},
       adsurl = {https://ui.adsabs.harvard.edu/abs/2012JCAP...07..028F}
}

@ARTICLE{2013JCAP...05..018F,
       author = {{Font-Ribera}, Andreu and {Arnau}, Eduard and {Miralda-Escud{\'e}}, Jordi and {Rollinde}, Emmanuel and {Brinkmann}, J. and {Brownstein}, Joel R. and {Lee}, Khee-Gan and {Myers}, Adam D. and {Palanque-Delabrouille}, Nathalie and {P{\^a}ris}, Isabelle and {Petitjean}, Patrick and {Rich}, James and {Ross}, Nicholas P. and {Schneider}, Donald P. and {White}, Martin},
        title = "{The large-scale quasar-Lyman {\ensuremath{\alpha}} forest cross-correlation from BOSS}",
      journal = {\jcap},
         year = 2013,
        month = may,
       volume = {2013},
       number = {5},
          eid = {018},
        pages = {018},
          doi = {10.1088/1475-7516/2013/05/018},
archivePrefix = {arXiv},
       eprint = {1303.1937},
 primaryClass = {astro-ph.CO},
       adsurl = {https://ui.adsabs.harvard.edu/abs/2013JCAP...05..018F}
}

@ARTICLE{2013ApJ...769..146R,
       author = {{Rudie}, Gwen C. and {Steidel}, Charles C. and {Shapley}, Alice E. and {Pettini}, Max},
        title = "{The Column Density Distribution and Continuum Opacity of the Intergalactic and Circumgalactic Medium at Redshift langzrang = 2.4}",
      journal = {\apj},
         year = 2013,
        month = jun,
       volume = {769},
       number = {2},
          eid = {146},
        pages = {146},
          doi = {10.1088/0004-637X/769/2/146},
archivePrefix = {arXiv},
       eprint = {1304.6719},
 primaryClass = {astro-ph.CO},
       adsurl = {https://ui.adsabs.harvard.edu/abs/2013ApJ...769..146R}
}

@ARTICLE{2022ApJ...935..121Y,
       author = {{Yang}, Li and {Zheng}, Zheng and {du Mas des Bourboux}, H{\'e}lion and {Dawson}, Kyle and {Pieri}, Matthew M. and {Rossi}, Graziano and {Schneider}, Donald P. and {Macorra}, Axel de la and {Guti{\'e}rrez}, Adrean Mu{\~n}oz},
        title = "{Metal Lines Associated with the Ly{\ensuremath{\alpha}} Forest from eBOSS Data}",
      journal = {\apj},
         year = 2022,
        month = aug,
       volume = {935},
       number = {2},
          eid = {121},
        pages = {121},
          doi = {10.3847/1538-4357/ac7b2e},
archivePrefix = {arXiv},
       eprint = {2206.11385},
 primaryClass = {astro-ph.CO},
       adsurl = {https://ui.adsabs.harvard.edu/abs/2022ApJ...935..121Y}
}

@ARTICLE{2018JCAP...05..029B,
       author = {{Blomqvist}, Michael and {Pieri}, Matthew M. and {du Mas des Bourboux}, H{\'e}lion and {Busca}, Nicol{\'a}s G. and {Slosar}, An{\v{z}}e and {Bautista}, Julian E. and {Brinkmann}, Jonathan and {Brownstein}, Joel R. and {Dawson}, Kyle and {de Sainte Agathe}, Victoria and {Guy}, Julien and {Percival}, Will J. and {P{\'e}rez-R{\`a}fols}, Ignasi and {Rich}, James and {Schneider}, Donald P.},
        title = "{The triply-ionized carbon forest from eBOSS: cosmological correlations with quasars in SDSS-IV DR14}",
      journal = {\jcap},
         year = 2018,
        month = may,
       volume = {2018},
       number = {5},
          eid = {029},
        pages = {029},
          doi = {10.1088/1475-7516/2018/05/029},
archivePrefix = {arXiv},
       eprint = {1801.01852},
 primaryClass = {astro-ph.CO},
       adsurl = {https://ui.adsabs.harvard.edu/abs/2018JCAP...05..029B}
}

@ARTICLE{2012JCAP...11..059F,
       author = {{Font-Ribera}, Andreu and {Miralda-Escud{\'e}}, Jordi and {Arnau}, Eduard and {Carithers}, Bill and {Lee}, Khee-Gan and {Noterdaeme}, Pasquier and {P{\^a}ris}, Isabelle and {Petitjean}, Patrick and {Rich}, James and {Rollinde}, Emmanuel and {Ross}, Nicholas P. and {Schneider}, Donald P. and {White}, Martin and {York}, Donald G.},
        title = "{The large-scale cross-correlation of Damped Lyman alpha systems with the Lyman alpha forest: first measurements from BOSS}",
      journal = {\jcap},
         year = 2012,
        month = nov,
       volume = {2012},
       number = {11},
          eid = {059},
        pages = {059},
          doi = {10.1088/1475-7516/2012/11/059},
archivePrefix = {arXiv},
       eprint = {1209.4596},
 primaryClass = {astro-ph.CO},
       adsurl = {https://ui.adsabs.harvard.edu/abs/2012JCAP...11..059F}
}

@ARTICLE{2023arXiv230904129P,
       author = {{Park}, Hyunbae and {Luki{\'c}}, Zarija and {Sexton}, Jean and {Alvarez}, Marcelo},
        title = "{Impact of Self-shielding Minihalos on the Ly$\alpha$ Forest at High Redshift}",
      journal = {arXiv e-prints},
         year = 2023,
        month = sep,
          eid = {arXiv:2309.04129},
        pages = {arXiv:2309.04129},
          doi = {10.48550/arXiv.2309.04129},
archivePrefix = {arXiv},
       eprint = {2309.04129},
 primaryClass = {astro-ph.CO},
       adsurl = {https://ui.adsabs.harvard.edu/abs/2023arXiv230904129P}
}

@software{2021ascl.soft06018D,
       author = {{du Mas des Bourboux}, H{\'e}lion and {Rich}, James and {Font-Ribera}, Andreu and {de Sainte Agathe}, Victoria and {Farr}, James and {Etourneau}, Thomas and {Le Goff}, Jean-Marc and {Cuceu}, Andrei and {Balland}, Christophe and {Bautista}, Julian E. and {Blomqvist}, Michael and {Brinkmann}, Jonathan and {Brownstein}, Joel R. and {Chabanier}, Sol{\`e}ne and {Chaussidon}, Edmond and {Dawson}, Kyle and {Gonz{\'a}lez-Morales}, Alma X. and {Guy}, Julien and {Lyke}, Brad W. and {de la Macorra}, Axel and {Mueller}, Eva-Maria and {Myers}, Adam D. and {Nitschelm}, Christian and {Mu{\~n}oz Guti{\'e}rrez}, Andrea and {Palanque-Delabrouille}, Nathalie and {Parker}, James and {Percival}, Will J. and {P{\'e}rez-R{\`a}fols}, Ignasi and {Petitjean}, Patrick and {Pieri}, Matthew M. and {Ravoux}, Corentin and {Rossi}, Graziano and {Schneider}, Donald P. and {Seo}, Hee-Jong and {Slosar}, An{\r{A}}{\textthreequarters}e and {Stermer}, Julianna and {Vivek}, M. and {Y{\`e}che}, Christophe and {Youles}, Samantha},
        title = "{picca: Package for Igm Cosmological-Correlations Analyses}",
 howpublished = {Astrophysics Source Code Library, record ascl:2106.018},
         year = 2021,
        month = jun,
          eid = {ascl:2106.018},
       adsurl = {https://ui.adsabs.harvard.edu/abs/2021ascl.soft06018D}
}

@unpublished{igm_pysr,
    author = {{Montero-Camacho}, Paulo}, 
    title = {Modeling the Post-Reionization Era with Symbolic Regression},
    note = {In preparation},
    year = {in prep.}
}

@ARTICLE{2011JCAP...09..001S,
       author = {{Slosar}, An{\v{z}}e and {Font-Ribera}, Andreu and {Pieri}, Matthew M. and {Rich}, James and {Le Goff}, Jean-Marc and {Aubourg}, {\'E}ric and {Brinkmann}, Jon and {Busca}, Nicolas and {Carithers}, Bill and {Charlassier}, Romain and {Cort{\^e}s}, Marina and {Croft}, Rupert and {Dawson}, Kyle S. and {Eisenstein}, Daniel and {Hamilton}, Jean-Christophe and {Ho}, Shirley and {Lee}, Khee-Gan and {Lupton}, Robert and {McDonald}, Patrick and {Medolin}, Bumbarija and {Muna}, Demitri and {Miralda-Escud{\'e}}, Jordi and {Myers}, Adam D. and {Nichol}, Robert C. and {Palanque-Delabrouille}, Nathalie and {P{\^a}ris}, Isabelle and {Petitjean}, Patrick and {Pi{\v{s}}kur}, Yodovina and {Rollinde}, Emmanuel and {Ross}, Nicholas P. and {Schlegel}, David J. and {Schneider}, Donald P. and {Sheldon}, Erin and {Weaver}, Benjamin A. and {Weinberg}, David H. and {Yeche}, Christophe and {York}, Donald G.},
        title = "{The Lyman-{\ensuremath{\alpha}} forest in three dimensions: measurements of large scale flux correlations from BOSS 1st-year data}",
      journal = {\jcap},
         year = 2011,
        month = sep,
       volume = {2011},
       number = {9},
          eid = {001},
        pages = {001},
          doi = {10.1088/1475-7516/2011/09/001},
archivePrefix = {arXiv},
       eprint = {1104.5244},
 primaryClass = {astro-ph.CO},
       adsurl = {https://ui.adsabs.harvard.edu/abs/2011JCAP...09..001S}
}

@ARTICLE{2023MNRAS.519.6162P,
       author = {{Puchwein}, Ewald and {Bolton}, James S. and {Keating}, Laura C. and {Molaro}, Margherita and {Gaikwad}, Prakash and {Kulkarni}, Girish and {Haehnelt}, Martin G. and {Ir{\v{s}}i{\v{c}}}, Vid and {{\v{S}}oltinsk{\'y}}, Tom{\'a}{\v{s}} and {Viel}, Matteo and {Aubert}, Dominique and {Becker}, George D. and {Meiksin}, Avery},
        title = "{The Sherwood-Relics simulations: overview and impact of patchy reionization and pressure smoothing on the intergalactic medium}",
      journal = {\mnras},
         year = 2023,
        month = mar,
       volume = {519},
       number = {4},
        pages = {6162-6183},
          doi = {10.1093/mnras/stac3761},
archivePrefix = {arXiv},
       eprint = {2207.13098},
 primaryClass = {astro-ph.CO},
       adsurl = {https://ui.adsabs.harvard.edu/abs/2023MNRAS.519.6162P}
}

@ARTICLE{2025MNRAS.536.1645M,
       author = {{Montero-Camacho}, Paulo and {Morales-Guti{\'e}rrez}, Catalina and {Zhang}, Yao and {Long}, Heyang and {Mao}, Yi},
        title = "{Reionization relics in the cross-correlation between the Ly{\ensuremath{\alpha}} forest and 21 cm intensity mapping in the post-reionization era}",
      journal = {\mnras},
         year = 2025,
        month = jan,
       volume = {536},
       number = {2},
        pages = {1645-1659},
          doi = {10.1093/mnras/stae2684},
archivePrefix = {arXiv},
       eprint = {2409.11613},
       adsurl = {https://ui.adsabs.harvard.edu/abs/2025MNRAS.536.1645M}
}

@ARTICLE{2021MNRAS.506.4389G,
       author = {{Gaikwad}, Prakash and {Srianand}, Raghunathan and {Haehnelt}, Martin G. and {Choudhury}, Tirthankar Roy},
        title = "{A consistent and robust measurement of the thermal state of the IGM at 2 {\ensuremath{\leq}} z {\ensuremath{\leq}} 4 from a large sample of Ly {\ensuremath{\alpha}} forest spectra: evidence for late and rapid He II reionization}",
      journal = {\mnras},
         year = 2021,
        month = sep,
       volume = {506},
       number = {3},
        pages = {4389-4412},
          doi = {10.1093/mnras/stab2017},
archivePrefix = {arXiv},
       eprint = {2009.00016},
 primaryClass = {astro-ph.CO},
       adsurl = {https://ui.adsabs.harvard.edu/abs/2021MNRAS.506.4389G}
}

@ARTICLE{2015MNRAS.447.2503G,
       author = {{Greig}, Bradley and {Bolton}, James S. and {Wyithe}, J. Stuart B.},
        title = "{The impact of temperature fluctuations on the large-scale clustering of the Ly{\ensuremath{\alpha}} forest}",
      journal = {\mnras},
         year = 2015,
        month = mar,
       volume = {447},
       number = {3},
        pages = {2503-2511},
          doi = {10.1093/mnras/stu2624},
archivePrefix = {arXiv},
       eprint = {1411.1687},
 primaryClass = {astro-ph.CO},
       adsurl = {https://ui.adsabs.harvard.edu/abs/2015MNRAS.447.2503G}
}

@ARTICLE{2023JCAP...10..037B,
       author = {{Bird}, Simeon and {Fernandez}, Martin and {Ho}, Ming-Feng and {Qezlou}, Mahdi and {Monadi}, Reza and {Ni}, Yueying and {Chen}, Nianyi and {Croft}, Rupert and {Di Matteo}, Tiziana},
        title = "{PRIYA: a new suite of Lyman-{\ensuremath{\alpha}} forest simulations for cosmology}",
      journal = {\jcap},
         year = 2023,
        month = oct,
       volume = {2023},
       number = {10},
          eid = {037},
        pages = {037},
          doi = {10.1088/1475-7516/2023/10/037},
archivePrefix = {arXiv},
       eprint = {2306.05471},
 primaryClass = {astro-ph.CO},
       adsurl = {https://ui.adsabs.harvard.edu/abs/2023JCAP...10..037B}
}

@ARTICLE{2024JCAP...07..029F,
       author = {{Fernandez}, M.~A. and {Bird}, Simeon and {Ho}, Ming-Feng},
        title = "{Cosmological constraints from the eBOSS Lyman-{\ensuremath{\alpha}} forest using the PRIYA simulations}",
      journal = {\jcap},
         year = 2024,
        month = jul,
       volume = {2024},
       number = {7},
          eid = {029},
        pages = {029},
          doi = {10.1088/1475-7516/2024/07/029},
archivePrefix = {arXiv},
       eprint = {2309.03943},
 primaryClass = {astro-ph.CO},
       adsurl = {https://ui.adsabs.harvard.edu/abs/2024JCAP...07..029F}
}

@ARTICLE{2020ApJS..250....8L,
       author = {{Lyke}, Brad W. and {Higley}, Alexandra N. and {McLane}, J.~N. and {Schurhammer}, Danielle P. and {Myers}, Adam D. and {Ross}, Ashley J. and {Dawson}, Kyle and {Chabanier}, Sol{\`e}ne and {Martini}, Paul and {Busca}, Nicol{\'a}s G. and {Mas des Bourboux}, H{\'e}lion du and {Salvato}, Mara and {Streblyanska}, Alina and {Zarrouk}, Pauline and {Burtin}, Etienne and {Anderson}, Scott F. and {Bautista}, Julian and {Bizyaev}, Dmitry and {Brandt}, W.~N. and {Brinkmann}, Jonathan and {Brownstein}, Joel R. and {Comparat}, Johan and {Green}, Paul and {de la Macorra}, Axel and {Mu{\~n}oz Guti{\'e}rrez}, Andrea and {Hou}, Jiamin and {Newman}, Jeffrey A. and {Palanque-Delabrouille}, Nathalie and {P{\^a}ris}, Isabelle and {Percival}, Will J. and {Petitjean}, Patrick and {Rich}, James and {Rossi}, Graziano and {Schneider}, Donald P. and {Smith}, Alexander and {Vivek}, M. and {Weaver}, Benjamin Alan},
        title = "{The Sloan Digital Sky Survey Quasar Catalog: Sixteenth Data Release}",
      journal = {\apjs},
         year = 2020,
        month = sep,
       volume = {250},
       number = {1},
          eid = {8},
        pages = {8},
          doi = {10.3847/1538-4365/aba623},
archivePrefix = {arXiv},
       eprint = {2007.09001},
 primaryClass = {astro-ph.GA},
       adsurl = {https://ui.adsabs.harvard.edu/abs/2020ApJS..250....8L}
}

@ARTICLE{2024MNRAS.528.6666R,
       author = {{Ram{\'\i}rez-P{\'e}rez}, C{\'e}sar and {P{\'e}rez-R{\`a}fols}, Ignasi and {Font-Ribera}, Andreu and {Karim}, M. Abdul and {Armengaud}, E. and {Bautista}, J. and {Beltran}, S.~F. and {Cabayol-Garcia}, L. and {Cai}, Z. and {Chabanier}, S. and {Chaussidon}, E. and {Chaves-Montero}, J. and {Cuceu}, A. and {de la Cruz}, R. and {Garc{\'\i}a-Bellido}, J. and {Gonzalez-Morales}, A.~X. and {Gordon}, C. and {Herrera-Alcantar}, H.~K. and {Ir{\v{s}}i{\v{c}}}, V. and {Ishak}, M. and {Kara{\c{c}}ayl{\i}}, N.~G. and {Luki{\'c}}, Zarija and {Manser}, C.~J. and {Montero-Camacho}, P. and {Napolitano}, L. and {Niz}, G. and {Pieri}, M.~M. and {Ravoux}, C. and {Sinigaglia}, F. and {Tan}, T. and {Walther}, M. and {Wang}, B. and {Aguilar}, J. and {Ahlen}, S. and {Bailey}, S. and {Brooks}, D. and {Claybaugh}, T. and {Dawson}, K. and {de la Macorra}, A. and {Dhungana}, G. and {Doel}, P. and {Fanning}, K. and {Forero-Romero}, J.~E. and {Gontcho}, S. Gontcho A. and {Guy}, J. and {Honscheid}, K. and {Kehoe}, R. and {Kisner}, T. and {Landriau}, M. and {Le Guillou}, L. and {Levi}, M.~E. and {Magneville}, C. and {Martini}, P. and {Meisner}, A. and {Miquel}, R. and {Moustakas}, J. and {Mueller}, E. and {Mu{\~n}oz-Guti{\'e}rrez}, A. and {Nie}, J. and {Palanque-Delabrouille}, N. and {Percival}, W.~J. and {Rossi}, G. and {Sanchez}, E. and {Schlafly}, E.~F. and {Schlegel}, D. and {Seo}, H. and {Tarl{\'e}}, G. and {Weaver}, B.~A. and {Y{\'e}che}, C. and {Zhou}, Z.},
        title = "{The Lyman-{\ensuremath{\alpha}} forest catalogue from the Dark Energy Spectroscopic Instrument Early Data Release}",
      journal = {\mnras},
         year = 2024,
        month = mar,
       volume = {528},
       number = {4},
        pages = {6666-6679},
          doi = {10.1093/mnras/stad3781},
archivePrefix = {arXiv},
       eprint = {2306.06312},
 primaryClass = {astro-ph.CO},
       adsurl = {https://ui.adsabs.harvard.edu/abs/2024MNRAS.528.6666R}
}

@ARTICLE{2022AJ....164..207D,
       author = {{DESI Collaboration} and {Abareshi}, B. and {Aguilar}, J. and {Ahlen}, S. and {Alam}, Shadab and {Alexander}, David M. and {Alfarsy}, R. and {Allen}, L. and {Allende Prieto}, C. and {Alves}, O. and {Ameel}, J. and {Armengaud}, E. and {Asorey}, J. and {Aviles}, Alejandro and {Bailey}, S. and {Balaguera-Antol{\'\i}nez}, A. and {Ballester}, O. and {Baltay}, C. and {Bault}, A. and {Beltran}, S.~F. and {Benavides}, B. and {BenZvi}, S. and {Berti}, A. and {Besuner}, R. and {Beutler}, Florian and {Bianchi}, D. and {Blake}, C. and {Blanc}, P. and {Blum}, R. and {Bolton}, A. and {Bose}, S. and {Bramall}, D. and {Brieden}, S. and {Brodzeller}, A. and {Brooks}, D. and {Brownewell}, C. and {Buckley-Geer}, E. and {Cahn}, R.~N. and {Cai}, Z. and {Canning}, R. and {Capasso}, R. and {Carnero Rosell}, A. and {Carton}, P. and {Casas}, R. and {Castander}, F.~J. and {Cervantes-Cota}, J.~L. and {Chabanier}, S. and {Chaussidon}, E. and {Chuang}, C. and {Circosta}, C. and {Cole}, S. and {Cooper}, A.~P. and {da Costa}, L. and {Cousinou}, M. -C. and {Cuceu}, A. and {Davis}, T.~M. and {Dawson}, K. and {de la Cruz-Noriega}, R. and {de la Macorra}, A. and {de Mattia}, A. and {Della Costa}, J. and {Demmer}, P. and {Derwent}, M. and {Dey}, A. and {Dey}, B. and {Dhungana}, G. and {Ding}, Z. and {Dobson}, C. and {Doel}, P. and {Donald-McCann}, J. and {Donaldson}, J. and {Douglass}, K. and {Duan}, Y. and {Dunlop}, P. and {Edelstein}, J. and {Eftekharzadeh}, S. and {Eisenstein}, D.~J. and {Enriquez-Vargas}, M. and {Escoffier}, S. and {Evatt}, M. and {Fagrelius}, P. and {Fan}, X. and {Fanning}, K. and {Fawcett}, V.~A. and {Ferraro}, S. and {Ereza}, J. and {Flaugher}, B. and {Font-Ribera}, A. and {Forero-Romero}, J.~E. and {Frenk}, C.~S. and {Fromenteau}, S. and {G{\"a}nsicke}, B.~T. and {Garcia-Quintero}, C. and {Garrison}, L. and {Gazta{\~n}aga}, E. and {Gerardi}, F. and {Gil-Mar{\'\i}n}, H. and {Gontcho A Gontcho}, S. and {Gonzalez-Morales}, Alma X. and {Gonzalez-de-Rivera}, G. and {Gonzalez-Perez}, V. and {Gordon}, C. and {Graur}, O. and {Green}, D. and {Grove}, C. and {Gruen}, D. and {Gutierrez}, G. and {Guy}, J. and {Hahn}, C. and {Harris}, S. and {Herrera}, D. and {Herrera-Alcantar}, Hiram K. and {Honscheid}, K. and {Howlett}, C. and {Huterer}, D. and {Ir{\v{s}}i{\v{c}}}, V. and {Ishak}, M. and {Jelinsky}, P. and {Jiang}, L. and {Jimenez}, J. and {Jing}, Y.~P. and {Joyce}, R. and {Jullo}, E. and {Juneau}, S. and {Kara{\c{c}}ayl{\i}}, N.~G. and {Karamanis}, M. and {Karcher}, A. and {Karim}, T. and {Kehoe}, R. and {Kent}, S. and {Kirkby}, D. and {Kisner}, T. and {Kitaura}, F. and {Koposov}, S.~E. and {Kov{\'a}cs}, A. and {Kremin}, A. and {Krolewski}, Alex and {L'Huillier}, B. and {Lahav}, O. and {Lambert}, A. and {Lamman}, C. and {Lan}, Ting-Wen and {Landriau}, M. and {Lane}, S. and {Lang}, D. and {Lange}, J.~U. and {Lasker}, J. and {Le Guillou}, L. and {Leauthaud}, A. and {Le Van Suu}, A. and {Levi}, Michael E. and {Li}, T.~S. and {Magneville}, C. and {Manera}, M. and {Manser}, Christopher J. and {Marshall}, B. and {Martini}, Paul and {McCollam}, W. and {McDonald}, P. and {Meisner}, Aaron M. and {Mena-Fern{\'a}ndez}, J. and {Meneses-Rizo}, J. and {Mezcua}, M. and {Miller}, T. and {Miquel}, R. and {Montero-Camacho}, P. and {Moon}, J. and {Moustakas}, J. and {Mueller}, E. and {Mu{\~n}oz-Guti{\'e}rrez}, Andrea and {Myers}, Adam D. and {Nadathur}, S. and {Najita}, J. and {Napolitano}, L. and {Neilsen}, E. and {Newman}, Jeffrey A. and {Nie}, J.~D. and {Ning}, Y. and {Niz}, G. and {Norberg}, P. and {Noriega}, Hern{\'a}n E. and {O'Brien}, T. and {Obuljen}, A. and {Palanque-Delabrouille}, N. and {Palmese}, A. and {Zhiwei}, P. and {Pappalardo}, D. and {PENG}, X. and {Percival}, W.~J. and {Perruchot}, S. and {Pogge}, R. and {Poppett}, C. and {Porredon}, A. and {Prada}, F. and {Prochaska}, J. and {Pucha}, R. and {P{\'e}rez-Fern{\'a}ndez}, A. and {P{\'e}rez-R{\`a}fols}, I. and {Rabinowitz}, D. and {Raichoor}, A. and {Ramirez-Solano}, S. and {Ram{\'\i}rez-P{\'e}rez}, C{\'e}sar and {Ravoux}, C. and {Reil}, K. and {Rezaie}, M. and {Rocher}, A. and {Rockosi}, C. and {Roe}, N.~A. and {Roodman}, A. and {Ross}, A.~J. and {Rossi}, G. and {Ruggeri}, R. and {Ruhlmann-Kleider}, V. and {Sabiu}, C.~G. and {Gaines}, S. and {Said}, K. and {Saintonge}, A. and {Salas Catonga}, Javier and {Samushia}, L. and {Sanchez}, E. and {Saulder}, C. and {Schaan}, E. and {Schlafly}, E. and {Schlegel}, D. and {Schmoll}, J. and {Scholte}, D. and {Schubnell}, M. and {Secroun}, A. and {Seo}, H. and {Serrano}, S. and {Sharples}, Ray M. and {Sholl}, Michael J. and {Silber}, Joseph Harry and {Silva}, D.~R. and {Sirk}, M. and {Siudek}, M. and {Smith}, A. and {Sprayberry}, D. and {Staten}, R. and {Stupak}, B. and {Tan}, T. and {Tarl{\'e}}, Gregory and {Tie}, Suk Sien and {Tojeiro}, R. and {Ure{\~n}a-L{\'o}pez}, L.~A. and {Valdes}, F. and {Valenzuela}, O. and {Valluri}, M. and {Vargas-Maga{\~n}a}, M. and {Verde}, L. and {Walther}, M. and {Wang}, B. and {Wang}, M.~S. and {Weaver}, B.~A. and {Weaverdyck}, C. and {Wechsler}, R. and {Wilson}, Michael J. and {Yang}, J. and {Yu}, Y. and {Yuan}, S. and {Y{\`e}che}, Christophe and {Zhang}, H. and {Zhang}, K. and {Zhao}, Cheng and {Zhou}, Rongpu and {Zhou}, Zhimin and {Zou}, H. and {Zou}, J. and {Zou}, S. and {Zu}, Y.},
        title = "{Overview of the Instrumentation for the Dark Energy Spectroscopic Instrument}",
      journal = {\aj},
         year = 2022,
        month = nov,
       volume = {164},
       number = {5},
          eid = {207},
        pages = {207},
          doi = {10.3847/1538-3881/ac882b},
archivePrefix = {arXiv},
       eprint = {2205.10939},
 primaryClass = {astro-ph.IM},
       adsurl = {https://ui.adsabs.harvard.edu/abs/2022AJ....164..207D}
}

@ARTICLE{2024arXiv240403002D,
       author = {{DESI Collaboration} and {Adame}, A.~G. and {Aguilar}, J. and {Ahlen}, S. and {Alam}, S. and {Alexander}, D.~M. and {Alvarez}, M. and {Alves}, O. and {Anand}, A. and {Andrade}, U. and {Armengaud}, E. and {Avila}, S. and {Aviles}, A. and {Awan}, H. and {Bahr-Kalus}, B. and {Bailey}, S. and {Baltay}, C. and {Bault}, A. and {Behera}, J. and {BenZvi}, S. and {Bera}, A. and {Beutler}, F. and {Bianchi}, D. and {Blake}, C. and {Blum}, R. and {Brieden}, S. and {Brodzeller}, A. and {Brooks}, D. and {Buckley-Geer}, E. and {Burtin}, E. and {Calderon}, R. and {Canning}, R. and {Carnero Rosell}, A. and {Cereskaite}, R. and {Cervantes-Cota}, J.~L. and {Chabanier}, S. and {Chaussidon}, E. and {Chaves-Montero}, J. and {Chen}, S. and {Chen}, X. and {Claybaugh}, T. and {Cole}, S. and {Cuceu}, A. and {Davis}, T.~M. and {Dawson}, K. and {de la Macorra}, A. and {de Mattia}, A. and {Deiosso}, N. and {Dey}, A. and {Dey}, B. and {Ding}, Z. and {Doel}, P. and {Edelstein}, J. and {Eftekharzadeh}, S. and {Eisenstein}, D.~J. and {Elliott}, A. and {Fagrelius}, P. and {Fanning}, K. and {Ferraro}, S. and {Ereza}, J. and {Findlay}, N. and {Flaugher}, B. and {Font-Ribera}, A. and {Forero-S{\'a}nchez}, D. and {Forero-Romero}, J.~E. and {Frenk}, C.~S. and {Garcia-Quintero}, C. and {Gazta{\~n}aga}, E. and {Gil-Mar{\'\i}n}, H. and {Gontcho}, S. Gontcho A and {Gonzalez-Morales}, A.~X. and {Gonzalez-Perez}, V. and {Gordon}, C. and {Green}, D. and {Gruen}, D. and {Gsponer}, R. and {Gutierrez}, G. and {Guy}, J. and {Hadzhiyska}, B. and {Hahn}, C. and {Hanif}, M.~M. S and {Herrera-Alcantar}, H.~K. and {Honscheid}, K. and {Howlett}, C. and {Huterer}, D. and {Ir{\v{s}}i{\v{c}}}, V. and {Ishak}, M. and {Juneau}, S. and {Kara{\c{c}}ayl{\i}}, N.~G. and {Kehoe}, R. and {Kent}, S. and {Kirkby}, D. and {Kremin}, A. and {Krolewski}, A. and {Lai}, Y. and {Lan}, T. -W. and {Landriau}, M. and {Lang}, D. and {Lasker}, J. and {Le Goff}, J.~M. and {Le Guillou}, L. and {Leauthaud}, A. and {Levi}, M.~E. and {Li}, T.~S. and {Linder}, E. and {Lodha}, K. and {Magneville}, C. and {Manera}, M. and {Margala}, D. and {Martini}, P. and {Maus}, M. and {McDonald}, P. and {Medina-Varela}, L. and {Meisner}, A. and {Mena-Fern{\'a}ndez}, J. and {Miquel}, R. and {Moon}, J. and {Moore}, S. and {Moustakas}, J. and {Mudur}, N. and {Mueller}, E. and {Mu{\~n}oz-Guti{\'e}rrez}, A. and {Myers}, A.~D. and {Nadathur}, S. and {Napolitano}, L. and {Neveux}, R. and {Newman}, J.~A. and {Nguyen}, N.~M. and {Nie}, J. and {Niz}, G. and {Noriega}, H.~E. and {Padmanabhan}, N. and {Paillas}, E. and {Palanque-Delabrouille}, N. and {Pan}, J. and {Penmetsa}, S. and {Percival}, W.~J. and {Pieri}, M.~M. and {Pinon}, M. and {Poppett}, C. and {Porredon}, A. and {Prada}, F. and {P{\'e}rez-Fern{\'a}ndez}, A. and {P{\'e}rez-R{\`a}fols}, I. and {Rabinowitz}, D. and {Raichoor}, A. and {Ram{\'\i}rez-P{\'e}rez}, C. and {Ramirez-Solano}, S. and {Ravoux}, C. and {Rashkovetskyi}, M. and {Rezaie}, M. and {Rich}, J. and {Rocher}, A. and {Rockosi}, C. and {Roe}, N.~A. and {Rosado-Marin}, A. and {Ross}, A.~J. and {Rossi}, G. and {Ruggeri}, R. and {Ruhlmann-Kleider}, V. and {Samushia}, L. and {Sanchez}, E. and {Saulder}, C. and {Schlafly}, E.~F. and {Schlegel}, D. and {Schubnell}, M. and {Seo}, H. and {Shafieloo}, A. and {Sharples}, R. and {Silber}, J. and {Slosar}, A. and {Smith}, A. and {Sprayberry}, D. and {Tan}, T. and {Tarl{\'e}}, G. and {Taylor}, P. and {Trusov}, S. and {Ure{\~n}a-L{\'o}pez}, L.~A. and {Vaisakh}, R. and {Valcin}, D. and {Valdes}, F. and {Vargas-Maga{\~n}a}, M. and {Verde}, L. and {Walther}, M. and {Wang}, B. and {Wang}, M.~S. and {Weaver}, B.~A. and {Weaverdyck}, N. and {Wechsler}, R.~H. and {Weinberg}, D.~H. and {White}, M. and {Yu}, J. and {Yu}, Y. and {Yuan}, S. and {Y{\`e}che}, C. and {Zaborowski}, E.~A. and {Zarrouk}, P. and {Zhang}, H. and {Zhao}, C. and {Zhao}, R. and {Zhou}, R. and {Zhuang}, T. and {Zou}, H.},
        title = "{DESI 2024 VI: Cosmological Constraints from the Measurements of Baryon Acoustic Oscillations}",
      journal = {arXiv e-prints},
         year = 2024,
        month = apr,
          eid = {arXiv:2404.03002},
        pages = {arXiv:2404.03002},
          doi = {10.48550/arXiv.2404.03002},
archivePrefix = {arXiv},
       eprint = {2404.03002},
 primaryClass = {astro-ph.CO},
       adsurl = {https://ui.adsabs.harvard.edu/abs/2024arXiv240403002D}
}

@ARTICLE{2024arXiv240715640D,
       author = {{Du}, Guo-Hong and {Wu}, Peng-Ju and {Li}, Tian-Nuo and {Zhang}, Xin},
        title = "{Impacts of dark energy on weighing neutrinos after DESI BAO}",
      journal = {arXiv e-prints},
         year = 2024,
        month = jul,
          eid = {arXiv:2407.15640},
        pages = {arXiv:2407.15640},
          doi = {10.48550/arXiv.2407.15640},
archivePrefix = {arXiv},
       eprint = {2407.15640},
 primaryClass = {astro-ph.CO},
       adsurl = {https://ui.adsabs.harvard.edu/abs/2024arXiv240715640D}
}

@ARTICLE{2017JCAP...06..047Y,
       author = {{Y{\`e}che}, Christophe and {Palanque-Delabrouille}, Nathalie and {Baur}, Julien and {du Mas des Bourboux}, H{\'e}lion},
        title = "{Constraints on neutrino masses from Lyman-alpha forest power spectrum with BOSS and XQ-100}",
      journal = {\jcap},
         year = 2017,
        month = jun,
       volume = {2017},
       number = {6},
          eid = {047},
        pages = {047},
          doi = {10.1088/1475-7516/2017/06/047},
archivePrefix = {arXiv},
       eprint = {1702.03314},
 primaryClass = {astro-ph.CO},
       adsurl = {https://ui.adsabs.harvard.edu/abs/2017JCAP...06..047Y}
}

@ARTICLE{2005PhRvD..71f3534V,
       author = {{Viel}, Matteo and {Lesgourgues}, Julien and {Haehnelt}, Martin G. and {Matarrese}, Sabino and {Riotto}, Antonio},
        title = "{Constraining warm dark matter candidates including sterile neutrinos and light gravitinos with WMAP and the Lyman-{\ensuremath{\alpha}} forest}",
      journal = {\prd},
         year = 2005,
        month = mar,
       volume = {71},
       number = {6},
          eid = {063534},
        pages = {063534},
          doi = {10.1103/PhysRevD.71.063534},
archivePrefix = {arXiv},
       eprint = {astro-ph/0501562},
 primaryClass = {astro-ph},
       adsurl = {https://ui.adsabs.harvard.edu/abs/2005PhRvD..71f3534V}
}

@ARTICLE{2021PhRvD.103h3533A,
       author = {{Alam}, Shadab and {Aubert}, Marie and {Avila}, Santiago and {Balland}, Christophe and {Bautista}, Julian E. and {Bershady}, Matthew A. and {Bizyaev}, Dmitry and {Blanton}, Michael R. and {Bolton}, Adam S. and {Bovy}, Jo and {Brinkmann}, Jonathan and {Brownstein}, Joel R. and {Burtin}, Etienne and {Chabanier}, Sol{\`e}ne and {Chapman}, Michael J. and {Choi}, Peter Doohyun and {Chuang}, Chia-Hsun and {Comparat}, Johan and {Cousinou}, Marie-Claude and {Cuceu}, Andrei and {Dawson}, Kyle S. and {de la Torre}, Sylvain and {de Mattia}, Arnaud and {Agathe}, Victoria de Sainte and {des Bourboux}, H{\'e}lion du Mas and {Escoffier}, Stephanie and {Etourneau}, Thomas and {Farr}, James and {Font-Ribera}, Andreu and {Frinchaboy}, Peter M. and {Fromenteau}, Sebastien and {Gil-Mar{\'\i}n}, H{\'e}ctor and {Le Goff}, Jean-Marc and {Gonzalez-Morales}, Alma X. and {Gonzalez-Perez}, Violeta and {Grabowski}, Kathleen and {Guy}, Julien and {Hawken}, Adam J. and {Hou}, Jiamin and {Kong}, Hui and {Parker}, James and {Klaene}, Mark and {Kneib}, Jean-Paul and {Lin}, Sicheng and {Long}, Daniel and {Lyke}, Brad W. and {de la Macorra}, Axel and {Martini}, Paul and {Masters}, Karen and {Mohammad}, Faizan G. and {Moon}, Jeongin and {Mueller}, Eva-Maria and {Mu{\~n}oz-Guti{\'e}rrez}, Andrea and {Myers}, Adam D. and {Nadathur}, Seshadri and {Neveux}, Richard and {Newman}, Jeffrey A. and {Noterdaeme}, Pasquier and {Oravetz}, Audrey and {Oravetz}, Daniel and {Palanque-Delabrouille}, Nathalie and {Pan}, Kaike and {Paviot}, Romain and {Percival}, Will J. and {P{\'e}rez-R{\`a}fols}, Ignasi and {Petitjean}, Patrick and {Pieri}, Matthew M. and {Prakash}, Abhishek and {Raichoor}, Anand and {Ravoux}, Corentin and {Rezaie}, Mehdi and {Rich}, James and {Ross}, Ashley J. and {Rossi}, Graziano and {Ruggeri}, Rossana and {Ruhlmann-Kleider}, Vanina and {S{\'a}nchez}, Ariel G. and {S{\'a}nchez}, F. Javier and {S{\'a}nchez-Gallego}, Jos{\'e} R. and {Sayres}, Conor and {Schneider}, Donald P. and {Seo}, Hee-Jong and {Shafieloo}, Arman and {Slosar}, An{\v{z}}e and {Smith}, Alex and {Stermer}, Julianna and {Tamone}, Amelie and {Tinker}, Jeremy L. and {Tojeiro}, Rita and {Vargas-Maga{\~n}a}, Mariana and {Variu}, Andrei and {Wang}, Yuting and {Weaver}, Benjamin A. and {Weijmans}, Anne-Marie and {Y{\`e}che}, Christophe and {Zarrouk}, Pauline and {Zhao}, Cheng and {Zhao}, Gong-Bo and {Zheng}, Zheng},
        title = "{Completed SDSS-IV extended Baryon Oscillation Spectroscopic Survey: Cosmological implications from two decades of spectroscopic surveys at the Apache Point Observatory}",
      journal = {\prd},
         year = 2021,
        month = apr,
       volume = {103},
       number = {8},
          eid = {083533},
        pages = {083533},
          doi = {10.1103/PhysRevD.103.083533},
archivePrefix = {arXiv},
       eprint = {2007.08991},
 primaryClass = {astro-ph.CO},
       adsurl = {https://ui.adsabs.harvard.edu/abs/2021PhRvD.103h3533A}
}

@ARTICLE{2024arXiv240218009B,
       author = {{Bault}, Abby and {Kirkby}, David and {Guy}, Julien and {Brodzeller}, Allyson and {Aguilar}, J. and {Ahlen}, S. and {Bailey}, S. and {Brooks}, D. and {Cabayol-Garcia}, L. and {Chaves-Montero}, J. and {Claybaugh}, T. and {Cuceu}, A. and {Dawson}, K. and {de la Cruz}, R. and {de la Macorra}, A. and {Dey}, A. and {Doel}, P. and {Filbert}, S. and {Font-Ribera}, A. and {Forero-Romero}, J.~E. and {Gazta{\~n}aga}, E. and {Gontcho}, S. Gontcho A and {Gordon}, C. and {Herrera-Alcantar}, H.~K. and {Honscheid}, K. and {Ir{\v{s}}i{\v{c}}}, V. and {Kara{\c{c}}ayl{\i}}, N.~G. and {Kehoe}, R. and {Kisner}, T. and {Kremin}, A. and {Lambert}, A. and {Landriau}, M. and {Le Guillou}, L. and {Levi}, M.~E. and {Manera}, M. and {Martini}, P. and {Meisner}, A. and {Miquel}, R. and {Montero-Camacho}, P. and {Moustakas}, J. and {Mu{\~n}oz-Guti{\'e}rrez}, A. and {Nie}, J. and {Niz}, G. and {Palanque-Delabrouille}, N. and {Percival}, W.~J. and {Poppett}, C. and {Prada}, F. and {P{\'e}rez-R{\`a}fols}, I. and {Ram{\'\i}rez-P{\'e}rez}, C. and {Ravoux}, C. and {Rezaie}, M. and {Rossi}, G. and {Sanchez}, E. and {Schlafly}, E.~F. and {Schlegel}, D. and {Schubnell}, M. and {Silber}, J. and {Tan}, T. and {Tarl{\'e}}, G. and {Walther}, M. and {Weaver}, B.~A. and {Zhou}, Z.},
        title = "{Impact of Systematic Redshift Errors on the Cross-correlation of the Lyman-$\alpha$ Forest with Quasars at Small Scales Using DESI Early Data}",
      journal = {arXiv e-prints},
         year = 2024,
        month = feb,
          eid = {arXiv:2402.18009},
        pages = {arXiv:2402.18009},
          doi = {10.48550/arXiv.2402.18009},
archivePrefix = {arXiv},
       eprint = {2402.18009},
 primaryClass = {astro-ph.CO},
       adsurl = {https://ui.adsabs.harvard.edu/abs/2024arXiv240218009B}
}

@ARTICLE{2024arXiv240910617S,
       author = {{Saha}, Akash Kumar and {Singh}, Abhijeet and {Parashari}, Priyank and {Laha}, Ranjan},
        title = "{Hunting Primordial Black Hole Dark Matter in Lyman-$\alpha$ Forest}",
      journal = {arXiv e-prints},
         year = 2024,
        month = sep,
          eid = {arXiv:2409.10617},
        pages = {arXiv:2409.10617},
          doi = {10.48550/arXiv.2409.10617},
archivePrefix = {arXiv},
       eprint = {2409.10617},
 primaryClass = {astro-ph.CO},
       adsurl = {https://ui.adsabs.harvard.edu/abs/2024arXiv240910617S}
}

@ARTICLE{2024JCAP...05..088A,
       author = {{Abdul Karim}, Marie Lynn and {Armengaud}, Eric and {Mention}, Guillaume and {Chabanier}, Sol{\`e}ne and {Ravoux}, Corentin and {Luki{\'c}}, Zarija},
        title = "{Measurement of the small-scale 3D Lyman-{\ensuremath{\alpha}} forest power spectrum}",
      journal = {\jcap},
         year = 2024,
        month = may,
       volume = {2024},
       number = {5},
          eid = {088},
        pages = {088},
          doi = {10.1088/1475-7516/2024/05/088},
archivePrefix = {arXiv},
       eprint = {2310.09116},
 primaryClass = {astro-ph.CO},
       adsurl = {https://ui.adsabs.harvard.edu/abs/2024JCAP...05..088A}
}

@ARTICLE{2024MNRAS.533.3756D,
       author = {{de Belsunce}, Roger and {Philcox}, Oliver H.~E. and {Ir{\v{s}}i{\v{c}}}, Vid and {McDonald}, Patrick and {Guy}, Julien and {Palanque-Delabrouille}, Nathalie},
        title = "{The 3D Lyman-{\ensuremath{\alpha}} forest power spectrum from eBOSS DR16}",
      journal = {\mnras},
         year = 2024,
        month = sep,
       volume = {533},
       number = {3},
        pages = {3756-3770},
          doi = {10.1093/mnras/stae2035},
archivePrefix = {arXiv},
       eprint = {2403.08241},
 primaryClass = {astro-ph.CO},
       adsurl = {https://ui.adsabs.harvard.edu/abs/2024MNRAS.533.3756D}
}

@ARTICLE{2024arXiv240403001D,
       author = {{DESI Collaboration} and {Adame}, A.~G. and {Aguilar}, J. and {Ahlen}, S. and {Alam}, S. and {Alexander}, D.~M. and {Alvarez}, M. and {Alves}, O. and {Anand}, A. and {Andrade}, U. and {Armengaud}, E. and {Avila}, S. and {Aviles}, A. and {Awan}, H. and {Bailey}, S. and {Baltay}, C. and {Bault}, A. and {Bautista}, J. and {Behera}, J. and {BenZvi}, S. and {Beutler}, F. and {Bianchi}, D. and {Blake}, C. and {Blum}, R. and {Brieden}, S. and {Brodzeller}, A. and {Brooks}, D. and {Buckley-Geer}, E. and {Burtin}, E. and {Calderon}, R. and {Canning}, R. and {Carnero Rosell}, A. and {Cereskaite}, R. and {Cervantes-Cota}, J.~L. and {Chabanier}, S. and {Chaussidon}, E. and {Chaves-Montero}, J. and {Chen}, S. and {Chen}, X. and {Claybaugh}, T. and {Cole}, S. and {Cuceu}, A. and {Davis}, T.~M. and {Dawson}, K. and {de la Cruz}, R. and {de la Macorra}, A. and {de Mattia}, A. and {Deiosso}, N. and {Dey}, A. and {Dey}, B. and {Ding}, J. and {Ding}, Z. and {Doel}, P. and {Edelstein}, J. and {Eftekharzadeh}, S. and {Eisenstein}, D.~J. and {Elliott}, A. and {Fagrelius}, P. and {Fanning}, K. and {Ferraro}, S. and {Ereza}, J. and {Findlay}, N. and {Flaugher}, B. and {Font-Ribera}, A. and {Forero-S{\'a}nchez}, D. and {Forero-Romero}, J.~E. and {Garcia-Quintero}, C. and {Gazta{\~n}aga}, E. and {Gil-Mar{\'\i}n}, H. and {Gontcho}, S. Gontcho A and {Gonzalez-Morales}, A.~X. and {Gonzalez-Perez}, V. and {Gordon}, C. and {Green}, D. and {Gruen}, D. and {Gsponer}, R. and {Gutierrez}, G. and {Guy}, J. and {Hadzhiyska}, B. and {Hahn}, C. and {Hanif}, M.~M. S and {Herrera-Alcantar}, H.~K. and {Honscheid}, K. and {Howlett}, C. and {Huterer}, D. and {Ir{\v{s}}i{\v{c}}}, V. and {Ishak}, M. and {Juneau}, S. and {Kara{\c{c}}ayli}, N.~G. and {Kehoe}, R. and {Kent}, S. and {Kirkby}, D. and {Kremin}, A. and {Krolewski}, A. and {Lai}, Y. and {Lan}, T. -W. and {Landriau}, M. and {Lang}, D. and {Lasker}, J. and {Le Goff}, J.~M. and {Le Guillou}, L. and {Leauthaud}, A. and {Levi}, M.~E. and {Li}, T.~S. and {Linder}, E. and {Lodha}, K. and {Magneville}, C. and {Manera}, M. and {Margala}, D. and {Martini}, P. and {Maus}, M. and {McDonald}, P. and {Medina-Varela}, L. and {Meisner}, A. and {Mena-Fern{\'a}ndez}, J. and {Miquel}, R. and {Moon}, J. and {Moore}, S. and {Moustakas}, J. and {Mueller}, E. and {Mu{\~n}oz-Guti{\'e}rrez}, A. and {Myers}, A.~D. and {Nadathur}, S. and {Napolitano}, L. and {Neveux}, R. and {Newman}, J.~A. and {Nguyen}, N.~M. and {Nie}, J. and {Niz}, G. and {Noriega}, H.~E. and {Padmanabhan}, N. and {Paillas}, E. and {Palanque-Delabrouille}, N. and {Pan}, J. and {Penmetsa}, S. and {Percival}, W.~J. and {Pieri}, M.~M. and {Pinon}, M. and {Poppett}, C. and {Porredon}, A. and {Prada}, F. and {P{\'e}rez-Fern{\'a}ndez}, A. and {P{\'e}rez-R{\`a}fols}, I. and {Rabinowitz}, D. and {Raichoor}, A. and {Ram{\'\i}rez-P{\'e}rez}, C. and {Ramirez-Solano}, S. and {Rashkovetskyi}, M. and {Ravoux}, C. and {Rezaie}, M. and {Rich}, J. and {Rocher}, A. and {Rockosi}, C. and {Roe}, N.~A. and {Rosado-Marin}, A. and {Ross}, A.~J. and {Rossi}, G. and {Ruggeri}, R. and {Ruhlmann-Kleider}, V. and {Samushia}, L. and {Sanchez}, E. and {Saulder}, C. and {Schlafly}, E.~F. and {Schlegel}, D. and {Schubnell}, M. and {Seo}, H. and {Sharples}, R. and {Silber}, J. and {Sinigaglia}, F. and {Slosar}, A. and {Smith}, A. and {Sprayberry}, D. and {Tan}, T. and {Tarl{\'e}}, G. and {Trusov}, S. and {Vaisakh}, R. and {Valcin}, D. and {Valdes}, F. and {Vargas-Maga{\~n}a}, M. and {Verde}, L. and {Walther}, M. and {Wang}, B. and {Wang}, M.~S. and {Weaver}, B.~A. and {Weaverdyck}, N. and {Wechsler}, R.~H. and {Weinberg}, D.~H. and {White}, M. and {Yu}, J. and {Yu}, Y. and {Yuan}, S. and {Y{\`e}che}, C. and {Zaborowski}, E.~A. and {Zarrouk}, P. and {Zhang}, H. and {Zhao}, C. and {Zhao}, R. and {Zhou}, R. and {Zou}, H.},
        title = "{DESI 2024 IV: Baryon Acoustic Oscillations from the Lyman Alpha Forest}",
      journal = {arXiv e-prints},
         year = 2024,
        month = apr,
          eid = {arXiv:2404.03001},
        pages = {arXiv:2404.03001},
          doi = {10.48550/arXiv.2404.03001},
archivePrefix = {arXiv},
       eprint = {2404.03001},
 primaryClass = {astro-ph.CO},
       adsurl = {https://ui.adsabs.harvard.edu/abs/2024arXiv240403001D}
}

@ARTICLE{2024arXiv240403003G,
       author = {{Guy}, J. and {Gontcho}, S. Gontcho A and {Armengaud}, E. and {Brodzeller}, A. and {Cuceu}, A. and {Font-Ribera}, A. and {Herrera-Alcantar}, H.~K. and {Kara{\c{c}}ayl{\i}}, N.~G. and {Mu{\~n}oz-Guti{\'e}rrez}, A. and {Pieri}, M. and {P{\'e}rez-R{\`a}fols}, I. and {Ram{\'\i}rez-P{\'e}rez}, C. and {Ravoux}, C. and {Rich}, J. and {Walther}, M. and {Karim}, M. Abdul and {Aguilar}, J. and {Ahlen}, S. and {Bault}, A. and {Brooks}, D. and {Claybaugh}, T. and {de la Cruz}, R. and {de la Macorra}, A. and {Doel}, P. and {Fanning}, K. and {Forero-Romero}, J.~E. and {Gazta{\~n}aga}, E. and {Gonzalez-Morales}, A.~X. and {Gutierrez}, G. and {Hahn}, C. and {Honscheid}, K. and {Juneau}, S. and {Kehoe}, R. and {Kirkby}, D. and {Kisner}, T. and {Kremin}, A. and {Lambert}, A. and {Landriau}, M. and {Le Guillou}, L. and {Manera}, M. and {Martini}, P. and {Meisner}, A. and {Miquel}, R. and {Montero-Camacho}, P. and {Moustakas}, J. and {Mueller}, E. and {Myers}, A.~D. and {Nie}, J. and {Niz}, G. and {Palanque-Delabrouille}, N. and {Percival}, W.~J. and {Poppett}, C. and {Rezaie}, M. and {Rossi}, G. and {Sanchez}, E. and {Schlegel}, D. and {Schubnell}, M. and {Seo}, H. and {Silber}, J. and {Sprayberry}, D. and {Tan}, T. and {Tarl{\'e}}, G. and {Vargas-Maga{\~n}a}, M. and {Zou}, H.},
        title = "{Characterization of contaminants in the Lyman-alpha forest auto-correlation with DESI}",
      journal = {arXiv e-prints},
         year = 2024,
        month = apr,
          eid = {arXiv:2404.03003},
        pages = {arXiv:2404.03003},
          doi = {10.48550/arXiv.2404.03003},
archivePrefix = {arXiv},
       eprint = {2404.03003},
 primaryClass = {astro-ph.CO},
       adsurl = {https://ui.adsabs.harvard.edu/abs/2024arXiv240403003G}
}

@ARTICLE{2024arXiv240403004C,
       author = {{Cuceu}, Andrei and {Herrera-Alcantar}, Hiram K. and {Gordon}, Calum and {Martini}, Paul and {Guy}, Julien and {Font-Ribera}, Andreu and {Gonzalez-Morales}, Alma X. and {Karim}, M. Abdul and {Aguilar}, J. and {Ahlen}, S. and {Armengaud}, E. and {Bault}, A. and {Brooks}, D. and {Claybaugh}, T. and {de la Macorra}, A. and {Doel}, P. and {Fanning}, K. and {Ferraro}, S. and {Forero-Romero}, J.~E. and {Gazta{\~n}aga}, E. and {Gontcho}, S. Gontcho A and {Gutierrez}, G. and {Honscheid}, K. and {Howlett}, C. and {Kara{\c{c}}ayl{\i}}, N.~G. and {Kirkby}, D. and {Kremin}, A. and {Landriau}, M. and {Le Goff}, J.~M. and {Le Guillou}, L. and {Levi}, M.~E. and {Manera}, M. and {Meisner}, A. and {Miquel}, R. and {Moustakas}, J. and {Mu{\~n}oz-Guti{\'e}rrez}, A. and {Myers}, A.~D. and {Niz}, G. and {Palanque-Delabrouille}, N. and {Percival}, W.~J. and {Poppett}, C. and {Prada}, F. and {P{\'e}rez-R{\`a}fols}, I. and {Ram{\'\i}rez-P{\'e}rez}, C. and {Ravoux}, C. and {Rezaie}, M. and {Rossi}, G. and {Sanchez}, E. and {Schlegel}, D. and {Schubnell}, M. and {Seo}, H. and {Sprayberry}, D. and {Tan}, T. and {Tarl{\'e}}, G. and {Vargas-Maga{\~n}a}, M. and {Walther}, M. and {Weaver}, B.~A. and {Zhou}, R. and {Zou}, H.},
        title = "{Validation of the DESI 2024 Ly$\alpha$ forest BAO analysis using synthetic datasets}",
      journal = {arXiv e-prints},
         year = 2024,
        month = apr,
          eid = {arXiv:2404.03004},
        pages = {arXiv:2404.03004},
          doi = {10.48550/arXiv.2404.03004},
archivePrefix = {arXiv},
       eprint = {2404.03004},
 primaryClass = {astro-ph.CO},
       adsurl = {https://ui.adsabs.harvard.edu/abs/2024arXiv240403004C}
}

@ARTICLE{2016A&A...594A..13P,
       author = {{Planck Collaboration} and {Ade}, P.~A.~R. and {Aghanim}, N. and {Arnaud}, M. and {Ashdown}, M. and {Aumont}, J. and {Baccigalupi}, C. and {Banday}, A.~J. and {Barreiro}, R.~B. and {Bartlett}, J.~G. and {Bartolo}, N. and {Battaner}, E. and {Battye}, R. and {Benabed}, K. and {Beno{\^\i}t}, A. and {Benoit-L{\'e}vy}, A. and {Bernard}, J. -P. and {Bersanelli}, M. and {Bielewicz}, P. and {Bock}, J.~J. and {Bonaldi}, A. and {Bonavera}, L. and {Bond}, J.~R. and {Borrill}, J. and {Bouchet}, F.~R. and {Boulanger}, F. and {Bucher}, M. and {Burigana}, C. and {Butler}, R.~C. and {Calabrese}, E. and {Cardoso}, J. -F. and {Catalano}, A. and {Challinor}, A. and {Chamballu}, A. and {Chary}, R. -R. and {Chiang}, H.~C. and {Chluba}, J. and {Christensen}, P.~R. and {Church}, S. and {Clements}, D.~L. and {Colombi}, S. and {Colombo}, L.~P.~L. and {Combet}, C. and {Coulais}, A. and {Crill}, B.~P. and {Curto}, A. and {Cuttaia}, F. and {Danese}, L. and {Davies}, R.~D. and {Davis}, R.~J. and {de Bernardis}, P. and {de Rosa}, A. and {de Zotti}, G. and {Delabrouille}, J. and {D{\'e}sert}, F. -X. and {Di Valentino}, E. and {Dickinson}, C. and {Diego}, J.~M. and {Dolag}, K. and {Dole}, H. and {Donzelli}, S. and {Dor{\'e}}, O. and {Douspis}, M. and {Ducout}, A. and {Dunkley}, J. and {Dupac}, X. and {Efstathiou}, G. and {Elsner}, F. and {En{\ss}lin}, T.~A. and {Eriksen}, H.~K. and {Farhang}, M. and {Fergusson}, J. and {Finelli}, F. and {Forni}, O. and {Frailis}, M. and {Fraisse}, A.~A. and {Franceschi}, E. and {Frejsel}, A. and {Galeotta}, S. and {Galli}, S. and {Ganga}, K. and {Gauthier}, C. and {Gerbino}, M. and {Ghosh}, T. and {Giard}, M. and {Giraud-H{\'e}raud}, Y. and {Giusarma}, E. and {Gjerl{\o}w}, E. and {Gonz{\'a}lez-Nuevo}, J. and {G{\'o}rski}, K.~M. and {Gratton}, S. and {Gregorio}, A. and {Gruppuso}, A. and {Gudmundsson}, J.~E. and {Hamann}, J. and {Hansen}, F.~K. and {Hanson}, D. and {Harrison}, D.~L. and {Helou}, G. and {Henrot-Versill{\'e}}, S. and {Hern{\'a}ndez-Monteagudo}, C. and {Herranz}, D. and {Hildebrandt}, S.~R. and {Hivon}, E. and {Hobson}, M. and {Holmes}, W.~A. and {Hornstrup}, A. and {Hovest}, W. and {Huang}, Z. and {Huffenberger}, K.~M. and {Hurier}, G. and {Jaffe}, A.~H. and {Jaffe}, T.~R. and {Jones}, W.~C. and {Juvela}, M. and {Keih{\"a}nen}, E. and {Keskitalo}, R. and {Kisner}, T.~S. and {Kneissl}, R. and {Knoche}, J. and {Knox}, L. and {Kunz}, M. and {Kurki-Suonio}, H. and {Lagache}, G. and {L{\"a}hteenm{\"a}ki}, A. and {Lamarre}, J. -M. and {Lasenby}, A. and {Lattanzi}, M. and {Lawrence}, C.~R. and {Leahy}, J.~P. and {Leonardi}, R. and {Lesgourgues}, J. and {Levrier}, F. and {Lewis}, A. and {Liguori}, M. and {Lilje}, P.~B. and {Linden-V{\o}rnle}, M. and {L{\'o}pez-Caniego}, M. and {Lubin}, P.~M. and {Mac{\'\i}as-P{\'e}rez}, J.~F. and {Maggio}, G. and {Maino}, D. and {Mandolesi}, N. and {Mangilli}, A. and {Marchini}, A. and {Maris}, M. and {Martin}, P.~G. and {Martinelli}, M. and {Mart{\'\i}nez-Gonz{\'a}lez}, E. and {Masi}, S. and {Matarrese}, S. and {McGehee}, P. and {Meinhold}, P.~R. and {Melchiorri}, A. and {Melin}, J. -B. and {Mendes}, L. and {Mennella}, A. and {Migliaccio}, M. and {Millea}, M. and {Mitra}, S. and {Miville-Desch{\^e}nes}, M. -A. and {Moneti}, A. and {Montier}, L. and {Morgante}, G. and {Mortlock}, D. and {Moss}, A. and {Munshi}, D. and {Murphy}, J.~A. and {Naselsky}, P. and {Nati}, F. and {Natoli}, P. and {Netterfield}, C.~B. and {N{\o}rgaard-Nielsen}, H.~U. and {Noviello}, F. and {Novikov}, D. and {Novikov}, I. and {Oxborrow}, C.~A. and {Paci}, F. and {Pagano}, L. and {Pajot}, F. and {Paladini}, R. and {Paoletti}, D. and {Partridge}, B. and {Pasian}, F. and {Patanchon}, G. and {Pearson}, T.~J. and {Perdereau}, O. and {Perotto}, L. and {Perrotta}, F. and {Pettorino}, V. and {Piacentini}, F. and {Piat}, M. and {Pierpaoli}, E. and {Pietrobon}, D. and {Plaszczynski}, S. and {Pointecouteau}, E. and {Polenta}, G. and {Popa}, L. and {Pratt}, G.~W. and {Pr{\'e}zeau}, G. and {Prunet}, S. and {Puget}, J. -L. and {Rachen}, J.~P. and {Reach}, W.~T. and {Rebolo}, R. and {Reinecke}, M. and {Remazeilles}, M. and {Renault}, C. and {Renzi}, A. and {Ristorcelli}, I. and {Rocha}, G. and {Rosset}, C. and {Rossetti}, M. and {Roudier}, G. and {Rouill{\'e} d'Orfeuil}, B. and {Rowan-Robinson}, M. and {Rubi{\~n}o-Mart{\'\i}n}, J.~A. and {Rusholme}, B. and {Said}, N. and {Salvatelli}, V. and {Salvati}, L. and {Sandri}, M. and {Santos}, D. and {Savelainen}, M. and {Savini}, G. and {Scott}, D. and {Seiffert}, M.~D. and {Serra}, P. and {Shellard}, E.~P.~S. and {Spencer}, L.~D. and {Spinelli}, M. and {Stolyarov}, V. and {Stompor}, R. and {Sudiwala}, R. and {Sunyaev}, R. and {Sutton}, D. and {Suur-Uski}, A. -S. and {Sygnet}, J. -F. and {Tauber}, J.~A. and {Terenzi}, L. and {Toffolatti}, L. and {Tomasi}, M. and {Tristram}, M. and {Trombetti}, T. and {Tucci}, M. and {Tuovinen}, J. and {T{\"u}rler}, M. and {Umana}, G. and {Valenziano}, L. and {Valiviita}, J. and {Van Tent}, F. and {Vielva}, P. and {Villa}, F. and {Wade}, L.~A. and {Wandelt}, B.~D. and {Wehus}, I.~K. and {White}, M. and {White}, S.~D.~M. and {Wilkinson}, A. and {Yvon}, D. and {Zacchei}, A. and {Zonca}, A.},
        title = "{Planck 2015 results. XIII. Cosmological parameters}",
      journal = {\aap},
         year = 2016,
        month = sep,
       volume = {594},
          eid = {A13},
        pages = {A13},
          doi = {10.1051/0004-6361/201525830},
archivePrefix = {arXiv},
       eprint = {1502.01589},
 primaryClass = {astro-ph.CO},
       adsurl = {https://ui.adsabs.harvard.edu/abs/2016A&A...594A..13P}
}

@ARTICLE{2024arXiv240513680M,
       author = {{Montero-Camacho}, Paulo and {Li}, Yin and {Cranmer}, Miles},
        title = "{Five parameters are all you need (in $\Lambda$CDM)}",
      journal = {arXiv e-prints},
         year = 2024,
        month = may,
          eid = {arXiv:2405.13680},
        pages = {arXiv:2405.13680},
          doi = {10.48550/arXiv.2405.13680},
archivePrefix = {arXiv},
       eprint = {2405.13680},
 primaryClass = {astro-ph.CO},
       adsurl = {https://ui.adsabs.harvard.edu/abs/2024arXiv240513680M}
}

@ARTICLE{2024arXiv240506743T,
       author = {{Turner}, Wynne and {Martini}, Paul and {G{\"o}ksel Kara{\c{c}}ayl{\i}}, Naim and {Aguilar}, J. and {Ahlen}, S. and {Brooks}, D. and {Claybaugh}, T. and {de la Macorra}, A. and {Dey}, A. and {Doel}, P. and {Fanning}, K. and {Forero-Romero}, J.~E. and {Gontcho}, S. Gontcho A and {Gonzalez-Morales}, A.~X. and {Gutierrez}, G. and {Guy}, J. and {Herrera-Alcantar}, H.~K. and {Honscheid}, K. and {Juneau}, S. and {Kisner}, T. and {Kremin}, A. and {Lambert}, A. and {Landriau}, M. and {Le Guillou}, L. and {Meisner}, A. and {Miquel}, R. and {Moustakas}, J. and {Mueller}, E. and {Mu{\~n}oz-Guti{\'e}rrez}, A. and {Myers}, A.~D. and {Nie}, J. and {Niz}, G. and {Poppett}, C. and {Prada}, F. and {Rezaie}, M. and {Rossi}, G. and {Sanchez}, E. and {Schlafly}, E.~F. and {Schlegel}, D. and {Schubnell}, M. and {Seo}, H. and {Sprayberry}, D. and {Tarl{\'e}}, G. and {Weaver}, B.~A. and {Zou}, H.},
        title = "{New measurements of the Lyman-$\alpha$ forest continuum and effective optical depth with LyCAN and DESI Y1 data}",
      journal = {arXiv e-prints},
         year = 2024,
        month = may,
          eid = {arXiv:2405.06743},
        pages = {arXiv:2405.06743},
          doi = {10.48550/arXiv.2405.06743},
archivePrefix = {arXiv},
       eprint = {2405.06743},
 primaryClass = {astro-ph.CO},
       adsurl = {https://ui.adsabs.harvard.edu/abs/2024arXiv240506743T}
}

@ARTICLE{2024MNRAS.533.3312G,
       author = {{Greig}, Bradley and {Bosman}, S.~E.~I. and {Davies}, F.~B. and {{\v{D}}urov{\v{c}}{\'\i}kov{\'a}}, D. and {Fathivavsari}, H. and {Liu}, B. and {Meyer}, R.~A. and {Sun}, Z. and {D'Odorico}, V. and {Gallerani}, S. and {Mesinger}, A. and {Ting}, Y. -S.},
        title = "{Blind QSO reconstruction challenge: exploring methods to reconstruct the Ly {\ensuremath{\alpha}} emission line of QSOs}",
      journal = {\mnras},
         year = 2024,
        month = sep,
       volume = {533},
       number = {3},
        pages = {3312-3343},
          doi = {10.1093/mnras/stae1985},
archivePrefix = {arXiv},
       eprint = {2404.01556},
 primaryClass = {astro-ph.CO},
       adsurl = {https://ui.adsabs.harvard.edu/abs/2024MNRAS.533.3312G}
}

@ARTICLE{2022JCAP...09..070G,
       author = {{Givans}, Jahmour J. and {Font-Ribera}, Andreu and {Slosar}, An{\v{z}}e and {Seeyave}, Louise and {Pedersen}, Christian and {Rogers}, Keir K. and {Garny}, Mathias and {Blas}, Diego and {Ir{\v{s}}i{\v{c}}}, Vid},
        title = "{Non-linearities in the Lyman-{\ensuremath{\alpha}} forest and in its cross-correlation with dark matter halos}",
      journal = {\jcap},
         year = 2022,
        month = sep,
       volume = {2022},
       number = {9},
          eid = {070},
        pages = {070},
          doi = {10.1088/1475-7516/2022/09/070},
archivePrefix = {arXiv},
       eprint = {2205.00962},
 primaryClass = {astro-ph.CO},
       adsurl = {https://ui.adsabs.harvard.edu/abs/2022JCAP...09..070G}
}

@ARTICLE{2024arXiv241107970Z,
       author = {{Zhao}, Cheng and {Huang}, Song and {He}, Mengfan and {Montero-Camacho}, Paulo and {Liu}, Yu and {Renard}, Pablo and {Tang}, Yunyi and {Verdier}, Aurelien and {Xu}, Wenshuo and {Yang}, Xiaorui and {Yu}, Jiaxi and {Zhang}, Yao and {Zhao}, Siyi and {Zhou}, Xingchen and {He}, Shengyu and {Kneib}, Jean-Paul and {Li}, Jiayi and {Li}, Zhuoyang and {Wang}, Wen-Ting and {Xianyu}, Zhong-Zhi and {Zhang}, Yidian and {Gsponer}, Rafaela and {Li}, Xiao-Dong and {Rocher}, Antoine and {Zou}, Siwei and {Tan}, Ting and {Huang}, Zhiqi and {Wang}, Zhuoxiao and {Li}, Pei and {Rombach}, Maxime and {Dong}, Chenxing and {Forero-Sanchez}, Daniel and {Shan}, Huanyuan and {Wang}, Tao and {Li}, Yin and {Zhai}, Zhongxu and {Wang}, Yuting and {Zhao}, Gong-Bo and {Shi}, Yong and {Mao}, Shude and {Huang}, Lei and {Guo}, Liquan and {Cai}, Zheng},
        title = "{MUltiplexed Survey Telescope: Perspectives for Large-Scale Structure Cosmology in the Era of Stage-V Spectroscopic Survey}",
      journal = {arXiv e-prints},
         year = 2024,
        month = nov,
          eid = {arXiv:2411.07970},
        pages = {arXiv:2411.07970},
          doi = {10.48550/arXiv.2411.07970},
archivePrefix = {arXiv},
       eprint = {2411.07970},
 primaryClass = {astro-ph.CO},
       adsurl = {https://ui.adsabs.harvard.edu/abs/2024arXiv241107970Z}
}

@ARTICLE{2018JCAP...01..003F,
       author = {{Font-Ribera}, Andreu and {McDonald}, Patrick and {Slosar}, An{\v{z}}e},
        title = "{How to estimate the 3D power spectrum of the Lyman-{\ensuremath{\alpha}} forest}",
      journal = {\jcap},
         year = 2018,
        month = jan,
       volume = {2018},
       number = {1},
          eid = {003},
        pages = {003},
          doi = {10.1088/1475-7516/2018/01/003},
archivePrefix = {arXiv},
       eprint = {1710.11036},
 primaryClass = {astro-ph.CO},
       adsurl = {https://ui.adsabs.harvard.edu/abs/2018JCAP...01..003F}
}

@ARTICLE{2023PhRvL.130s1003C,
       author = {{Cuceu}, Andrei and {Font-Ribera}, Andreu and {Nadathur}, Seshadri and {Joachimi}, Benjamin and {Martini}, Paul},
        title = "{Constraints on the Cosmic Expansion Rate at Redshift 2.3 from the Lyman-{\ensuremath{\alpha}} Forest}",
      journal = {\prl},
         year = 2023,
        month = may,
       volume = {130},
       number = {19},
          eid = {191003},
        pages = {191003},
          doi = {10.1103/PhysRevLett.130.191003},
archivePrefix = {arXiv},
       eprint = {2209.13942},
 primaryClass = {astro-ph.CO},
       adsurl = {https://ui.adsabs.harvard.edu/abs/2023PhRvL.130s1003C}
}

@ARTICLE{2006SSRv..123..485G,
       author = {{Gardner}, Jonathan P. and {Mather}, John C. and {Clampin}, Mark and {Doyon}, Rene and {Greenhouse}, Matthew A. and {Hammel}, Heidi B. and {Hutchings}, John B. and {Jakobsen}, Peter and {Lilly}, Simon J. and {Long}, Knox S. and {Lunine}, Jonathan I. and {McCaughrean}, Mark J. and {Mountain}, Matt and {Nella}, John and {Rieke}, George H. and {Rieke}, Marcia J. and {Rix}, Hans-Walter and {Smith}, Eric P. and {Sonneborn}, George and {Stiavelli}, Massimo and {Stockman}, H.~S. and {Windhorst}, Rogier A. and {Wright}, Gillian S.},
        title = "{The James Webb Space Telescope}",
      journal = {\ssr},
         year = 2006,
        month = apr,
       volume = {123},
       number = {4},
        pages = {485-606},
          doi = {10.1007/s11214-006-8315-7},
archivePrefix = {arXiv},
       eprint = {astro-ph/0606175},
 primaryClass = {astro-ph},
       adsurl = {https://ui.adsabs.harvard.edu/abs/2006SSRv..123..485G}
}

@ARTICLE{2024MNRAS.535L..37M,
       author = {{Mu{\~n}oz}, Julian B. and {Mirocha}, Jordan and {Chisholm}, John and {Furlanetto}, Steven R. and {Mason}, Charlotte},
        title = "{Reionization after JWST: a photon budget crisis?}",
      journal = {\mnras},
         year = 2024,
        month = nov,
       volume = {535},
       number = {1},
        pages = {L37-L43},
          doi = {10.1093/mnrasl/slae086},
archivePrefix = {arXiv},
       eprint = {2404.07250},
 primaryClass = {astro-ph.CO},
       adsurl = {https://ui.adsabs.harvard.edu/abs/2024MNRAS.535L..37M}
}

@ARTICLE{2023ApJ...950...68E,
       author = {{Eilers}, Anna-Christina and {Simcoe}, Robert A. and {Yue}, Minghao and {Mackenzie}, Ruari and {Matthee}, Jorryt and {{\v{D}}urov{\v{c}}{\'\i}kov{\'a}}, Dominika and {Kashino}, Daichi and {Bordoloi}, Rongmon and {Lilly}, Simon J.},
        title = "{EIGER. III. JWST/NIRCam Observations of the Ultraluminous High-redshift Quasar J0100+2802}",
      journal = {\apj},
         year = 2023,
        month = jun,
       volume = {950},
       number = {1},
          eid = {68},
        pages = {68},
          doi = {10.3847/1538-4357/acd776},
archivePrefix = {arXiv},
       eprint = {2211.16261},
 primaryClass = {astro-ph.GA},
       adsurl = {https://ui.adsabs.harvard.edu/abs/2023ApJ...950...68E}
}

@ARTICLE{2023MNRAS.518.4755A,
       author = {{Adams}, N.~J. and {Conselice}, C.~J. and {Ferreira}, L. and {Austin}, D. and {Trussler}, J.~A.~A. and {Juod{\v{z}}balis}, I. and {Wilkins}, S.~M. and {Caruana}, J. and {Dayal}, P. and {Verma}, A. and {Vijayan}, A.~P.},
        title = "{Discovery and properties of ultra-high redshift galaxies (9 < z < 12) in the JWST ERO SMACS 0723 Field}",
      journal = {\mnras},
         year = 2023,
        month = jan,
       volume = {518},
       number = {3},
        pages = {4755-4766},
          doi = {10.1093/mnras/stac3347},
archivePrefix = {arXiv},
       eprint = {2207.11217},
 primaryClass = {astro-ph.GA},
       adsurl = {https://ui.adsabs.harvard.edu/abs/2023MNRAS.518.4755A}
}

@ARTICLE{2022MNRAS.514...55B,
       author = {{Bosman}, Sarah E.~I. and {Davies}, Frederick B. and {Becker}, George D. and {Keating}, Laura C. and {Davies}, Rebecca L. and {Zhu}, Yongda and {Eilers}, Anna-Christina and {D'Odorico}, Valentina and {Bian}, Fuyan and {Bischetti}, Manuela and {Cristiani}, Stefano V. and {Fan}, Xiaohui and {Farina}, Emanuele P. and {Haehnelt}, Martin G. and {Hennawi}, Joseph F. and {Kulkarni}, Girish and {Mesinger}, Andrei and {Meyer}, Romain A. and {Onoue}, Masafusa and {Pallottini}, Andrea and {Qin}, Yuxiang and {Ryan-Weber}, Emma and {Schindler}, Jan-Torge and {Walter}, Fabian and {Wang}, Feige and {Yang}, Jinyi},
        title = "{Hydrogen reionization ends by z = 5.3: Lyman-{\ensuremath{\alpha}} optical depth measured by the XQR-30 sample}",
      journal = {\mnras},
         year = 2022,
        month = jul,
       volume = {514},
       number = {1},
        pages = {55-76},
          doi = {10.1093/mnras/stac1046},
archivePrefix = {arXiv},
       eprint = {2108.03699},
 primaryClass = {astro-ph.CO},
       adsurl = {https://ui.adsabs.harvard.edu/abs/2022MNRAS.514...55B}
}

@ARTICLE{2023MNRAS.525.6036L,
       author = {{Long}, Heyang and {Morales-Guti{\'e}rrez}, Catalina and {Montero-Camacho}, Paulo and {Hirata}, Christopher M.},
        title = "{Impact of inhomogeneous reionization on post-reionization 21-cm intensity mapping measurement of cosmological parameters}",
      journal = {\mnras},
         year = 2023,
        month = nov,
       volume = {525},
       number = {4},
        pages = {6036-6049},
          doi = {10.1093/mnras/stad2639},
archivePrefix = {arXiv},
       eprint = {2210.02385},
 primaryClass = {astro-ph.CO},
       adsurl = {https://ui.adsabs.harvard.edu/abs/2023MNRAS.525.6036L}
}

@ARTICLE{2024MNRAS.528.5845W,
       author = {{Wells}, Alexandra and {Robinson}, David and {Avestruz}, Camille and {Gnedin}, Nickolay Y.},
        title = "{Emergence of the temperature-density relation in the low-density intergalactic medium}",
      journal = {\mnras},
         year = 2024,
        month = mar,
       volume = {528},
       number = {4},
        pages = {5845-5851},
          doi = {10.1093/mnras/stae401},
archivePrefix = {arXiv},
       eprint = {2310.15226},
 primaryClass = {astro-ph.CO},
       adsurl = {https://ui.adsabs.harvard.edu/abs/2024MNRAS.528.5845W}
}

@ARTICLE{2025arXiv250104770C,
       author = {{Chudaykin}, Anton and {Ivanov}, Mikhail M.},
        title = "{Lyman Alpha Forest - Halo Cross-Correlations in Effective Field Theory}",
      journal = {arXiv e-prints},
         year = 2025,
        month = jan,
          eid = {arXiv:2501.04770},
        pages = {arXiv:2501.04770},
          doi = {10.48550/arXiv.2501.04770},
archivePrefix = {arXiv},
       eprint = {2501.04770},
 primaryClass = {astro-ph.CO},
       adsurl = {https://ui.adsabs.harvard.edu/abs/2025arXiv250104770C}
}

@article{kass1995bayes,
  title={Bayes factors},
  author={Kass, Robert E and Raftery, Adrian E},
  journal={Journal of the american statistical association},
  volume={90},
  number={430},
  pages={773--795},
  year={1995},
  publisher={Taylor \& Francis}
}

@ARTICLE{1100705,
  author={Akaike, H.},
  journal={IEEE Transactions on Automatic Control}, 
  title={A new look at the statistical model identification}, 
  year={1974},
  volume={19},
  number={6},
  pages={716-723},
  doi={10.1109/TAC.1974.1100705}}

@ARTICLE{2025arXiv250315595K,
       author = {{Khan}, Nabendu Kumar and {Ray}, Anupam and {Kulkarni}, Girish and {Dasgupta}, Basudeb},
        title = "{Stronger Constraints on Primordial Black Holes as Dark Matter Derived from the Thermal Evolution of the Intergalactic Medium over the Last Twelve Billion Years}",
      journal = {arXiv e-prints},
         year = 2025,
        month = mar,
          eid = {arXiv:2503.15595},
        pages = {arXiv:2503.15595},
          doi = {10.48550/arXiv.2503.15595},
archivePrefix = {arXiv},
       eprint = {2503.15595},
 primaryClass = {astro-ph.CO},
       adsurl = {https://ui.adsabs.harvard.edu/abs/2025arXiv250315595K}
}

@ARTICLE{2009ApJ...694..842M,
       author = {{McQuinn}, Matthew and {Lidz}, Adam and {Zaldarriaga}, Matias and {Hernquist}, Lars and {Hopkins}, Philip F. and {Dutta}, Suvendra and {Faucher-Gigu{\`e}re}, Claude-Andr{\'e}},
        title = "{He II Reionization and its Effect on the Intergalactic Medium}",
      journal = {\apj},
         year = 2009,
        month = apr,
       volume = {694},
       number = {2},
        pages = {842-866},
          doi = {10.1088/0004-637X/694/2/842},
archivePrefix = {arXiv},
       eprint = {0807.2799},
 primaryClass = {astro-ph},
       adsurl = {https://ui.adsabs.harvard.edu/abs/2009ApJ...694..842M}
}

@ARTICLE{2023PhRvD.108b3502V,
       author = {{Villasenor}, Bruno and {Robertson}, Brant and {Madau}, Piero and {Schneider}, Evan},
        title = "{New constraints on warm dark matter from the Lyman-{\ensuremath{\alpha}} forest power spectrum}",
      journal = {\prd},
         year = 2023,
        month = jul,
       volume = {108},
       number = {2},
          eid = {023502},
        pages = {023502},
          doi = {10.1103/PhysRevD.108.023502},
archivePrefix = {arXiv},
       eprint = {2209.14220},
 primaryClass = {astro-ph.CO},
       adsurl = {https://ui.adsabs.harvard.edu/abs/2023PhRvD.108b3502V}
}

@ARTICLE{2024PhRvD.109d3511I,
       author = {{Ir{\v{s}}i{\v{c}}}, Vid and {Viel}, Matteo and {Haehnelt}, Martin G. and {Bolton}, James S. and {Molaro}, Margherita and {Puchwein}, Ewald and {Boera}, Elisa and {Becker}, George D. and {Gaikwad}, Prakash and {Keating}, Laura C. and {Kulkarni}, Girish},
        title = "{Unveiling dark matter free streaming at the smallest scales with the high redshift Lyman-alpha forest}",
      journal = {\prd},
         year = 2024,
        month = feb,
       volume = {109},
       number = {4},
          eid = {043511},
        pages = {043511},
          doi = {10.1103/PhysRevD.109.043511},
archivePrefix = {arXiv},
       eprint = {2309.04533},
 primaryClass = {astro-ph.CO},
       adsurl = {https://ui.adsabs.harvard.edu/abs/2024PhRvD.109d3511I}
}

@ARTICLE{2024MNRAS.535.1035M,
       author = {{Meiksin}, Avery and {Puchwein}, Ewald},
        title = "{The effect of helium reionization on the Ly {\ensuremath{\alpha}} forest hydrogen flux statistics}",
      journal = {\mnras},
         year = 2024,
        month = nov,
       volume = {535},
       number = {1},
        pages = {1035-1051},
          doi = {10.1093/mnras/stae2438},
archivePrefix = {arXiv},
       eprint = {2410.21023},
 primaryClass = {astro-ph.CO},
       adsurl = {https://ui.adsabs.harvard.edu/abs/2024MNRAS.535.1035M}
}

@ARTICLE{2019MNRAS.486.4075O,
       author = {{O{\~n}orbe}, Jose and {Davies}, F.~B. and {Luki{\'c}} and {}, Z. and {Hennawi}, J.~F. and {Sorini}, D.},
        title = "{Inhomogeneous reionization models in cosmological hydrodynamical simulations}",
      journal = {\mnras},
         year = 2019,
        month = jul,
       volume = {486},
       number = {3},
        pages = {4075-4097},
          doi = {10.1093/mnras/stz984},
archivePrefix = {arXiv},
       eprint = {1810.11683},
 primaryClass = {astro-ph.CO},
       adsurl = {https://ui.adsabs.harvard.edu/abs/2019MNRAS.486.4075O}
}

@ARTICLE{2017ApJ...847...63O,
       author = {{O{\~n}orbe}, J. and {Hennawi}, J.~F. and {Luki{\'c}}, Z. and {Walther}, M.},
        title = "{Constraining Reionization with the z {\ensuremath{\sim}} 5-6 Ly{\ensuremath{\alpha}} Forest Power Spectrum: The Outlook after Planck}",
      journal = {\apj},
         year = 2017,
        month = sep,
       volume = {847},
       number = {1},
          eid = {63},
        pages = {63},
          doi = {10.3847/1538-4357/aa898d},
archivePrefix = {arXiv},
       eprint = {1703.08633},
 primaryClass = {astro-ph.CO},
       adsurl = {https://ui.adsabs.harvard.edu/abs/2017ApJ...847...63O}
}

@ARTICLE{2018MNRAS.477.5501K,
       author = {{Keating}, Laura C. and {Puchwein}, Ewald and {Haehnelt}, Martin G.},
        title = "{Spatial fluctuations of the intergalactic temperature-density relation after hydrogen reionization}",
      journal = {\mnras},
         year = 2018,
        month = jul,
       volume = {477},
       number = {4},
        pages = {5501-5516},
          doi = {10.1093/mnras/sty968},
archivePrefix = {arXiv},
       eprint = {1709.05351},
 primaryClass = {astro-ph.CO},
       adsurl = {https://ui.adsabs.harvard.edu/abs/2018MNRAS.477.5501K}
}

@ARTICLE{2025ApJ...981L..27G,
       author = {{Gao}, Anning and {Prochaska}, J. Xavier and {Cai}, Zheng and {Zou}, Siwei and {Zhao}, Cheng and {Sun}, Zechang and {Ahlen}, S. and {Bianchi}, D. and {Brooks}, D. and {Claybaugh}, T. and {de la Macorra}, A. and {Dey}, Arjun and {Doel}, P. and {Forero-Romero}, J.~E. and {Gazta{\~n}aga}, E. and {Gontcho A Gontcho}, S. and {Gutierrez}, G. and {Honscheid}, K. and {Juneau}, S. and {Kremin}, A. and {Martini}, P. and {Meisner}, A. and {Miquel}, R. and {Moustakas}, J. and {Mu{\~n}oz-Guti{\'e}rrez}, A. and {Newman}, J.~A. and {P{\'e}rez-R{\`a}fols}, I. and {Rossi}, G. and {Sanchez}, E. and {Schubnell}, M. and {Sprayberry}, D. and {Tarl{\'e}}, G. and {Weaver}, B.~A. and {Zou}, H.},
        title = "{Measuring the Mean Free Path of H I Ionizing Photons at 3.2 {\ensuremath{\leq}} z {\ensuremath{\leq}} 4.6 with DESI Y1 Quasars}",
      journal = {\apjl},
         year = 2025,
        month = mar,
       volume = {981},
       number = {2},
          eid = {L27},
        pages = {L27},
          doi = {10.3847/2041-8213/adb48f},
archivePrefix = {arXiv},
       eprint = {2411.15838},
 primaryClass = {astro-ph.CO},
       adsurl = {https://ui.adsabs.harvard.edu/abs/2025ApJ...981L..27G}
}

@ARTICLE{2021ApJ...917L..37C,
       author = {{Cain}, Christopher and {D'Aloisio}, Anson and {Gangolli}, Nakul and {Becker}, George D.},
        title = "{A Short Mean Free Path at z = 6 Favors Late and Rapid Reionization by Faint Galaxies}",
      journal = {\apjl},
         year = 2021,
        month = aug,
       volume = {917},
       number = {2},
          eid = {L37},
        pages = {L37},
          doi = {10.3847/2041-8213/ac1ace},
archivePrefix = {arXiv},
       eprint = {2105.10511},
 primaryClass = {astro-ph.CO},
       adsurl = {https://ui.adsabs.harvard.edu/abs/2021ApJ...917L..37C}
}

@ARTICLE{2003ApJ...596..768S,
       author = {{Schaye}, Joop and {Aguirre}, Anthony and {Kim}, Tae-Sun and {Theuns}, Tom and {Rauch}, Michael and {Sargent}, Wallace L.~W.},
        title = "{Metallicity of the Intergalactic Medium Using Pixel Statistics. II. The Distribution of Metals as Traced by C IV}",
      journal = {\apj},
         year = 2003,
        month = oct,
       volume = {596},
       number = {2},
        pages = {768-796},
          doi = {10.1086/378044},
archivePrefix = {arXiv},
       eprint = {astro-ph/0306469},
 primaryClass = {astro-ph},
       adsurl = {https://ui.adsabs.harvard.edu/abs/2003ApJ...596..768S}
}

@ARTICLE{2022MNRAS.513..117L,
       author = {{Long}, Heyang and {Givans}, Jahmour J. and {Hirata}, Christopher M.},
        title = "{Streaming velocity effects on the post-reionization 21-cm baryon acoustic oscillation signal}",
      journal = {\mnras},
         year = 2022,
        month = jun,
       volume = {513},
       number = {1},
        pages = {117-128},
          doi = {10.1093/mnras/stac658},
archivePrefix = {arXiv},
       eprint = {2107.07615},
 primaryClass = {astro-ph.CO},
       adsurl = {https://ui.adsabs.harvard.edu/abs/2022MNRAS.513..117L}
}

@ARTICLE{2023ApJ...942...59J,
       author = {{Jin}, Xiangyu and {Yang}, Jinyi and {Fan}, Xiaohui and {Wang}, Feige and {Ba{\~n}ados}, Eduardo and {Bian}, Fuyan and {Davies}, Frederick B. and {Eilers}, Anna-Christina and {Farina}, Emanuele Paolo and {Hennawi}, Joseph F. and {Pacucci}, Fabio and {Venemans}, Bram and {Walter}, Fabian},
        title = "{(Nearly) Model-independent Constraints on the Neutral Hydrogen Fraction in the Intergalactic Medium at z   5-7 Using Dark Pixel Fractions in Ly{\ensuremath{\alpha}} and Ly{\ensuremath{\beta}} Forests}",
      journal = {\apj},
         year = 2023,
        month = jan,
       volume = {942},
       number = {2},
          eid = {59},
        pages = {59},
          doi = {10.3847/1538-4357/aca678},
archivePrefix = {arXiv},
       eprint = {2211.12613},
 primaryClass = {astro-ph.CO},
       adsurl = {https://ui.adsabs.harvard.edu/abs/2023ApJ...942...59J}
}

@ARTICLE{2015MNRAS.446..566M,
       author = {{Mesinger}, Andrei and {Aykutalp}, Aycin and {Vanzella}, Eros and {Pentericci}, Laura and {Ferrara}, Andrea and {Dijkstra}, Mark},
        title = "{Can the intergalactic medium cause a rapid drop in Ly{\ensuremath{\alpha}} emission at z > 6?}",
      journal = {\mnras},
         year = 2015,
        month = jan,
       volume = {446},
       number = {1},
        pages = {566-577},
          doi = {10.1093/mnras/stu2089},
archivePrefix = {arXiv},
       eprint = {1406.6373},
 primaryClass = {astro-ph.CO},
       adsurl = {https://ui.adsabs.harvard.edu/abs/2015MNRAS.446..566M}
}

@ARTICLE{2015MNRAS.453.1843S,
       author = {{Sobacchi}, Emanuele and {Mesinger}, Andrei},
        title = "{The clustering of Lyman {\ensuremath{\alpha}} emitters at z {\ensuremath{\approx}} 7: implications for reionization and host halo masses}",
      journal = {\mnras},
         year = 2015,
        month = oct,
       volume = {453},
       number = {2},
        pages = {1843-1854},
          doi = {10.1093/mnras/stv1751},
archivePrefix = {arXiv},
       eprint = {1505.02787},
 primaryClass = {astro-ph.CO},
       adsurl = {https://ui.adsabs.harvard.edu/abs/2015MNRAS.453.1843S}
}

@ARTICLE{2018ApJ...856....2M,
       author = {{Mason}, Charlotte A. and {Treu}, Tommaso and {Dijkstra}, Mark and {Mesinger}, Andrei and {Trenti}, Michele and {Pentericci}, Laura and {de Barros}, Stephane and {Vanzella}, Eros},
        title = "{The Universe Is Reionizing at z {\ensuremath{\sim}} 7: Bayesian Inference of the IGM Neutral Fraction Using Ly{\ensuremath{\alpha}} Emission from Galaxies}",
      journal = {\apj},
         year = 2018,
        month = mar,
       volume = {856},
       number = {1},
          eid = {2},
        pages = {2},
          doi = {10.3847/1538-4357/aab0a7},
archivePrefix = {arXiv},
       eprint = {1709.05356},
 primaryClass = {astro-ph.CO},
       adsurl = {https://ui.adsabs.harvard.edu/abs/2018ApJ...856....2M}
}

%%%%%%%%%%%%%%%%%%%%%%%%%%%%%%%%%%%%%%%%%%%%%%%%%%

%%%%%%%%%%%%%%%%% APPENDICES %%%%%%%%%%%%%%%%%%%%%

\appendix
\section{Yukawa template parameter values}
\label{app:yuk-par}
Table \ref{tab:yuk-par} presents the Yukawa template parameters from Eq.~(\ref{eq:yukawa}) at the effective redshift of the correlation functions. The parameter $A_{\rm re}$ is the amplitude of the reionization relics, while $\alpha^{-1}_{\rm re}$ corresponds to the effective range of the Yukawa-like interaction \citep{2023MNRAS.520.4853M}. This screening mechanism reflects the expected behavior of reionization relics, which are coupled to the characteristic scales of reionization bubbles -- $\mathcal{O}({\rm a \ few \ Mpc})$ -- and are significantly suppressed at smaller scales. The coefficient $\beta_{\rm re}$ governs the large-scale shape of the template. Note that earlier reionization leads to smaller (larger) values of $A_{\rm re} \, {\rm and} \, \alpha_{\rm re}$ ($\beta_{\rm re}$). 

\begin{table}
    \centering
    \caption{Template parameters for the different reionization scenarios within the Yukawa model evaluated at $z_{\rm eff}$.}
    \begin{tabular}{c|ccc}
    \hline\hline
    Reionization model & $A_{\rm re}$ & $\alpha_{\rm re}$ & $\beta_{\rm re}$ \\
    \hline
    Yukawa early & 0.2088 & 0.7239 & 2.3656 \\
    Yukawa mid & 0.4775 & 1.5094 & 2.1774 \\
    Yukawa late & 1.5050 & 3.0237 & 1.8354 \\
    \hline\hline
    \end{tabular}
    \label{tab:yuk-par}
\end{table}

\section{Reionization templates: PySR vs Yukawa}
\label{app:pysr}
Here, we evaluate the performance of our two methodologies designed to account for the impact of reionization on the post-reionization \lya forest, i.e. the PySR template \citep[Eq.~\ref{eq:pysr-template};][]{igm_pysr}, and the Yukawa-like template \citep[Eq.~\ref{eq:yukawa};][]{2023MNRAS.520.4853M}. Our comparison begins with assessing the templates' ability to reproduce data from simulations. We focus on high-redshift data $(z=4)$, where the imprints of reionization in the \lya forest are stronger.

\begin{figure}
    \centering
    \includegraphics[width=\linewidth]{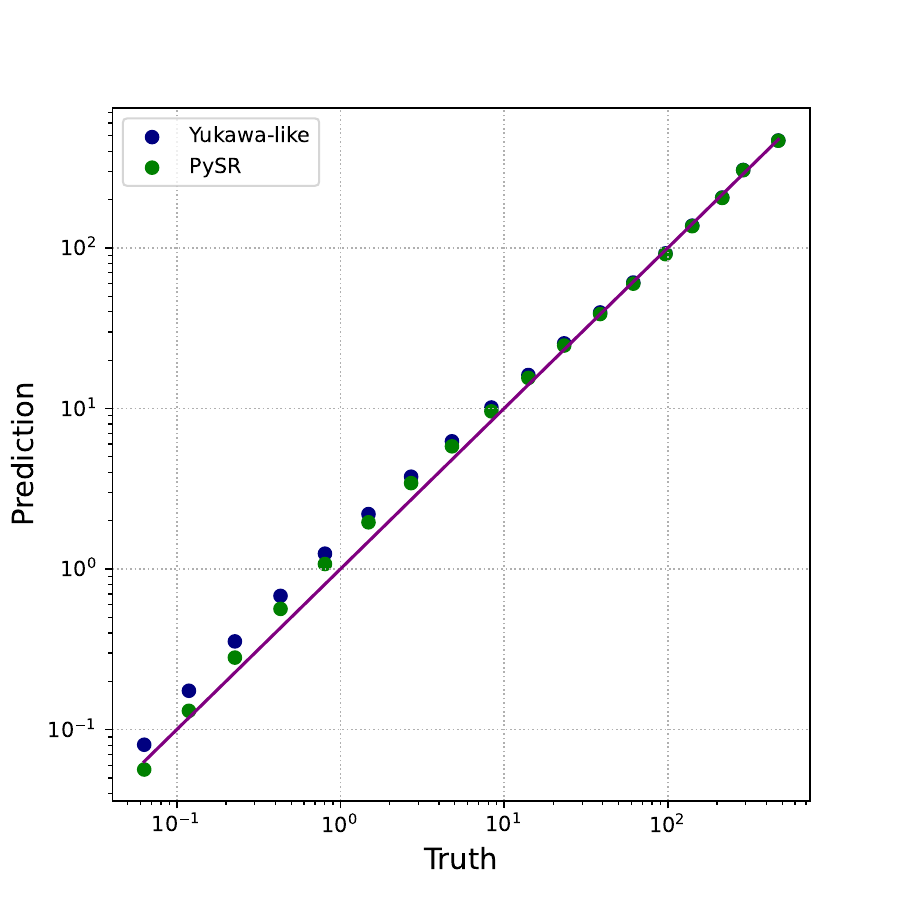}
    \caption{Performance comparison of the reionization templates: the symbolic regression-based template derived using {\sc PySR} (green dots) and the Yukawa-like template from \citealt{2023MNRAS.520.4853M} (blue dots). The purple line corresponds to the diagonal where predictions perfectly match the ground truth.}
    \label{fig:pysr-perf}
\end{figure}

Figure \ref{fig:pysr-perf} demonstrates the performance of Eq.~(\ref{eq:pysr-template}) in capturing the memory of reionization within the \lya forest 3D power spectrum  -- see \cite{igm_pysr} for further details. The symbolic regression-based template derived using {\sc PySR} outperforms the analytical Yukawa-like template in accuracy. Additionally, the PySR template achieves a marginally higher $R^{2}$ metric, with a value of 0.9982 compared to 0.9981 for the Yukawa-like template.

\begin{figure}
    \centering
    \includegraphics[width=\linewidth]{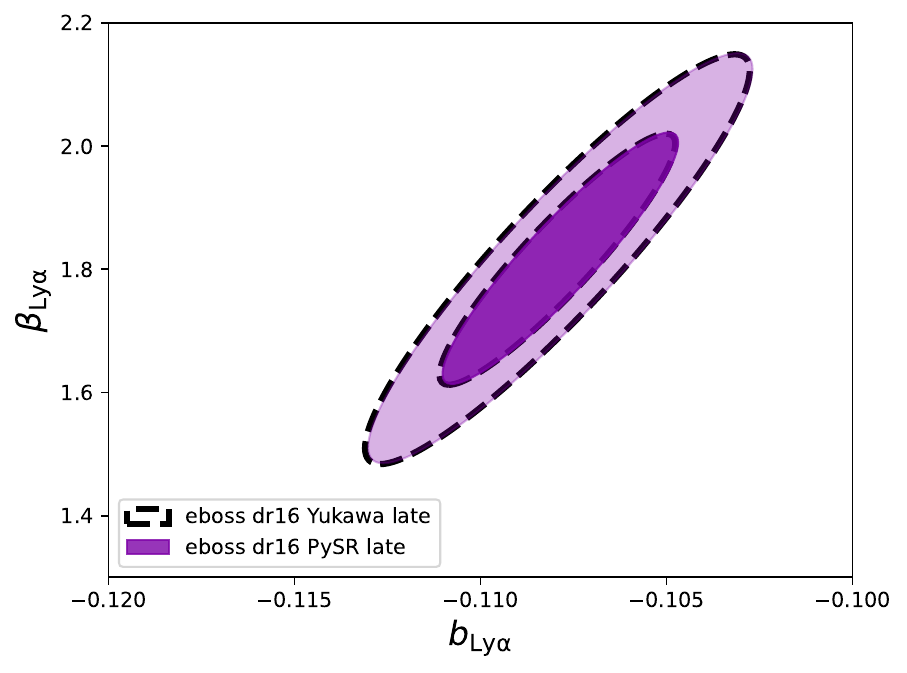}
    \caption{Contour plot of $\beta_{\rm Ly\alpha}$ and $b_{\rm Ly\alpha}$ for the Yukawa late model and the PySR late model. The near-complete overlap of the two contours suggests that the models yield largely equivalent results.}
    \label{fig:contour-pysryuk}
\end{figure}

Considering the small degree of deviation between both templates and their similarly high predictive accuracy, we anticipate negligible bias between models utilizing either template. This is a promising outcome, as the {\sc PySR}-based template achieves this performance without introducing additional degrees of freedom for fitting across different redshift bins -- as long as the redshift range is covered by the training data. 

To facilitate comparison between the two approaches applied to real data, we tabulate the results of the Yukawa analyses of the eBOSS DR16 \lya correlations in the last three columns of Table \ref{tab:freeA-yukawa(+)}. Comparing these results with those discussed in the main body of the paper reveals excellent agreement in the shifts of the BAO parameters between the two templates. Furthermore, these shifts remain consistent with the fiducial model without reionization relics, all within $1\sigma$. This consistency arises because the effects of reionization do not couple with the BAO scale, as illustrated in Figure \ref{fig:xi_comp}.

To verify that the different templates yield consistent results regarding the impact of reionization on non-BAO parameters that are more sensitive to such broadband perturbations, we plot the confidence ellipses for the \lya bias and the \lya RSD parameter in Figure \ref{fig:contour-pysryuk} for the Yukawa late and PySR late templates. Figure \ref{fig:contour-pysryuk} demonstrates the remarkable agreement between the two models.

\section{Yukawa template with one free parameter}
\label{app:yuk-free}
Our main analysis leverages reionization templates to model the lasting impact of reionization on the \lya forest. These templates are derived from a simulation suite \citep{2020MNRAS.499.1640M,2023MNRAS.520.4853M} encompassing three distinct reionization scenarios, centered on the reionization midpoint inferred from CMB data \citep{2020A&A...641A...6P}. To increase the flexibility of our templates and accommodate a broader range of reionization histories, we incorporate a Yukawa-like template (see Eq.~\ref{eq:yukawa}) in which the amplitude parameter $A_{\rm re}$ is allowed to vary, while the parameters $\alpha_{\rm re}$ and $\beta_{\rm re}$ are fixed to their respective values for each reionization scenario (described in Table \ref{tab:yuk-par}). Since $A_{\rm re}$ represents an amplitude of a power spectrum, we impose a physically motivated prior of $A_{\rm re} \geq 0$. To distinguish these models from the Yukawa templates with fixed amplitude values, we designate them with a ``(+)'' notation, e.g. Yukawa early(+). 

Besides enabling a broader exploration of reionization histories within the parameter space, the amplitude $A_{\rm re}$ is also intrinsically related to the underlying cosmology. For instance, higher values of the matter density $\Omega_{\rm m}$ will hasten the reionization process \citep{2024arXiv240513680M}, leading to lower values of $A_{\rm re}$. Thus, if the analysis of the data returned a best-fit result that deviates from expectations associated with earlier reionization, it may also indicate a preference for a particular region within the cosmological parameter space.

We present our results with the Yukawa template with one free parameter in Table \ref{tab:freeA-yukawa(+)}. The auto-correlation is best fitted by a model with $A_{\rm re}$ values roughly consistent with an early reionization scenario. Interestingly, allowing the amplitude to vary produces similar values of $\beta_{{\rm Ly}\alpha}$ and $b_{{\rm Ly}\alpha}$ across different reionization models, while simultaneously maintaining the previously observe trends: a decrease in $\alpha_\parallel$ and a positive shift in $\alpha_\perp$ relative to the no reionization model. The consistency in bias and RSD parameters may be interpreted as another sign of a compensatory mechanisms when using the other templates. Specifically, there is a decrease (in absolute value) in $b_{{\rm Ly}\alpha}$ from early to late reionization scenarios, resulting in reduced power in the \lya power spectrum. Meanwhile, $\beta_{{\rm Ly}\alpha}$ appears to counteract this trend by shifting in the opposite direction to modulate the overall strength, as shown in Figure \ref{fig:contour-pysr-fid}.

The inclusion of a free parameter in the reionization model can improve the fit of the cross-correlation, but this improvement is primarily due to effectively turning off the reionization model. Therefore, the cross-correlation between \lya and quasars still shows a preference for the fiducial model. As discussed in \S\ref{sec:re}, the cross-correlation is less constraining than the auto-correlation, and statistical fluctuations may drive this preference. More interestingly, this behavior may be indicative of imprecision in the modeling. Nonlinearities in the quasar modeling could play a significant role \citep{2022JCAP...09..070G}, but it is also possible that our assumption of no reionization imprints in the quasar model is incorrect. The assumption stems from the expectation for the baryonic modulation effect of reionization in shallow potential wells \citep{2023MNRAS.520..948L,2025MNRAS.536.1645M}. However, quasars likely form in denser environments, which could make them more resilient to reionization effects. Another intriguing possibility is that quasars may be better described by a different cosmological model. Table 9 of \cite{2020ApJ...901..153D} hints that the cross-correlation in eBOSS DR16 is better fit by an alternative cosmology. Nevertheless, the most plausible explanation for these trends is that they originate from the weaker constraining power.

For the combination of all correlations, the best fit gives $A_{\rm re} \approx 0.19$, which is in good agreement with the value corresponding to our early reionization model and consistent with the $1\sigma$ range from \cite{2020A&A...641A...6P}. The free $A_{\rm re}$ templates also recover the same trends for the shifts in the BAO, bias, and RSD parameters relative to the fiducial model. Notably, both the Yukawa mid(+) and Yukawa late(+) models show a systematic preference for lower values of $A_{\rm re}$, and thus signaling a preference for reionization to occur earlier across the board. 

\begin{table*}
    \centering
    \caption{Similar to Table \ref{tab:uber}, this table presents the best-fit parameters for the auto-correlation function of the \lya forest, the cross-correlation between the forest and quasars, and the combination of all four correlations. For each correlation, an additional row displays the best-fit value of $A_{\rm re}$, which represents the amplitude of Yukawa-like template. This parameter $A_{\rm re}$ is treated as a free parameter that is allowed to vary and be fitted in the models marked with a ``(+)'' notation, or otherwise is fixed. We show the results for the three different reionization scenarios.}
    \begin{tabular}{l|l|l|l|l|l|l}
    \hline
    \hline
    Parameter &  Yukawa early(+) & Yukawa mid(+) & Yukawa late(+) & Yukawa early & Yukawa mid & Yukawa late \\
    \hline
    \multicolumn{7}{c}{\lya auto-correlation}\\
    \hline
    % \midrule
    Prob. & 0.334 & 0.327 & 0.313 & 0.335 & 0.333 & 0.318\\
    $\chi^2_{\rm min}$ & 1599.49 & 1600.63 & 1602.85 & 1600.33 & 1600.66 & 1603.12\\
    $\alpha_\parallel$ & 1.0089 $\pm$ 0.0441 & 1.0111 $\pm$ 0.0440 & 1.0173 $\pm$ 0.0426 & 1.0234 $\pm$ 0.0378 & 1.0143 $\pm$ 0.0400 & 1.0081 $\pm$ 0.0405 \\
    $\alpha_\perp$ & 1.0076 $\pm$ 0.0483 & 1.0050 $\pm$ 0.0483 & 0.9991 $\pm$ 0.0495 & 0.9933 $\pm$ 0.0486 & 1.0021 $\pm$ 0.0463 & 1.0073 $\pm$ 0.0452 \\
    $\beta_{\rm{Ly}\alpha}$ & 1.7641 $\pm$ 0.1100 & 1.7734 $\pm$ 0.1141 & 1.7785 $\pm$ 0.1223 & 1.7200 $\pm$ 0.0942 & 1.7634 $\pm$ 0.0980 & 1.8164 $\pm$ 0.1027\\
    $b_{\rm{Ly\alpha}}$ & -0.1106 $\pm$ 0.0022 & -0.1102 $\pm$ 0.0022 & -0.1100 $\pm$ 0.0022 & -0.1132 $\pm$ 0.0022 & -0.1107 $\pm$ 0.0021 & -0.1079 $\pm$ 0.0021\\
    $b_{\eta,\rm{Ly\alpha}}$ & -0.2011 $\pm$ 0.0039 & -0.2013 $\pm$ 0.0039 & -0.2016 $\pm$ 0.0040 & -0.2006 $\pm$ 0.0039 & -0.2012 $\pm$ 0.0039 & -0.2020 $\pm$ 0.0039\\
    $10^3 b_{\eta, \rm{CIV(eff)}}$ & -5.1 $\pm$ 2.6 & -5.1 $\pm$ 2.6 & -5.2 $\pm$ 2.6 & -5.2 $\pm$ 2.6 & -5.1 $\pm$ 2.6 & -5.2 $\pm$ 2.6\\
    $10^3 b_{\eta, \rm{SiII(119)}}$ & -2.8 $\pm$ 0.5 & -2.8 $\pm$ 0.5 & -2.8 $\pm$ 0.5 & -2.9 $\pm$ 0.5 & -2.8 $\pm$ 0.5 & -2.8 $\pm$ 0.5\\ 
    $10^3 b_{\eta, \rm{SiII(119.3)}}$ & -1.8 $\pm$ 0.5 & -1.7 $\pm$ 0.5 & -1.7 $\pm$ 0.5 & -1.7 $\pm$ 0.5 & -1.7 $\pm$ 0.5 & -1.7 $\pm$ 0.5\\ 
    $10^3 b_{\eta, \rm{SiII(126)}}$ & -2.1 $\pm$ 0.6 & -2.1 $\pm$ 0.6 & -2.1 $\pm$ 0.6 & -2.1 $\pm$ 0.6 & -2.1 $\pm$ 0.6 & -2.1 $\pm$ 0.6\\ 
    $10^3 b_{\eta, \rm{SiII(120.7)}}$ & -5.0 $\pm$ 0.6 & -5.0 $\pm$ 0.6 & -5.0 $\pm$ 0.6 & -4.8 $\pm$ 0.5 & -5.0 $\pm$ 0.5 & -5.2 $\pm$ 0.5\\
    $b_{\rm HCD}$ & -0.0443 $\pm$ 0.0053 & -0.0455 $\pm$ 0.0052 & -0.0477 $\pm$ 0.0049 & -0.0468 $\pm$ 0.0045 & -0.0459 $\pm$ 0.0045 & -0.0467 $\pm$ 0.0045\\
    $\beta_{\rm HCD}$ & 0.5894 $\pm$ 0.0847 & 0.5937 $\pm$ 0.0844 & 0.6010 $\pm$ 0.0838 & 0.5953 $\pm$ 0.0840 & 0.5937 $\pm$ 0.0844 & 0.6010 $\pm$ 0.0838\\
    $10^2 A_{\rm sky,auto}$ & 0.94 $\pm$ 0.6 & 0.94 $\pm$ 0.6 & 0.94 $\pm$ 0.6 & 0.94 $\pm$ 0.6 & 0.94 $\pm$ 0.6 & 0.94 $\pm$ 0.6\\
    $\sigma_{\rm sky,auto}$ & 31.97 $\pm$ 1.70 & 31.93 $\pm$ 1.70 & 31.85 $\pm$ 1.70 & 31.82 $\pm$ 1.69 & 31.91 $\pm$ 1.69 & 31.94 $\pm$ 1.69 \\
    $A_{\rm re}$ & 0.3097 $\pm$ 0.1136 & 0.5122 $\pm$ 0.1998 & 1.2100 $\pm$ 0.5496 & 0.2088 & 0.4775 & 1.5050\\
    \hline
    \multicolumn{7}{c}{\lya $\times$ quasar cross-correlation}\\
    \hline
    Prob. & 0.196 & 0.196 & 0.196 & 0.183 & 0.175 & 0.168\\
    $\chi^2_{\rm min}$ & 3236.92 & 3236.92 & 3236.92 & 3241.77 & 3244.34 & 3246.51\\
    $\alpha_\parallel$ & 1.0617 $\pm$ 0.0321 & 1.0617 $\pm$ 0.0321 & 1.0616 $\pm$ 0.0321 & 1.0537 $\pm$ 0.0343 & 1.0505 $\pm$ 0.0350 & 1.0484 $\pm$ 0.0354 \\
    $\alpha_\perp$ & 0.9289 $\pm$ 0.0381 & 0.9290 $\pm$ 0.0381 & 0.9290 $\pm$ 0.0381 & 0.9296 $\pm$ 0.0419 & 0.9328 $\pm$ 0.0433 & 0.9366 $\pm$ 0.0447 \\
    $\beta_{\rm{Ly}\alpha}$ & 1.9313 $\pm$ 0.1446 & 1.9312 $\pm$ 0.1445 & 1.9312 $\pm$ 0.1446 & 2.1325 $\pm$ 0.1681 & 2.2238 $\pm$ 0.1762 & 2.3146 $\pm$ 0.1853\\
    $b_{\rm{Ly\alpha}}$ & -0.1127 $\pm$ 0.0052 & -0.1127 $\pm$ 0.0052 & -0.1127 $\pm$ 0.0052 & -0.1027 $\pm$ 0.0049 & -0.0993 $\pm$ 0.0047 & -0.0969 $\pm$ 0.0045\\
    $b_{\eta,\rm{Ly\alpha}}$ & -0.2242 $\pm$ 0.0104 & -0.2242 $\pm$ 0.0104 & -0.2242 $\pm$ 0.0104 & -0.2257 $\pm$ 0.0107 & -0.2276 $\pm$ 0.0107 & -0.2310 $\pm$ 0.0108\\
    $10^3 b_{\eta, \rm{CIV(eff)}}$ & -4.8 & -4.8 & -4.8 & -4.8 & -4.8 & -4.8\\
    $10^3 b_{\eta, \rm{SiII(119)}}$ & -4.7 $\pm$ 1.2 & -4.7 $\pm$ 1.2 & -4.7 $\pm$ 1.2 & -4.6 $\pm$ 1.2 & -4.6 $\pm$ 1.2 & -4.6 $\pm$ 1.2\\ 
    $10^3 b_{\eta, \rm{SiII(119.3)}}$ & 2.2 $\pm$ 1.1 & 2.2 $\pm$ 1.1 & 2.2 $\pm$ 1.1 & 2.2 $\pm$ 1.2 & 2.2 $\pm$ 1.2 & 2.2 $\pm$ 1.2\\ 
    $10^3 b_{\eta, \rm{SiII(126)}}$ & -1.9 $\pm$ 0.8 & -1.9 $\pm$ 0.8 & -1.9 $\pm$ 0.8 & -1.8 $\pm$ 0.8 & -1.8 $\pm$ 0.8 & -1.8 $\pm$ 0.8\\ 
    $10^3 b_{\eta, \rm{SiII(120.7)}}$ & -1.0 $\pm$ 1.0 & -1.0 $\pm$ 1.0 & -1.0 $\pm$ 1.0 & -1.0 $\pm$ 1.0 & -1.0 $\pm$ 1.0 & -1.0 $\pm$ 1.0\\
    $b_{\rm HCD}$ & -0.0501 & -0.0501 & -0.0501 & -0.0501 & -0.0501 & -0.0501\\
    $\beta_{\rm HCD}$ & 0.7031 & 0.7031 & 0.7031 & 0.7031 & 0.7031 & 0.7031\\
    $\beta_{\rm QSO}$ & 0.2602 & 0.2602 & 0.2602 & 0.2602 & 0.2602 & 0.2602\\
    $\xi_0^{\rm TP}$ & 0.7386 & 0.7386 & 0.7386 & 0.7386 & 0.7386 & 0.7386 \\
    $\Delta r_{\rm \parallel, QSO}(h^{-1} \rm{Mpc})$ & 0.2255 $\pm$ 0.1256 & 0.2263 $\pm$ 0.1256 & 0.2264 $\pm$ 0.1256 & 0.2256 $\pm$ 0.1258 & 0.2244 $\pm$ 0.1259 & 0.2218 $\pm$ 0.1259 \\
    $\sigma_{\rm \nu}$ & 7.7006 $ \pm$ 0.4484 & 7.7003 $\pm$ 0.4484 & 7.6997 $\pm$ 0.4484 & 7.9234 $\pm$ 0.4823 & 8.0635 $\pm$ 0.4937 & 8.2599 $\pm$ 0.5079 \\
    $A_{\rm re}$ & 0.0000 $\pm$ 0.0700 & 0.0000 $\pm$ 0.1049 & 0.0000 $\pm$ 0.2319 & 0.2088 & 0.4775 & 1.5050 \\
    \hline
    \multicolumn{7}{c}{All combined}\\
    \hline
    Prob. & 0.198 & 0.194 & 0.188 & 0.200 & 0.191 & 0.176\\
    $\chi^2_{\rm min}$ & 9637.11 & 9639.07 & 9642.20 & 9637.18 & 9641.28 & 9649.43\\
    $\alpha_\parallel$ & 1.0333 $\pm$ 0.0239 & 1.0348 $\pm$ 0.0236 & 1.0382 $\pm$ 0.0231 & 1.0318 $\pm$ 0.0235 & 1.0262 $\pm$ 0.0240 & 1.0222 $\pm$ 0.0245 \\
    $\alpha_\perp$ & 0.9560 $\pm$ 0.0320 & 0.9558 $\pm$ 0.0314 & 0.9555 $\pm$ 0.0303 & 0.9569 $\pm$ 0.0321 & 0.9621 $\pm$ 0.0328 & 0.9675 $\pm$ 0.0335 \\
    $\beta_{\rm{Ly}\alpha}$ & 1.7619 $\pm$ 0.0858 & 1.7586 $\pm$ 0.0873 & 1.7395 $\pm$ 0.0891 & 1.7722 $\pm$ 0.0783 & 1.8241 $\pm$ 0.0821 & 1.8841 $\pm$ 0.0865\\
    $b_{\rm{Ly\alpha}}$ & -0.1114 $\pm$ 0.0018 & -0.1116 $\pm$ 0.0018 & -0.1127 $\pm$ 0.0018 & -0.1108 $\pm$ 0.0018 & -0.1080 $\pm$ 0.0017 & -0.1050 $\pm$ 0.0017\\
    $b_{\eta,\rm{Ly\alpha}}$ & -0.2022 $\pm$ 0.0032 & -0.2022 $\pm$ 0.0032 & -0.2020 $\pm$ 0.0033 & -0.2024 $\pm$ 0.0032 & -0.2031 $\pm$ 0.0032 & -0.2039 $\pm$ 0.0032\\
    $10^3 b_{\eta, \rm{CIV(eff)}}$ & -4.8 $\pm$ 2.6 & -4.9 $\pm$ 2.6 & -4.9 $\pm$ 2.6 & -4.8 $\pm$ 2.6 & -4.8 $\pm$ 2.6 & -4.8 $\pm$ 2.6\\
    $10^3 b_{\eta, \rm{SiII(119)}}$ & -2.6 $\pm$ 0.4 & -2.6 $\pm$ 0.4 & -2.6 $\pm$ 0.4 & -2.6 $\pm$ 0.4 & -2.5 $\pm$ 0.4 & -2.5 $\pm$ 0.4\\ 
    $10^3 b_{\eta, \rm{SiII(119.3)}}$ & -1.0 $\pm$ 0.4 & -1.0 $\pm$ 0.4 & -1.0 $\pm$ 0.4 & -1.0 $\pm$ 0.4 & -1.0 $\pm$ 0.4 & -1.0 $\pm$ 0.4\\ 
    $10^3 b_{\eta, \rm{SiII(126)}}$ & -2.2 $\pm$ 0.4 & -2.2 $\pm$ 0.4 & -2.2 $\pm$ 0.4 & -2.2 $\pm$ 0.4 & -2.2 $\pm$ 0.4 & -2.2 $\pm$ 0.4\\ 
    $10^3 b_{\eta, \rm{SiII(120.7)}}$ & -3.9 $\pm$ 0.4 & -3.9 $\pm$ 0.4 & -3.9 $\pm$ 0.4 & -4.0 $\pm$ 0.4 & -4.1 $\pm$ 0.4 & -4.3 $\pm$ 0.4\\
    $b_{\rm HCD}$ & -0.0464 $\pm$ 0.0039 & -0.0473 $\pm$ 0.0038 & -0.0488 $\pm$ 0.0037 & -0.0460 $\pm$ 0.0036 & -0.0455 $\pm$ 0.0036 & -0.0464 $\pm$ 0.0036\\
    $\beta_{\rm HCD}$ & 0.6926 $\pm$ 0.0808 & 0.6972 $\pm$ 0.0805 & 0.7040 $\pm$ 0.0800 & 0.6908 $\pm$ 0.0808 & 0.6911 $\pm$ 0.0809 & 0.6994 $\pm$ 0.0807\\
    $\beta_{\rm QSO}$ & 0.2611 $\pm$ 0.0059 & 0.2607 $\pm$ 0.0059 & 0.2601 $\pm$ 0.0059 & 0.2613 $\pm$ 0.0059 & 0.2618 $\pm$ 0.0059 & 0.2617 $\pm$ 0.0059\\
    $10^2 A_{\rm sky,auto}$ & 0.93 $\pm$ 0.06 & 0.93 $\pm$ 0.06 & 0.93 $\pm$ 0.06 & 0.93 $\pm$ 0.06 & 0.92 $\pm$ 0.06 & 0.92 $\pm$ 0.06 \\
    $10^2 A_{\rm sky,autoLyb}$ & 1.32 $\pm$ 0.09 & 1.32 $\pm$ 0.09 & 1.32 $\pm$ 0.09 & 1.31 $\pm$ 0.09 & 1.31 $\pm$ 0.09 & 1.31 $\pm$ 0.09 \\
    $\sigma_{\rm sky,auto}$ & 31.83 $\pm$ 1.71 & 31.78 $\pm$ 1.70 & 31.64 $\pm$ 1.70 & 31.88 $\pm$ 1.70 & 31.98 $\pm$ 1.70 & 32.00 $\pm$ 1.70 \\
    $\sigma_{\rm sky,autoLyb}$ & 34.54 $\pm$ 2.32 & 34.50 $\pm$ 2.32 & 34.40 $\pm$ 2.31 & 34.57 $\pm$ 2.32 & 34.64 $\pm$ 2.32 & 34.64 $\pm$ 2.32 \\
    $\xi_0^{\rm TP}$ & 0.7317 $\pm$ 0.0922 & 0.7367 $\pm$ 0.0922 & 0.7460 $\pm$ 0.0923 & 0.7288 $\pm$ 0.0915 & 0.7217 $\pm$ 0.0913 & 0.7226 $\pm$ 0.0912 \\
    $\Delta r_{\rm \parallel, QSO}(h^{-1} \rm{Mpc})$ & 0.0896 $\pm$ 0.1091 & 0.0899 $\pm$ 0.1091 & 0.0938 $\pm$ 0.1091 & 0.0875 $\pm$ 0.1089 & 0.0781 $\pm$ 0.1089 & 0.0680 $\pm$ 0.1089 \\
    $\sigma_{\rm \nu}$ & 7.0055 $\pm$ 0.2898 & 6.9860 $\pm$ 0.2890 & 6.9367 $\pm$ 0.2864 & 7.0257 $\pm$ 0.2826 & 7.0972 $\pm$ 0.2879 & 7.1537 $\pm$ 0.2918 \\
    $A_{\rm re}$ & 0.1898 $\pm$ 0.0707 &0.2899 $\pm$ 0.1238 & 0.5501 $\pm$ 0.3413 & 0.2088 & 0.4775 & 1.5050 \\
    \hline
    \hline
    \end{tabular}
    \label{tab:freeA-yukawa(+)}
\end{table*}

%%%%%%%%%%%%%%%%%%%%%%%%%%%%%%%%%%%%%%%%%%%%%%%%%%

% Don't change these lines
\bsp	% typesetting comment
\label{lastpage}
\end{document}